\documentclass[12pt]{article}

\usepackage{amsmath}

\usepackage{url}

\def\C{{\rm\kern.24em \vrule width.02em
 height1.4ex depth-.05ex
\kern-.26em C}}
\def\R{{\rm I\kern-.20em R}}
\def\Z{{\rm\kern.26em \vrule width.02em height0.5ex depth0ex
 \kern.04em
 \vrule  width.02em height1.47ex depth-1ex \kern-.34em Z}}
\def\N{{\rm I\kern-.20em N}}
\def\Q{{\rm\kern.24em \vrule width.02em
 height1.4ex depth-.05ex \kern-.26em Q}}
\def\beq{\begin{equation}}
\def\eeq{\end{equation}}
\def\beqa{\begin{eqnarray}}
\def\eeqa{\end{eqnarray}}

\newcommand{\D}{\ensuremath{\mathsf{D}}}
\newcommand{\Id}{\ensuremath{\mathsf{I}}}

\newcommand{\bq}{\begin{equation}}
\newcommand{\eq}{\end{equation}}
\begin{document}
 \begin{titlepage}
 .
 \vskip 3.5cm
 \begin{center}
  {\bf \Large Direct Methods and Symbolic Software for} \\
  {\bf \Large Conservation Laws of Nonlinear Equations} \\
\vskip 1.3cm
          Willy Hereman, Paul J.\ Adams, Holly L.\ Eklund \\
          Department of Mathematical and Computer Sciences \\
          Colorado School of Mines \\
          Golden, CO 80401-1887, U.S.A. \\
\vskip 1cm
          Mark S.\ Hickman \\
          Department of Mathematics and Statistics \\
          University of Canterbury, Private Bag 4800 \\
          8140 Christchurch, New Zealand \\
\vskip 1cm
          Barend M.\ Herbst \\
          Department of Mathematical Sciences \\
          Applied Mathematics Division \\
          General Engineering Building \\
          University of Stellenbosch \\
          Private Bag X1, 7602 Stellenbosch, South Africa. \\
\end{center}
\vskip 0.3cm
 \begin{center}
 {\sc In Memory of Martin D.\ Kruskal (1925-2006)}
 \end{center}
\vskip 1cm
\begin{center}
{\rm Research Supported in Part by
the South African National Research Foundation (NRF)
under Grant No.\ FA2007032500003.} \\
\vskip 1cm
\end{center}
\begin{center}
Manuscript prepared for Advances in Nonlinear Waves and Symbolic Computation \\
Ed.: Zhenya Yan, Nova Science Publishers, New York, U.S.A.\ (2008).
\end{center}
\begin{center}
2000 Mathematics Subject Classification \\
Primary: 37K05, 37J35, 37K10; 
Secondary: 37Q35, 35Q58, 37K40, 37K60
\end{center}
\end{titlepage}
\vfill
\newpage
%
%
\section*{Abstract}
We present direct methods, algorithms, and symbolic software for the
computation of conservation laws of nonlinear partial differential
equations (PDEs) and differential-difference equations (DDEs).

Our method for PDEs is based on calculus, linear algebra, and variational
calculus.
First, we compute the dilation symmetries of the given nonlinear system.
Next, we build a candidate density as a linear combination with undetermined
coefficients of terms that are scaling invariant.
The variational derivative (Euler operator) is used to derive a linear
system for the undetermined coefficients.
This system is then analyzed and solved.
Finally, we compute the flux with the homotopy operator.

The method is applied to nonlinear PDEs in $(1+1)$ dimensions
with polynomial nonlinearities which include the Korteweg-de Vries (KdV), 
Boussinesq, and Drinfel'd-Sokolov-Wilson equations.
An adaptation of the method is applied to PDEs with transcendental 
nonlinearities.
Examples include the sine-Gordon, sinh-Gordon, and Liouville equations.
For equations in laboratory coordinates, the coefficients of the
candidate density are undetermined functions which must satisfy 
a mixed linear system of algebraic and ordinary differential equations.

For the computation of conservation laws of nonlinear DDEs we use a 
splitting of the identity operator. 
This method is more efficient that an approach based on the discrete Euler 
and homotopy operators. 
We apply the method of undetermined coefficients to the Kac-van Moerbeke,
Toda, and Ablowitz-Ladik lattices.
To overcome the shortcomings of the undetermined coefficient technique, 
we designed a new method that first calculates the leading order term and
then the required terms of lower order.   
That method, which is no longer restricted to polynomial conservation laws, 
is applied to discretizations of the KdV and modified KdV equations, 
and a combination thereof. 
Additional examples include lattices due to Bogoyavlenskii,
Belov-Chaltikian, and Blaszak-Marciniak.

The undetermined coefficient methods for PDEs and DDEs have been implemented
in {\tt Mathematica}.
The code {\tt TransPDEDensityFlux.m} computes densities and fluxes of
systems of PDEs with or without transcendental nonlinearities.
The code {\tt DDEDensityFlux.m} does the same for polynomial nonlinear DDEs.
Starting from the leading order terms, the new {\tt Maple} library 
{\tt discrete} computes densities and fluxes of nonlinear DDEs.

The software can be used to answer integrability questions and to gain
insight in the physical and mathematical properties of nonlinear models.
When applied to nonlinear systems with parameters, the software computes
the conditions on the parameters for conservation laws to exist.
The existence of a hierarchy of conservation laws is a predictor for
complete integrability of the system and its solvability with the
Inverse Scattering Transform.
%
\tableofcontents
%
%
\vspace{-3mm}
\section{Introduction}
\label{introduction}
This chapter focuses on symbolic methods to compute polynomial conservation
laws of partial differential equations (PDEs) in $(1+1)$ dimensions and
differential-difference equations (DDEs), which are semi-discrete lattices.
For the latter we treat systems where time is continuous and the spatial
variable has been discretized.

Nonlinear PDEs that admit conservation laws arise in many disciplines 
of the applied sciences including physical chemistry, fluid mechanics, 
particle and quantum physics, plasma physics, elasticity, gas dynamics, 
electromagnetism, magneto-hydro-dynamics, nonlinear optics, 
and the bio-sciences.
Conservation laws are fundamental laws of physics that maintain that a 
certain quantity will not change in time during physical processes.
Familiar conservation laws include conservation of momentum, mass (matter),
electric charge, or energy.
The continuity equation of electromagnetic theory is an example of a 
conservation law which relates charge to current. 
In fluid dynamics, the continuity equation expresses conservation of mass, 
and in quantum mechanics the conservation of probability of the density and 
flux functions also yields a continuity equation. 

There are many reasons to compute conserved densities and fluxes of PDEs
explicitly.
Invariants often lead to new discoveries as was the case in soliton theory.
One may want to verify if conserved quantities of physical importance
(e.g.\ momentum, energy, Hamiltonians, entropy, density, charge) are
intact after constitutive relations have been added to close a system.
For PDEs with arbitrary parameters one may wish to compute conditions on the
parameters so that the model admits conserved quantities.
Conserved densities also facilitate the study of qualitative properties
of PDEs \cite{JSandJW1997,JWphdthesis1998}, such as recursion operators,
bi- or tri-Hamiltonian structures, and the like.
They often guide the choice of solution methods or reveal the nature of
special solutions.
For example, an infinite sequence of conserved densities is a predictor of 
the existence of solitons \cite{MAandHSbook1981} and complete integrability 
\cite{MAandPCbook1991} which means that the PDE can be solved with the 
Inverse Scattering Transform (IST) method \cite{MAandPCbook1991}.

Conserved densities aid in the design of numerical solvers for PDEs 
\cite{JSjcp1982,JSnm1997} and their stability analysis (see references in
\cite{ACcpc2007}).
Indeed, semi-discretizations that conserve discrete conserved quantities 
lead to {\em stable} numerical schemes that are free of nonlinear 
instabilities and blowup.
While solving DDEs, which arise in nonlinear networks and as 
semi-discretizations of PDEs, one should check that their conserved 
quantities indeed remain unchanged as time steps are taken.

Computer algebra systems (CAS) like {\tt Mathematica}, {\tt Maple}, and
{\tt REDUCE}, can greatly assist the computation of conservation laws of
nonlinear PDEs and DDEs.
Using CAS interactively, one can make a judicious guess (ansatz) and find 
a few simple densities and fluxes.
Yet, that approach is fruitless for complicated systems with nontrivial
conservation laws with increasing complexity.
Furthermore, completely integrable equations PDEs 
\cite{MAandPCbook1991,MAandHSbook1981,AMetalinbook1991,VSandASinbook1984} 
and DDEs \cite{VAetaltmp2000,AMetal1987} admit infinitely many independent 
conservation laws.
Computing them is a challenging task.
It involves tedious computations which are prone to error if done with pen
and paper.
Kruskal and collaborators demonstrated the complexities of calculating 
conservation laws in their seminal papers 
\cite{MKetalVjmp1970,RMSIAMRev1976,RMetalIIjmp1968} on the Korteweg-de Vries 
(KdV) equation from soliton theory 
\cite{MAandPCbook1991,MAandHSbook1981,PDandRJbook1989}. 
We use this historical example to introduce the method of undetermined 
coefficients.

In the first part of this chapter we cover the symbolic computation of
conservation laws of completely integrable PDEs in $(1+1)$ dimensions
(with independent variables $x$ and $t).$ 
Our approach \cite{PAthesis2003,UGandWHjsc1997,WHetalbirkhauser2005,WHandUGwester1999}
uses the concept of dilation (scaling) invariance and the method of 
undetermined coefficients. 
Our method proceeds as follows. 
First, build a candidate density as a linear combination 
(with undetermined coefficients) of ``building blocks'' 
that are homogeneous under the scaling symmetry of the PDE.
If no such symmetry exists, construct one by introducing parameters 
with scaling.
Next, use the Euler operator (variational derivative) to derive a linear 
algebraic system for the undetermined coefficients.
After the system is analyzed and solved, use the homotopy operator to
compute the flux.
When applied to systems with parameters, our codes can determine the
conditions on the parameters so that a sequence of conserved densities
exists.

The method is applied to nonlinear PDEs in $(1+1)$ dimensions
with polynomial terms which include the KdV, Boussinesq, and 
Drinfel'd-Sokolov-Wilson equations.
An adaptation of the method is applied to PDEs with transcendental 
nonlinearities.
Examples include the sine-Gordon, sinh-Gordon, and Liouville equations.
For equations written in laboratory coordinates, the coefficients of the
candidate density are undetermined functions which must satisfy a mixed  
linear system of algebraic and ordinary differential equations (ODEs).

Capitalizing on the analogy between PDEs and DDEs, the second part of this 
chapter deals with the symbolic computation of conservation laws of
nonlinear DDEs
\cite{HEthesis2003,UGandWHpd1998,UGetalpla1997,WHetalbirkhauser2005,WHetalcrm2004,MHandWHprsa2003}.
Again, we use scaling symmetries and the method of undetermined coefficients.
One could use discrete versions of the Euler operator (to verify exactness)
and the homotopy operator (to invert the forward difference).
Although these operators might be valuable in theory, they are highly 
inefficient as tools for the symbolic computation of conservation laws of DDEs.
We advocate the use of a ``splitting and shifting" technique, which allows
us to compute densities and fluxes simultaneously at minimal cost.
The undetermined coefficient method for DDEs is illustrated with the 
Kac-van Moerbeke, Toda, and Ablowitz-Ladik lattices.

There is a fundamental difference between the continuous and discrete cases
in the way densities are constructed.
The total derivative has a weight whereas the shift operator does not.
Consequently, a density of a PDE is bounded in order with respect 
to the space variable.
Unfortunately, there is no {\em a priori} bound on the number of shifts
in the density, unless a leading order analysis is carried out.
To overcome this difficulty and other shortcomings of the undetermined
coefficient method, we present a new method to compute conserved densities
of DDEs.
That method no longer uses dilation invariance and is no longer restricted
to polynomial conservation laws.
Instead of building a candidate density with undetermined coefficients,
one first computes the leading order term in the density and, secondly,
generates the required terms of lower order.
The method is fast and efficient since unnecessary terms are never computed.
The new method is illustrated using a modified Volterra lattice as an
example.
The new method performs exceedingly well when applied to lattices due to
Bogoyavlenskii, Belov-Chaltikian, and Blaszak-Marciniak.
The new method is also applied to completely integrable discretizations
of the KdV and modified KdV (mKdV) equations, and a combination thereof, 
known as the Gardner equation.
Starting from a discretized eigenvalue problem, we first derive the 
Gardner lattice and then compute conservation laws.

There are several methods (see \cite{WHetalbirkhauser2005}) to compute
conservation laws of nonlinear PDEs and DDEs.
Some methods use a generating function \cite{MAandPCbook1991,MAandHSbook1981},
which requires the knowledge of key pieces of the IST.
Another common approach uses the link between conservation laws and
symmetries as stated in Noether's theorem
\cite{IA1992,IAbook2004,IKandAVbook1998,PObook1993}.
However, the computation of generalized (variational) symmetries,
though algorithmic, is as daunting a task as the direct computation of
conservation laws.
Most of the more algorithmic methods
\cite{SAandGB2002a,SAandGB2002b,GBandSA2002,ACwebsite2008,AKandFM2000,TW2002},
require the solution of a determining system of ODEs or PDEs.
Despite their power, only a few of these methods have been implemented in CAS.
We devote a section to symbolic software for the computation of
conservation laws.
Additional reviews can be found in
\cite{UGandWHjsc1997,WHetalbirkhauser2005,TW2002}.

Over the past decade, in collaboration with students and researchers,
we have designed and implemented direct algorithms for the computation of
conservation laws of nonlinear PDEs and DDEs.
We purposely avoid Noether's theorem, pre-knowledge of symmetries, and a
Lagrangian formulation.
Neither do we use differential forms or advanced differential-geometric tools.
Instead, we concentrate on the undetermined coefficient method for PDEs
and DDEs, which uses tools from calculus, linear algebra, and the
variational calculus.
Therefore, the method is easy to implement in {\tt Mathematica}
and easy to use by scientists and engineers.
The code {\tt TransPDEDensityFlux.m} computes densities and fluxes of
systems of PDEs with or without transcendental nonlinearities.
The code {\tt DDEDensityFlux.m} does the same for polynomial nonlinear DDEs.
Starting from the leading order terms, the new {\tt Maple} library 
{\tt discrete} computes densities and fluxes of nonlinear DDEs very 
efficiently.
The software can thus be used to answer integrability questions and to gain
insight in the physical and mathematical properties of nonlinear models.

Our software is in the public domain.
The {\tt Mathematica} packages and notebooks are available at
\cite{WHwebsite2004} and Hickman's code in {\tt Maple} is available at 
\cite{HickmanDiscreteCode}.
We are currently working on a comprehensive package to compute conservation
laws of PDEs in multiple space dimensions
\cite{WHIJQC2006,WHetalbirkhauser2005,DPphdthesis2008}.
\vskip 15pt
\noindent
{\Large \bf Part I: Partial Differential Equations in $(1+1)$ Dimensions}
\vskip 8pt
\noindent
In this first part we cover PDEs in $(1+1)$ dimensions, that is, PDEs in one 
space variable and time. 
Starting from a historical example, we introduce the concept of dilation 
invariance and use the method of undetermined coefficients to compute 
conservation laws of evolution equations.
Later on, we adapt the method of undetermined coefficients to cover PDEs 
with transcendental terms.
\vspace{-3mm}
\section{The Most Famous Example in Historical Perspective}
\label{history}
The story of conservation laws for nonlinear PDEs begins with the discovery
of an infinite number of conservation laws of the ubiquitous
Korteweg-de Vries equation which models a variety of nonlinear wave
phenomena, including shallow water waves \cite{WHencyclop2008} 
and ion-acoustic waves in plasmas 
\cite{MAandPCbook1991,MAandHSbook1981,PDandRJbook1989}.
The KdV equation can be recast in dimensionless variables as
\begin{equation}
\label{kdv}
u_t + \alpha u u_x + u_{3x} = 0,
\end{equation}
where the subscripts denote partial derivatives, i.e.\
$u_t = \frac{\partial u}{\partial t}, u_x = \frac{\partial u}{\partial x},$
and $ u_{3x} = \frac{\partial^3 u}{\partial x^3}.$
The parameter $\alpha$ can be scaled to any real number.
Commonly used values are $\alpha = \pm 1$ or $\alpha = \pm 6.$

Equation (\ref{kdv}) is an example of a scalar $(1+1)-$dimensional
evolution equation,
\begin{equation}
\label{scalarpde}
u_t = F(x,t,u,u_x,u_{2x}, \cdots, u_{nx}),
\end{equation}
of order $n$ in the independent space variable $x$ and of first order
in time $t.$
Obviously, the dependent variable is $u(x,t).$
If parameters are present in (\ref{scalarpde}), they will be denoted by 
lower-case Greek letters.
A {\em conservation law} of (\ref{scalarpde}) is of the form
\begin{equation}
\label{conslawscalarpde}
\D_{t} \, \rho + \D_{x} \, J = 0,
\end{equation}
which is satisfied for all solutions $u(x,t)$ of the PDE.
In physics, $\rho$ is called the {\em conserved density\/} (or charge);
$J$ is the associated {\em flux\/} (or current).
In general, both are differential functions (functionals),
i.e.\ functions of $x, t, u,$ and partial derivatives of $u$ with
respect to $x.$
In (\ref{conslawscalarpde}), 
$\D_{x}$ denotes the total derivative with respect to $x,$ that is,
\begin{equation}
\label{kdvDx}
\D_x J = \frac{\partial J}{\partial x}
+ \sum_{k=0}^{N} \frac{\partial J}{\partial u_{kx}} u_{(k+1)x},
\end{equation}
where $N$ is the order of $J,$ and $\D_{t}$ is the total derivative with 
respect to $t,$ defined by 
\begin{equation}
\label{kdvDtrho3}
\D_t \, \rho = \frac{\partial \rho}{\partial t} + \rho^{\prime}[u_t]
= \frac{\partial \rho}{\partial t} +
\sum_{k=0}^{M} \frac{\partial \rho}{\partial u_{kx}} \D_x^k u_t,
\end{equation}
where $\rho^{\prime}[u_t]$ is the Fr\'echet derivative of
$\rho$ in the direction of $u_t$ and $M$ is the order of $\rho.$

The densities $\rho^{(1)} = u$ and $\rho^{(2)} = u^2$ of (\ref{kdv}) were 
long known.
In 1965, Whitham  \cite{GWprsa1965} had found a third density,
$\rho^{(3)} = u^3 - \frac{3}{\alpha} u_x^2,$ which, in the context of water
waves, corresponds to Boussinesq's moment of instability \cite{JMjfm1981}.
One can readily verify that
\begin{eqnarray}
\label{kdvconslaw1}
&& \D_{t}(u) + \D_{x} (\frac{1}{2} \alpha u^2 + u_{2x}) = 0, \\
\label{kdvconslaw2}
&& \D_{t}(u^2) +
\D_{x} (\frac{2}{3} \alpha u^3 - u_x^2 + 2 u u_{2x}) = 0, \\
\label{kdvconslaw3}
&& \D_{t}(u^3 - \frac{3}{\alpha} u_x^2) +
\D_{x} (\frac{3}{4} \alpha u^4 - 6 u u_x^2 + 3 u^2 u_{2x}
+ \frac{3}{\alpha} u_{2x}^2 - \frac{6}{\alpha} u_x u_{3x}) = 0.
\end{eqnarray}
Indeed, (\ref{kdvconslaw1}) is the KdV equation written as a conservation
law; (\ref{kdvconslaw2}) is obtained after multiplying (\ref{kdv}) by
$2 u;$ (\ref{kdvconslaw3}) requires more work.
Hence, the first three density-flux pairs of (\ref{kdv}) are
\begin{eqnarray}
\label{kdvrho1J1}
\rho^{(1)} \!&\!=\!&\! u, \quad \quad \quad \quad \;\;
J^{(1)} = \frac{1}{2} \alpha u^2 + u_{2x},  \\
\label{kdvrho1J2}
\rho^{(2)} \!&\!=\!&\! u^2, \quad \quad \quad \quad \;
J^{(2)} = \frac{2}{3} \alpha u^3 - u_x^2 + 2 u u_{2x}, \\
\label{kdvrho3J3}
\rho^{(3)} \!&\!=\!&\! u^3 - \frac{3}{\alpha} u_x^2, \quad
J^{(3)} = \frac{3}{4} \alpha u^4 - 6 u u_x^2 + 3 u^2 u_{2x}
+ \frac{3}{\alpha} u_{2x}^2 - \frac{6}{\alpha} u_x u_{3x}.
\end{eqnarray}
Integrals of motion readily follow from the densities.
Indeed, assuming that $J$ vanishes at infinity 
(for example due to sufficiently fast decay of $u$ and its $x$ derivatives), 
upon integration of (\ref{conslawscalarpde}) with respect to $x$ 
one obtains that
\begin{equation}
\label{conservedquantity}
P = \int_{-\infty}^{\infty} \rho \, dx
\end{equation}
is {\em constant} in time.
Such constants of motion also arise when $u$ is periodic, in which case one
integrates over the finite period.
Depending on the physical setting, the first few constants of motion
(i.e.\ integrals (\ref{conservedquantity})) express conservation of mass,
momentum, and energy.

Martin Kruskal and postdoctoral fellow Norman Zabusky discovered the fourth
and fifth densities for the KdV equation \cite{NZchaos2005}.
However, they failed in finding a sixth conservation law due to an algebraic
mistake in their computations.
Kruskal asked Robert Miura, also postdoctoral fellow at the Princeton
Plasma Physics Laboratory at New Jersey, to search for further conservation
laws of the KdV equation.
Miura \cite{RMSIAMRev1976} computed the seventh conservation law.
After correcting the mistake mentioned before, he also found the sixth
and eventually three additional conservation laws.
Rumor \cite{ANbook1985} has it that in the summer of 1966 Miura went up 
into the Canadian Rockies and returned from the mountains with the first 
10 conservation laws of the KdV equation engraved in his notebook.
This biblical metaphor probably does not do justice to Miura's intense and
tedious work with pen and paper.

With ten conservation laws in hand, it was conjectured that the KdV equation 
had an infinite sequence of conservation laws, later proven to be true 
\cite{MKetalVjmp1970,RMetalIIjmp1968}.
Aficionados of explicit formulas can find the first ten densities
(and seven of the associated fluxes) in \cite{RMetalIIjmp1968}
and the eleventh density (with 45 terms) in \cite{MKetalVjmp1970},
where a recursion formula is given to generate all further conserved
densities.
As an aside, in 1966 the first five conserved densities were computed on
an IBM 7094 computer with FORMAC, an early CAS.
The sixth density could no longer be computed because the available storage
space was exceeded.
In contrast, using a method of undetermined coefficients, the first eleven
densities were computed in 1969 on a AEC CDC-6600 computer in a record time
of 2.2 seconds.
Due to limitations in handling large integers, the computer could not
correctly produce any further densities.

Undoubtedly, the discovery of conservation laws played a pivotal role in
the comprehensive study of the properties and solutions of nonlinear
completely integrable PDEs (like the KdV equation) and the development of
the IST (see e.g.\ \cite{ANbook1985} for the history).
Clifford Gardner, John Greene, Martin Kruskal, and Robert Miura
received the 2006 Leroy P.\ Steele Prize \cite{NoticesAMS2006}, 
awarded by the American Mathematical Society, for their seminal 
contribution to research on the KdV equation.
In turn, Martin Kruskal has received numerous honors and awards 
\cite{rutgerswebsite2006} for his
fundamental contributions to the understanding of integrable systems and
soliton theory. 
This chapter is dedicated to Martin Kruskal (1925-2006).
\vspace{-3mm}
\section{The Method of Undetermined Coefficients}
\label{continuousmuc}
We now sketch the method of undetermined coefficients to compute conservation
laws \cite{UGandWHjsc1997,FVandWH1994}, which draws on ideas and
observations in before mentioned work by Kruskal and collaborators.
\vspace{-3mm}
\subsection{Dilation Invariance of Nonlinear PDEs}
\label{dilationinvariancePDEs}

Crucial to the computation of conservation laws is that
(\ref{conslawscalarpde}) must hold on the PDE.
This is achieved by substituting $u_t$ (and $u_{tx}, u_{txx},$ etc.)
from (\ref{kdv}) in the evaluation of
(\ref{kdvconslaw1})-(\ref{kdvconslaw3}) and in all subsequent conservation
laws of degree larger than 3.
The elimination of all $t-$derivatives of $u$ in favor of $x$ derivatives has
two important consequences:
(i)  any symmetry of the PDE, in particular, the dilation symmetry,
     will be adopted by the conservation law,
(ii) once $\D_t$ is computed and evaluated on the PDE,
     $t$ becomes a parameter in the computation of the flux.

We will first investigate the dilation (scaling) symmetry of evolution
equations.
The KdV equation is {\em dilation invariant} under the scaling symmetry
\begin{equation}
\label{kdvscale}
(t, x, u) \rightarrow ({\lambda}^{-3} t, \lambda^{-1} x, {\lambda}^{2} u),
\end{equation}
where $\lambda$ is an arbitrary parameter.
Indeed, after a change of variables with
$\tilde{t} = {\lambda}^{-3} t, \tilde{x} = {\lambda}^{-1} x,
\tilde{u} = {\lambda}^{2} u,$
and cancellation of a common factor $\lambda^5,$ the KdV for
$\tilde{u}(\tilde{x},\tilde{t})$ arises.
The dilation symmetry of (\ref{kdv}) can be expressed as
\begin{equation}
\label{kdvscale2}
u \sim \frac{\partial^2}{\partial x^2}, \qquad
\frac{\partial}{\partial t} \sim \frac{\partial^3}{\partial x^3},
\end{equation}
which means that $u$ corresponds to two $x-$derivatives and the time
derivative corresponds to three $x-$derivatives.
If we define the {\em weight}, $W,$ of a variable (or operator)
as the exponent of $\lambda$ in (\ref{kdvscale}), then
$W(x) = -1$ or $W(\frac{\partial}{\partial x}) = 1; W(t) = -3$ or
$W(\frac{\partial}{\partial t}) = 3,$ and $W(u) = 2.$

All weights of dependent variables and the weights of
${\partial/\partial x}, {\partial/\partial t},$ are assumed to be
non-negative and rational.
The {\em rank} of a monomial is defined as the total weight of the monomial.
Such monomials may involve the independent and dependent variables 
and the operators $\frac{\partial}{\partial x}, \D_x,
\frac{\partial}{\partial t},$ and $\D_t.$
Ranks must be positive integers or positive rational numbers.

An expression (or equation) is {\em uniform in rank} if its
monomial terms have equal rank.
For example, (\ref{kdv}) is uniform in rank since each of the three terms
has rank $5.$

Conversely, if one does not know the dilation symmetry of (\ref{kdv}),
then it can be readily computed by requiring that (\ref{kdv})
is uniform in rank.
Indeed, setting $W({\partial/\partial x}) = 1$ and equating the ranks of the
three terms in (\ref{kdv}) gives
\begin{equation}
\label{kdvweightequations}
W(u) + W(\frac{\partial}{\partial t}) = 2 W(u) + 1 = W(u) + 3,
\end{equation}
which yields
$W(u) = 2, W({\partial/\partial t}) = 3,$ and, in turn,
confirms (\ref{kdvscale}).
So, requiring {\em uniformity in rank} of a PDE allows one to compute the
weights of the variables (and thus the scaling symmetry) with linear algebra.

Dilation symmetries, which are special Lie-point symmetries, are common to
many nonlinear PDEs.
Needless to say, not every PDE is dilation invariant, but non-uniform PDEs
can be made uniform by extending the set of dependent variables with auxiliary
parameters with appropriate weights.
Upon completion of the computations one can set these parameters to one.
In what follows, we set $W({\partial/\partial x}) = W(\D_x) = 1$ and
$W({\partial/\partial t}) = W(\D_t).$
Applied to (\ref{kdvconslaw1}),
${\rm rank} \, \rho^{(1)} = 2$, ${\rm rank} \, J^{(1)} = 4.$
Hence,
${\rm rank} \, ( \D_t \, \rho^{(1)} )
= {\rm rank} \, ( \D_x \, J^{(1)} ) = 5.$
Therefore, (\ref{kdvconslaw1}) is uniform of rank $5.$
In (\ref{kdvconslaw2}), ${\rm rank} \, \rho^{(2)} = 4$ and
${\rm rank} \, J^{(2)} = 6,$ consequently, (\ref{kdvconslaw2}) is uniform
of rank $7.$

In (\ref{kdvrho3J3}), each term in $\rho^{(3)}$ has rank $6$ and each term
in $J^{(3)}$ has rank 8.
Consequently,
${\rm rank}\, ( \D_t \, \rho^{(3)} )
= {\rm rank}\, ( \D_x \, J^{(3)} ) = 9,$
which makes (\ref{kdvconslaw3}) is uniform of rank $9.$
All densities of (\ref{kdv}) are uniform in rank and so are the associated 
fluxes and the conservation laws.

Equation (\ref{kdv}) also has density-flux pairs that depend explicitly on 
$t$ and $x;$ for example,
\begin{equation}
\label{xtrhofluxkdv}
{\tilde \rho} = t u^2 + \frac{2}{\alpha} x u, \quad
{\tilde J} = t (\frac{2}{3} \alpha u^3 - u_x^2 + 2 u u_{2x} )
- x \left(u^2 - \frac{2}{\alpha} u_{2x} \right) + \frac{2}{\alpha} u_x.
\end{equation}
Since $W(x) = -1$ and $W(t) = -3,$ one has ${\rm rank}\, {\tilde \rho} = 1,$
and ${\rm rank}\, {\tilde J} = 3.$
The methods and algorithms discussed in subsequent sections have been adapted
to compute densities and fluxes explicitly dependent on $x$ and $t.$
Instead of addressing this issue in this chapter, we refer the reader to
\cite{UGandWHjsc1997,WHandUGwester1999}.
\vspace{-2mm}
\subsection{The Method of Undetermined Coefficients Applied to a Scalar 
Nonlinear PDE}
\label{mucPDEs}
We outline how densities and fluxes can be constructed for a scalar evolution 
equation (\ref{scalarpde}).
To keep matters transparent, we illustrate the steps for the KdV equation
resulting in $\rho^{(3)}$ of rank $R = 6$ with associated flux $J^{(3)}$
of rank $8,$ both listed in (\ref{kdvrho3J3}).
The tools needed for the computations will be presented in the next section.
\vskip 3pt
\noindent
$\bullet$
Select the rank $R$ of $\rho.$
Make a list, ${\cal R},$ of all monomials in $u$ and its $x$-derivatives
so that each monomial has rank $R.$
This can be done as follows.
Starting from the set ${\cal V}$ of dependent variables
(including parameters with weight, when applicable), make a set
${\cal M}$ of all non-constant monomials of rank $R$ or less
(but without $x-$derivatives).
Next, for each term in ${\cal M},$ introduce the right number of
$x-$derivatives to adjust the rank of that term.
Distribute the $x-$derivatives, strip off the numerical coefficients,
and gather the resulting terms in a set ${\cal R}.$
For the KdV equation and $R = 6$, ${\cal V} = \{ u \}$ and
${\cal M} = \{ u^3, u^2, u \}.$
Since $u^3, u^2,$ and $u$ have ranks $6, 4$ and $2,$ respectively, one
computes
\begin{equation}
\label{buildingblockskdv}
\frac{\partial^0 u^3}{\partial x^0} = u^3, \quad
\frac{\partial^2 u^2}{\partial x^2} = 2 u_x^2 + 2 u u_{2x}, \quad
\frac{\partial^4 u}{\partial x^4}   = u_{4x}.
\end{equation}
Ignoring numerical coefficients in the right hand sides of the equations in
(\ref{buildingblockskdv}), one gets
${\cal R} = \{ u^3, u_x^2, u u_{2x}, u_{4x} \}.$
\vskip 3pt
\noindent
$\bullet$
Remove from ${\cal R}$ all monomials that are total $x-$derivatives.
Also remove all ``equivalent'' monomials, i.e.\ the monomials that differ
from another by a total $x-$derivative, keeping the monomial of lowest order.
Call the resulting set ${\cal S}.$
In our example, $u_{4x}$ must be removed (because $u_{4x} = \D_x u_{3x})$
and $u u_{2x}$ must be removed since $u u_{2x}$ and $u_x^2$ are equivalent.
Indeed, $u u_{2x} = \D_x (u u_x) - u_x^2.$
Thus, ${\cal S} = \{ u^3, u_x^2 \}.$
\vskip 3pt
\noindent
$\bullet$
Linearly combine the monomials in ${\cal S}$ with constant undetermined
coefficients $c_i$ to obtain the candidate $\rho.$
Continuing with the example,
\begin{equation}
\label{kdvcandidaterho3}
\rho = c_1 \, u^3 + c_2 \, u_x^2,
\end{equation}
which is of first order in $x.$
\vskip 3pt
\noindent
$\bullet$ Using (\ref{kdvDtrho3}), compute $\D_t \, \rho.$
Applied to (\ref{kdvcandidaterho3}) where $M = 1,$ one gets
\begin{equation}
\label{kdvDtrho3explicit}
\D_t \, \rho = (3 c_1 u^2 \Id + 2 c_2 u_x \D_x) [u_t].
\end{equation}
As usual, $\D_x^0 = \Id$ is the identity operator.
\vskip 3pt
\noindent
$\bullet$
Evaluate $-\D_t \, \rho$ on the PDE (\ref{scalarpde}) by replacing $u_t$
by $F.$
The result is a differential function $E$ in which $t$ is a parameter.
For the KdV equation (\ref{kdv}), $F = - (\alpha u u_x + u_{3x}).$
After reversing the sign,  the evaluated form of (\ref{kdvDtrho3explicit}) is
\begin{eqnarray}
\label{kdvDtrho3evaluated}
E \!&\!=\!&\!
(3 c_1 u^2 \Id + 2 c_2 u_x \D_x) (\alpha u u_x + u_{3x})
\nonumber \\
\!&\!=\!&\! 3 c_1 \alpha u^3 u_x  + 2 c_2 \alpha u_x^3
+ 2 c_2 \alpha u u_x u_{2x} + 3 c_1 u^2 u_{3x} + 2 c_2 u_x u_{4x}.
\end{eqnarray}
\vskip 3pt
\noindent
$\bullet$
To obtain a conservation law, $E$ must be a total derivative.
Starting with highest orders, repeatedly integrate $E$ by parts.
Doing so, allows one to write $E$ as the sum of a total $x-$derivative,
$\D_x \, J,$ and a non-integrable part (i.e.\ the obstructing terms).
$J$ is the (candidate) flux with ${\rm rank}\, J = R + W(\D_t) - 1.$
Integration by parts of (\ref{kdvDtrho3evaluated}) gives
\begin{eqnarray}
\label{kdvEbyparts}
\!\!\!\!E \!\!&\!=\!&\!\!
\D_x ( \frac{3}{4} c_1 \alpha u^4 + 3 c_1 u^2 u_{2x}
+ c_2 \alpha u u_x^2 + 2 c_2 u_x u_{3x} )
- 6 c_1 u u_x u_{2x} - 2 c_2 u_{2x} u_{3x} + c_2 \alpha u_x^3
\nonumber \\
\!\!&\!=\!&\!\!
\D_x ( \frac{3}{4} c_1 \alpha u^4 - 3 c_1 u u_x^2
+ c_2 \alpha u u_x^2 + 3 c_1 u^2 u_{2x} + 2 c_2 u_x u_{3x} - c_2 u_{2x}^2)
\!+ (3 c_1 + c_2 \alpha) u_x^3.
\end{eqnarray}
The candidate flux therefore is 
\begin{equation}
\label{kdvcandidateJ3}
J = \frac{3}{4} c_1 \alpha u^4 - 3 c_1 u u_x^2 + c_2 \alpha u u_x^2
+ 3 c_1 u^2 u_{2x} + 2 c_2 u_x u_{3x}  - c_2 u_{2x}^2.
\end{equation}
\vskip 1pt
\noindent
$\bullet$
Equate the coefficients of the obstructing terms to zero.
Solve the linear system for the undetermined coefficients $c_i.$
In the example $(3 c_1 + c_2 \alpha) u_x^3$ is the only obstructing term
which vanishes for $c_2 = -\frac{3}{\alpha} c_1,$ where $c_1$ is arbitrary.
\vskip 3pt
\noindent
$\bullet$
Substitute the coefficients $c_i$ into $\rho$ and $J$ to obtain the final
forms of the density and associated flux (with a common arbitrary factor
which can be set to 1).
Setting $c_1 = 1$ and substituting $c_2 = -\frac{3}{\alpha}$ into
(\ref{kdvcandidaterho3}) and (\ref{kdvcandidateJ3}) yields $\rho^{(3)}$
and $J^{(3)}$ as given in (\ref{kdvrho3J3}).
\vskip 8pt
\noindent
Constructing ``minimal'' densities, i.e.\ densities which are free
of equivalent terms and total derivatives terms, becomes challenging if
the rank $R$ is high.
Furthermore, integration by parts is cumbersome and prone to mistakes if
done by hand.
Moreover, it would be advantageous if the integration by parts could be
postponed until the undetermined coefficients $c_i$ have been computed and
substituted in $E.$
Ideally, the computations of the density and the flux should be decoupled.
There is a need for computational tools to address these issues,
in particular, if one wants to compute conservation laws of systems of
evolution equations.
\vspace{-3mm}
\section{Tools from the Calculus of Variations and Differential Geometry}
\label{toolsPDEs}
A scalar differential function $E$ of order $M$ is called
{\em exact} (integrable) if and only if there exists a scalar
differential function $J$ of order $M-1$ such that $E = \D_x \, J.$
Obviously, $J = \D_x^{-1} E = \int E \, dx$ is then the primitive
(or integral) of $E.$
Two questions arise:
(i) How can one test whether or not $E$ is exact?
(ii) If $E$ is exact, how can one compute $J$ without using standard
integration by parts?
To answer the first question we will use the variational derivative
(Euler operator) from the calculus of variations.
To perform integration by parts we will use the homotopy operator from 
differential geometry. 
%
\subsection{The Continuous Variational Derivative (Euler Operator)}
\label{zeroeulercontinuous}
The continuous {\em variational derivative}, also called the
{\em Euler operator of order zero}, ${\cal L}^{(0)}_{u(x)},$
for variable $u(x)$ is defined \cite{PObook1993} by
\begin{eqnarray}
\label{zeroeulerscalarux}
\!\!\!\!\!\! {\cal L}^{(0)}_{u(x)} E
&=& \sum_{k=0}^{M} (-\D_x)^k \frac{\partial E}{\partial u_{kx} }
\nonumber \\
&=& \frac{\partial E}{\partial {u} }
- \D_x \frac{\partial E}{\partial {u_x} }
+ \D_{x}^2 \frac{\partial E}{\partial {u_{2x}} }
- \D_{x}^3 \frac{\partial E}{\partial {u_{3x}} }
+ \cdots
+ (-1)^M \D_{x}^M \frac{\partial E}{\partial {u_{Mx}} },
\end{eqnarray}
where $E$ is a differential function in $u(x)$ of order $M.$

A necessary and sufficient condition for a differential function $E$ to be
exact is that ${\cal L}_{u(x)}^{(0)} E \equiv 0.$
A proof of this statement is given in e.g.\ \cite{MKetalVjmp1970}.
If ${\cal L}^{(0)}_{u(x)} E \ne 0,$ then $E$ is not a total $x-$derivative
due to obstructing terms.
\vskip 5pt
\noindent
{\bf Application 1}.
Returning to (\ref{kdvDtrho3evaluated}), we now use the variational derivative
to determine $c_1$ and $c_2$ so that $E$ of order $M = 4$ will be exact.
Using nothing but differentiations, we readily compute
\begin{eqnarray}
\label{kdvcomputationzeroeulerux}
{\cal {L}}^{(0)}_{u(x)} E
\!&\!=\!&\!
\frac{\partial E}{\partial u}
- \D_x \frac{\partial E}{\partial u_x}
+ \D_{x}^2 \frac{\partial E}{\partial u_{2x}}
- \D_{x}^3 \frac{\partial E}{\partial u_{3x}}
+ \D_{x}^4 \frac{\partial E}{\partial u_{4x}}
\nonumber \\
\!&\!=\!&\!
9 c_1 \alpha u^2 u_x  + 2 c_2 \alpha u_x u_{2x} + 6 c_1 u u_{3x}
- \D_x (
3 c_1 \alpha u^3 + 6 c_2 \alpha u_x^2 + 2 c_2 \alpha u u_{2x} + 2 c_2 u_{4x} )
\nonumber \\
&& + \D_x^2 ( 2 c_2 \alpha u u_x ) - \D_x^3 ( 3 c_1 u^2 )
+ \D_x^4 ( 2 c_2 u_x )
\nonumber \\
\!&\!=\!&\!
( 9 c_1 \alpha u^2 u_x + 2 c_2 \alpha u_x u_{2x} + 6 c_1 u u_{3x} )
- (9 c_1 \alpha u^2 u_x + 14 c_2 \alpha u_x u_{2x} + 2 c_2 \alpha u u_{3x}
\nonumber \\
&& + 2 c_2 u_{5x})
+ ( 6 c_2 \alpha u_x u_{2x} + 2 c_2 \alpha u u_{3x} )
- ( 18 c_1 u_x u_{2x} + 6 c_1 u u_{3x} ) + ( 2 c_2 u_{5x} )
\nonumber \\
\!&\!=\!&\! - 6 (3 c_1 + c_2 \alpha) u_x u_{2x}.
\end{eqnarray}
Note that the terms in $u^2 u_x, u u_{3x},$ and $u_{5x}$ dropped out.
Hence, requiring that ${\cal {L}}^{(0)}_{u(x)} E \equiv 0$ leads to
$3 c_1 \!+\! c_2 \alpha = 0.$
Substituting $c_1 \!=\! 1, c_2 = -\frac{3}{\alpha},$ into
(\ref{kdvcandidaterho3}) yields $\rho^{(3)}$ in (\ref{kdvrho3J3}).
\vskip 5pt
\noindent
{\bf Application 2}.
It is paramount that the candidate density is free of total $x-$derivatives
and equivalent terms.
If such terms were present, they could be moved into the flux $J,$
and their coefficients $c_i$ would be arbitrary.
$\rho^{(1)}$ and $\rho^{(2)}$, are {\em equivalent} if and only if
$\rho^{(1)} + k \rho^{(2)} = \D_x \, J,$ for some $J$ and non-zero scalar $k$.
We write $\rho^{(1)} \equiv \rho^{(2)}$.
Clearly $\rho$ is equivalent to any non-zero multiple of itself and
$\rho \equiv 0$ if and only if $\rho$ is exact.

Instead of working with different densities, we investigate the equivalence
of terms $t_i$ in the same density.
For example, returning to the set
${\cal R} = \{ u^3, u_x^2, u u_{2x}, u_{4x} \},$
terms $t_2 = u_x^2$ and $t_3 = u u_{2x}$ are equivalent because
$t_3 + t_2 = u u_{2x} + u_x^2 = \D_x (u u_x).$

The variational derivative can be used to detect equivalent and exact terms.
Indeed, note that $v_1 = {\cal L}_{u(x)}^{(0)} (u u_{2x}) = 2 u_{2x}$
and $v_2 = {\cal L}_{u(x)}^{(0)} u_x^2 = -2 u_{2x}$ are linearly dependent.
Also, for $t_4 = u_{4x} = \D_x u_{3x}$ one gets
$v_3 = {\cal L}_{u(x)}^{(0)} u_{4x} = 0.$
To weed out the terms $t_i$ in ${\cal R}$ that are equivalent or total
derivatives, it suffices to check the linear independence of their
images $v_i$ under the Euler operator.

One can optimize this procedure by starting from a set ${\cal R}$ where
some of the equivalent and total derivatives terms have been removed
{\em a priori}.
Indeed, in view of (\ref{buildingblockskdv}), one can ignore the
highest-order terms (typically the last terms) in each of the right hand
sides.
Therefore, ${\cal R} = \{ u^3, u_x^2 \}$ and, for this example, 
no further reduction would be necessary.
Various algorithms are possible to construct minimal densities.
Details are given in \cite{UGandWHjsc1997,WHetalbirkhauser2005}.
\subsection{The Continuous Homotopy Operator}
\label{homotopycontinuous}
We now discuss the homotopy operator 
\cite{SAandGB2002a,SAandGB2002b,WHandBDandDPmcs2007,PObook1993} 
which will allow one to reduce the computation of 
$J = \D_x^{-1} E = \int E \, dx $ to a single
integral with respect to an auxiliary variable denoted by $\lambda$
(not to be confused with $\lambda$ in Section~\ref{dilationinvariancePDEs}).
Hence, the homotopy operator circumvents integration by parts and
reduces the inversion of $\D_x$ to a problem of single-variable calculus.

The {\em homotopy operator} \cite[p.\ 372]{PObook1993}
for variable $u(x),$ acting on an exact expression $E$ of order $M,$ 
is given by
\begin{equation}
\label{homotopyscalarux}
{\cal H}_{u(x)} E =
\int_{0}^{1} \left( I_u E \right) [\lambda u] \, \frac{d \lambda}{\lambda},
\end{equation}
where the integrand $I_u E$ is given by
\begin{equation}
\label{integrandhomotopyscalarux}
I_u E = \sum_{k=1}^{M} \left(
\sum_{i=0}^{k-1} u_{ix} (-\D_x)^{k-(i+1)} \right)
\frac{\partial E}{\partial u_{kx}}.
\end{equation}
In (\ref{homotopyscalarux}),
$(I_u E) [\lambda u]$ means that in $I_u E$ one replaces
$u \rightarrow \lambda u, \, u_x \rightarrow \lambda u_x,\, {\rm etc.}$
This is a special case of the homotopy,
$\lambda (u^{(1)} - u^{(0)}) + u^{(0)},$ between two points,
$u^{(0)} = (u^{0},u^{0}_x,u^{0}_{2x},\cdots, u^{0}_{Mx})$ and
$u^{(1)} = (u^{1}, u^{1}_x, u^{1}_{2x},\cdots, u^{1}_{Mx}),$ in the jet space.
For our purposes we set $u^{(0)} = (0,0,\cdots, 0)$ and
$u^{(1)} = (u, u_x, u_{2x},\cdots, u_{Mx}).$
Formula (\ref{integrandhomotopyscalarux}) is equivalent to the one in
\cite{WHandBDandDPmcs2007},
which in turn is equivalent to the formula in terms of higher Euler
operators \cite{WHIJQC2006,WHetalbirkhauser2005}.

Given an exact differential function $E$ of order $M$ one has
$J = \D_x^{-1} E =  \int E \, dx = {\cal H}_{u(x)} E.$
A proof of this statement can be found in
\cite{WHandBDandDPmcs2007}.
\vskip 3pt
\noindent
{\bf Application}.
After substituting $c_1 = 1$ and $c_2 = -\frac{3}{\alpha}$ into
(\ref{kdvDtrho3evaluated}) we obtain the exact expression
\begin{equation}
\label{Eexactkdv}
E = 3 \alpha u^3 u_x - 6 u_x^3 - 6 u u_x u_{2x} + 3 u^2 u_{3x}
- \frac{6}{\alpha} u_x u_{4x},
\end{equation}
of order $M = 4.$
First, using (\ref{integrandhomotopyscalarux}), we compute
\begin{eqnarray}
\label{kdvIuforE}
I_u E \!\!&\!=\!&\!\! \sum_{k=1}^{4} \left(
\sum_{i=0}^{k-1} u_{ix} (-\D_x)^{k-(i+1)} \right)
\frac{\partial E}{\partial u_{kx}}
\nonumber \\
\!\!&\!=\!&\!\!
( u \Id )(\frac{\partial E}{\partial u_x})
+ ( - u \D_x + u_x \Id ) (\frac{\partial E}{\partial u_{2x}})
+ ( u \D_x^2 - u_x \D_x + u_{2x} \Id) (\frac{\partial E}{\partial u_{3x}})
\nonumber \\
&&\!
+ (-u \D_x^3 + u_x \D_x^2 - u_{2x} \D_x + u_{3x} \Id )
(\frac{\partial E}{\partial u_{4x}}).
\end{eqnarray}
After substitution of (\ref{Eexactkdv}), one gets 
\begin{eqnarray}
\label{kdvIuforEpart2}
I_u E \!\!&\!=\!&\!\! ( u \Id )
(3 \alpha u^3 + 18 u_x^2 - 6 u u_{2x} - \frac{6}{\alpha} u_{4x})
+ (- u \D_x + u_x \Id ) (-6 u u_x)
\nonumber \\
&&\! + ( u \D_x^2 - u_x \D_x + u_{2x} \Id ) (3 u^2)
+ (-u \D_x^3 + u_x \D_x^2 - u_{2x} \D_x + u_{3x} \Id )
(-\frac{6}{\alpha} u_x)
\nonumber \\
\!\!&\!=\!&\!\!
3 \alpha u^4 - 18 u u_x^2 + 9 u^2 u_{2x} + \frac{6}{\alpha} u_{2x}^2
- \frac{12}{\alpha} u_x u_{3x},
\end{eqnarray}
which has the correct terms of $J^{(3)}$ but incorrect coefficients.
Finally, using (\ref{homotopyscalarux}),
\begin{eqnarray}
\label{Jkdv}
J \!\!&\!\!=\!\!&\!\! {\cal H}_{u(x)} E
= \int_0^1\! (I_u E)[\lambda u] \,\frac{d\lambda}{\lambda}
\nonumber \\
\!\!&\!\!=\!\!&\!\!\int_0^1\!\left(
3 \alpha \lambda^3 u^4 - 18 \lambda^2 u u_x^2 + 9 \lambda^2 u^2 u_{2x}
+ \frac{6}{\alpha} \lambda u_{2x}^2 - \frac{12}{\alpha}
\lambda u_x u_{3x} \right) d\lambda
\nonumber \\
\!\!&\!\!=\!\!&\!\!
\frac{3}{4} \alpha u^4 - 6 u u_x^2 + 3 u^2 u_{2x}
+ \frac{3}{\alpha} u_{2x}^2 - \frac{6}{\alpha} u_x u_{3x},
\end{eqnarray}
which matches $J^{(3)}$ in (\ref{kdvrho3J3}).

The crux of the homotopy operator method 
\cite{SAandGB2002a,SAandGB2002b,BDandMNmcs2008,PObook1993} 
is that the integration by parts of a differential expression like 
(\ref{Eexactkdv}), which involves an arbitrary function $u(x)$ and its 
$x-$derivatives, can be reduced to a standard integration of a polynomial 
in $\lambda.$
\vspace{-2mm}
\section{Conservation Laws of Nonlinear Systems of Polynomial PDEs}
\label{continuousexamples}

Thus far we have dealt with the computation of density-flux pairs of
scalar evolution equations, with the KdV equation as the leading example.
In this section we show how the method and tools can be generalized
to cover systems of evolution equations.
We will use the Drinfel'd-Sokolov-Wilson system and the Boussinesq equation
to illustrate the steps.
%
\subsection{Tools for Systems of Evolution Equations}
\label{toolssystems}
For differential functions (like densities and fluxes) of two dependent
variables $(u,v)$ and their $x-$derivatives, the total derivatives are
\begin{eqnarray}
\label{operatordtforuandv}
\D_t \rho
&=& \frac{\partial \rho}{\partial t}
+ \sum_{k=0}^{M_1} \frac{\partial \rho}{\partial u_{kx}} \D_x^k \, u_t
+ \sum_{k=0}^{M_2} \frac{\partial \rho}{\partial v_{kx}} \D_x^k \, v_t,
\\
\label{operatordxforuandv}
\D_x J
&=& \frac{\partial J}{\partial x}
+ \sum_{k=0}^{N_1} \frac{\partial J}{\partial u_{kx}} u_{(k+1)x}
+ \sum_{k=0}^{N_2} \frac{\partial J}{\partial v_{kx}} v_{(k+1)x},
\end{eqnarray}
where $M_1$ and $M_2$ are the (highest) orders of $u$ and $v$ in $\rho,$
and $N_1$ and $N_2$ are the (highest) orders of $u$ and $v$ in $J.$

To accommodate two dependent variables, we need Euler operators
${\cal L}^{(0)}_{u(x)}$ and ${\cal L}^{(0)}_{v(x)}$ for each dependent
variable separately.
For brevity, we will use vector notation, that is,
${\cal L}_{{\bf u}(x)}^{(0)} E
= ({\cal L}^{(0)}_{u(x)} E, {\cal L}^{(0)}_{v(x)} E).$
Likewise, the homotopy operator in (\ref{homotopyscalarux})
must be replaced by
\begin{equation}
\label{homotopyuvx}
{\cal H}_{{\bf u}(x)} E
= \int_{0}^{1} ( I_u E + I_v E )[\lambda {\bf u}]
\, \frac{d \lambda}{\lambda},
\end{equation}
where
\begin{equation}
\label{integrandhomotopyscalarux2}
I_u E = \sum_{k=1}^{M_1} \left(
\sum_{i=0}^{k-1} u_{ix} (-\D_x)^{k-(i+1)} \right)
\frac{\partial E}{\partial u_{kx}},
\end{equation}
and
\begin{equation}
\label{integrandhomotopyscalarvx}
I_v E = \sum_{k=1}^{M_2} \left(
\sum_{i=0}^{k-1} v_{ix} (-\D_x)^{k-(i+1)} \right)
\frac{\partial E}{\partial v_{kx}},
\end{equation}
where $M_1, M_2$ are the orders of $E$ in $u, v,$ respectively.
In (\ref{homotopyuvx}),
$u \rightarrow \lambda u, \, u_x \rightarrow \lambda u_x, \cdots,
v \rightarrow \lambda v, \, v_x \rightarrow \lambda v_x, {\rm etc.}$
\vspace{-1mm}
\subsection{The Drinfel'd-Sokolov-Wilson System: Dilation Invariance and
Conservation Laws}
\label{DSWsystem}
We consider a parameterized family of the
{\rm Drinfel'd-Sokolov-Wilson (DSW) equations}
\begin{equation}
\label{DSWsys}
u_t + 3 v v_x = 0, \quad
v_t + 2 u v_x + \alpha u_x v + 2 v_{3x} = 0,
\end{equation}
where $\alpha$ is a nonzero parameter.
The system with $\alpha = 1$ was first proposed by
Drinfel'd and Sokolov \cite{VDandVSsmd1981,VDandVSjsm1984}
and Wilson \cite{GWpla1982}.
It can be obtained \cite{RHandBGandARjmp1986} as a reduction of
the Kadomtsev-Petviashvili equation (i.e.\ a two-dimensional version of
the KdV equation) and is a completely integrable system.
In \cite{RYandZLbirkhauser2005}, Yao and Li computed conservation laws of
(\ref{DSWsys}), where they had introduced four arbitrary coefficients.
Using scales on $x,t,u$ and $v,$ all but one coefficients in
(\ref{DSWsys}) can be scaled to any real number.
Therefore, to cover the entire family of DSW equations it suffices to leave
one coefficient arbitrary, e.g.\ $\alpha$ in front of $u_x v.$

To compute the dilation symmetry of (\ref{DSWsys}), we assign weights,
$W(u)$ and $W(v),$ to both dependent variables and express that each
equation separately must be uniform in rank (i.e.\ the ranks of the equations
in (\ref{DSWsys}) may differ from each other).

For the DSW equations (\ref{DSWsys}), one has
\begin{eqnarray}
\label{DSWweightequations}
W(u) + W({\partial/\partial t}) \!&\!=\!&\! 2 W(v) + 1,
\nonumber \\
W(v) + W({\partial/\partial t}) \!&\!=\!&\! W(u) + W(v) + 1 = W(v) + 3,
\end{eqnarray}
which yields
$W(u) = W(v) = 2,\, W({\partial/\partial t}) = 3. $
The DSW system (\ref{DSWsys}) is thus invariant under the scaling symmetry
\begin{equation}
\label{DSWscale}
(x, t, u, v) \rightarrow
(\lambda^{-1} x, \lambda^{-3} t, \lambda^2 u, \lambda^2 v),
\end{equation}
where $\lambda$ is an arbitrary scaling parameter.

The first three density-flux pairs for the DSW equations (\ref{DSWsys}) are
\begin{eqnarray}
\label{DSWconslaw1}
\rho^{(1)} \!&\!=\!&\! u,
\quad\quad\quad\quad\quad\quad\quad\quad
J^{(1)} = \frac{3}{2} v^2,
\\
\label{DSWconslaw2}
\rho^{(2)} \!&\!=\!&\! v,
\quad\quad\quad\quad\quad\quad\quad\quad
J^{(2)} = 2 (u v + v_{2x}),\quad\; {\rm if} \; \alpha = 2,
\\
\label{DSWconslaw3}
\rho^{(3)} \!&\!=\!&\! (\alpha - 1) u^2 + \frac{3}{2} v^2,
\quad\;\;
J^{(3)} = 3 (\alpha u v^2 - v_x^2 + 2 v v_{2x}),
\end{eqnarray}
Both $\rho^{(1)}$ and $\rho^{(2)}$ have rank $2;$ their fluxes have rank $4.$
The pair $(\rho^{(1)},J^{(1)})$ exists for any $\alpha,$
whereas $(\rho^{(2)},J^{(2)})$ only exists if $\alpha = 2.$
Density $\rho^{(3)}$ of rank 4 and flux $J^{(3)}$ of rank $6$ are valid
for any $\alpha.$
At rank $R=6,$
\begin{equation}
\label{DSWconslaw4}
\rho^{(4)} = (\alpha + 1)(\alpha - 2) u^3 - \frac{9}{2} (\alpha + 1) u v^2
- \frac{3}{2} (\alpha - 2) u_x^2 - \frac{27}{2} v_x^2.
\end{equation}
The corresponding flux (not shown) has $7$ terms.
At rank $R=8,$
\begin{equation}
\label{DSWconslaw5}
\rho^{(5)} = u^4 - \frac{9}{2} u^2 v^2 - \frac{27}{8} v^4
- \frac{9}{2} u u_x^2 + \frac{3}{4} u_{2x}^2 + \frac{45}{2} v u_x v_x
+ 27 u v_x^2 - \frac{81}{4} v_{2x}^2,
\end{equation}
provided $\alpha = 1.$
The corresponding flux (not shown) has $15$ terms.
There exists a density-flux pair for all even ranks $R \le 10$ provided
$\alpha = 1,$ for which (\ref{DSWsys}) is completely integrable.
\vspace{-1mm}
\subsection{Computation of a Conservation Law of the Drinfel'd-Sokolov
Wilson System}
\label{applDSWsytem}
To illustrate how the presence of a parameter, like $\alpha,$ affects the
computation of densities, we compute ${\rho}^{(1)}$ and ${\rho}^{(2)}$
of rank $R = 2$ given in (\ref{DSWconslaw1}) and (\ref{DSWconslaw2}).
\vskip 5pt
\noindent
{\bf Step 1}: {\bf Construct the form of the density}
\vskip 4pt
\noindent
The set of dependent variables is ${\cal V} = \{ u, v\}.$
Both elements are of rank $2$ so, no $x-$derivatives are needed.
Thus, ${\cal M} = {\cal R} = {\cal S} = \{ u, v \}.$
Linearly combining the elements in ${\cal S}$ gives $\rho = c_1 u + c_2 v.$
\vskip 5pt
\noindent
{\bf Step 2}: {\bf Compute the undetermined coefficients $c_i$}
\vskip 4pt
\noindent
Evaluating $E = - \D_t \rho = - (c_1 u_t + c_2 v_t) $ on (\ref{DSWsys}),
yields
\begin{equation}
\label{EDSWsys}
E = 3 c_1 v v_x + c_2 (2 u v_x + \alpha u_x v + 2 v_{3x}),
\end{equation}
which will be exact if ${\cal L}^{(0)}_{{\bf u}(x)} E =
( {\cal L}^{(0)}_{u(x)} E, {\cal L}^{(0)}_{v(x)} E ) \equiv (0,0).$
Since $E$ is of order $M_1 = 1$ in $u$ and order $M_2 = 3$ in $v,$
\begin{equation}
\label{DSWsyscomputationzeroeulerux}
{\cal L}^{(0)}_{u(x)} E =
\frac{\partial E}{\partial u} - \D_x \frac{\partial E}{\partial u_x}
= 2 c_2 v_x - \D_x (c_2 \alpha v)
= (2 - \alpha) c_2 v_x,
\end{equation}
and
\begin{eqnarray}
\label{DSWsyscomputationzeroeulervx}
{\cal L}^{(0)}_{v(x)} E &=&
\frac{\partial E}{\partial v}
- \D_x \frac{\partial E}{\partial v_x}
+ \D_x^2 \frac{\partial E}{\partial v_{2x}}
- \D_x^3 \frac{\partial E}{\partial v_{3x}}
\nonumber \\
&=& 3 c_1 v_x + c_2 \alpha u_x - \D_x (3 c_1 v + 2 c_2 u)
- \D_x^3 (2 c_2)
= (\alpha - 2) c_2 u_x.
\end{eqnarray}
Both (\ref{DSWsyscomputationzeroeulerux}) and
     (\ref{DSWsyscomputationzeroeulervx}) will vanish identically
if and only if $(\alpha - 2) c_2 = 0.$
This equation (with unknowns $c_1$ and $c_2)$ is parameterized by
$\alpha \ne 0.$
The solution algorithm \cite{UGandWHjsc1997} considers all branches
of the solution and possible compatibility conditions.
Setting $c_1 = 1,$ leads to either (i) $c_2 = 0$ if $\alpha \ne 2,$ or
(ii) $c_2$ arbitrary if $\alpha = 2.$
Setting $c_2 = 1$ leads to the compatibility condition, $\alpha = 2,$ and
$c_1$ arbitrary.
Substituting the solutions into $\rho = c_1 u + c_2 v$ gives
$\rho = u$ which is valid for any $\alpha;$ and
$\rho = u + c_2 v$ or $\rho = c_1 u + v$ provided $\alpha = 2.$
In other words, $\rho^{(1)} = u$ is the only density of rank $2$ for
arbitrary values of $\alpha.$
For $\alpha = 2$ there exist two independent densities,
$\rho^{(1)} = u$ and $\rho^{(2)} = v.$
\vskip 5pt
\noindent
{\bf Step 3}: {\bf Compute the associated flux $J$}
\vskip 4pt
\noindent
As an example, we compute the flux in (\ref{DSWconslaw2}) associated
with $\rho^{(2)} = v$ and $\alpha = 2.$
In this case, $c_1 = 0, c_2 = 1,$ for which (\ref{EDSWsys}) simplifies into
\begin{equation}
\label{EDSWsyssimp}
E = 2 ( u v_x + u_x v + v_{3x} ),
\end{equation}
which is of order $M_1 = 1$ in $u$ and order $M_2 = 3$ in $v.$
Using (\ref{integrandhomotopyscalarux2}) and (\ref{integrandhomotopyscalarvx}),
we obtain
\begin{equation}
\label{DSWIuE}
I_u E = u \Id \frac{\partial E}{\partial u_x}
= u \Id (2 v)
= 2 u v,
\end{equation}
and
\begin{eqnarray}
\label{DSWIvE}
I_v E \!&\!=\!&\!
(v \Id) (\frac{\partial E}{\partial v_x})
+ (- v \D_x + v_x \Id) (\frac{\partial E}{\partial v_{2x}})
+ (v \D_x^2 - v_x \D_x + v_{2x} \Id)
  (\frac{\partial E}{\partial v_{3x}})
\nonumber \\
\!\!&\!=\!&\!\!
(v \Id) (2 u) + (v \D_x^2 - v_x \D_x + v_{2x} \Id) (2)
= 2 u v + 2 v_{2x}.
\end{eqnarray}
Hence, using (\ref{homotopyuvx}),
\begin{equation}
\label{JDSWsys}
J = {\cal H}_{u(x)} E
= \int_0^1 ( I_u E + I_v E )[\lambda {\bf u}] \,\frac{d\lambda}{\lambda}
= \int_0^1 ( 4 \lambda u v + 2 v_{2x} ) \, d\lambda
= 2 (u v + v_{2x}),
\end{equation}
which is $J^{(2)}$ in (\ref{DSWconslaw2}).
The integration of (\ref{EDSWsyssimp}) could easily be done by hand.
The homotopy operator method pays off if the expression to be integrated
has a large number of terms.
%
\subsection{The Boussinesq Equation: Dilation Invariance and Conservation Laws}
\label{boussinesqequation}
The wave equation,
\begin{equation}
\label{boussinesq}
u_{2t} - u_{2x} + 3 u_x^2 + 3 u u_{2x} + \alpha u_{4x} = 0,
\end{equation}
for $u(x,t)$ with real parameter $\alpha,$ was proposed by Boussinesq to
describe surface waves in shallow water \cite{MAandPCbook1991}.
For what follows, we rewrite (\ref{boussinesq}) as a system of
evolution equations,
\begin{equation}
\label{boussinesqsys}
u_t + v_x = 0, \quad v_t + u_x - 3 u u_x - \alpha u_{3x} = 0,
\end{equation}
where $v(x,t)$ is an auxiliary dependent variable.

The Boussinesq system (\ref{boussinesqsys}) is not uniform in rank
because the terms $u_x$ and $\alpha u_{3x}$ lead to an inconsistent system
of weight equations.
To circumvent the problem we introduce an auxiliary parameter $\beta$ with
(unknown) weight, and replace (\ref{boussinesqsys}) by
\begin{equation}
\label{boussinesqsystem}
u_t + v_x = 0, \quad
v_t + \beta u_x - 3 u u_x - \alpha u_{3x} = 0.
\end{equation}
Requiring uniformity in rank, we obtain (after some algebra)
\begin{equation}
\label{boussinesqweights}
W(u) = 2, \quad W(v) = 3, \quad W(\beta) = 2, \quad
W(\frac{\partial}{\partial t}) = 2.
\end{equation}
Therefore, (\ref{boussinesqsystem}) is invariant under the scaling symmetry
\begin{equation}
\label{boussinesqscale}
(x, t, u, v, \beta) \rightarrow
(\lambda^{-1} x, \lambda^{-2} t, \lambda^2 u, \lambda^3 v, \lambda^2 \beta).
\end{equation}
As the above example shows, a PDE that is not dilation invariant can be made
so by extending the set of dependent variables with one or more auxiliary
parameters with weights.
Upon completion of the computations one can set each of these parameters 
equal to $1.$

The Boussinesq equation (\ref{boussinesq}) has infinitely many conservation
laws and is completely integrable \cite{MAandPCbook1991,MAandHSbook1981}.
The first four density-flux pairs \cite{PAthesis2003} for
(\ref{boussinesqsystem}) are
\begin{eqnarray}
\label{boussinesqconslaw1}
\rho^{(1)} \!&\!=\!&\! u,
\quad\quad\quad\quad\quad\quad\quad\quad\quad\quad\;\,
J^{(1)} = v, \\
\label{boussinesqconslaw2}
\rho^{(2)} \!&\!=\!&\! v,
\quad\quad\quad\quad\quad\quad\quad\quad\quad\quad\;\,
J^{(2)} = \beta u - \frac{3}{2} u^2 - \alpha u_{2x}, \\
\label{boussinesqconslaw3}
\rho^{(3)} \!&\!=\!&\! u v,
\quad\quad\quad\quad\quad\quad\quad\quad\quad\;\;\;
J^{(3)} = \frac{1}{2} \beta u^2 - u^3 + \frac{1}{2}v^2
+ \frac{1}{2} \alpha u_x^2 - \alpha u u_{2x}, \\
\label{boussinesqconslaw4}
\rho^{(4)} \!&\!=\!&\! \beta u^2 - u^3 + v^2 + \alpha u_x^2,
\quad\quad
J^{(4)} = 2 \beta u v - 3 u^2 v + 2 \alpha u_x v_x- 2 \alpha u_{2x} v.
\end{eqnarray}
These densities are of ranks $2,3,5$ and $6,$ respectively.
The corresponding fluxes are of one rank higher.
After setting $\beta = 1$ we obtain the conserved quantities of
(\ref{boussinesqsys}) even though initially this system was not uniform
in rank.
%
\subsection{Computation of a Conservation Law for the Boussinesq System}
\label{applboussinesqsystem}
We show the computation of ${\rho}^{(4)}$ and $J^{(4)}$ of ranks $6$ and $7,$
respectively.
The presence of the auxiliary parameter $\beta$ with weight complicates
matters.
At a fixed rank $R,$ conserved densities corresponding to lower ranks might
appear in the result.
These lower-rank densities are easy to recognize for they are multiplied with
arbitrary coefficients $c_i.$
Consequently, when parameters with weight are introduced, the densities
corresponding to distinct ranks are no longer linearly independent.
As the example below will show, densities must be split into independent
pieces.
\vskip 5pt
\noindent
{\bf Step 1}: {\bf Construct the form of the density}
\vskip 4pt
\noindent
Augment the set of dependent variables with the parameter $\beta$
(with non-zero weight).
Hence, ${\cal V} = \{ u, v, \beta \}.$
Construct
${\cal M} =
\{ \beta^2 u, \beta u^2, \beta u, \beta v, u^3, u^2, u, v^2, v, u v \},$
which contains all non-constant monomials of (chosen) rank $6$ or less
(without derivatives).
Next, for each term in ${\cal M},$ introduce the right number of
$x$-derivatives so that each term has rank 6.
For example,
\begin{equation}
\label{buildingblocksboussinesq}
\frac{{\partial}^2 \beta u}{\partial{x^2}} \!=\! \beta u_{2x}, \quad
\frac{{\partial}^2 u^2}{\partial{x^2}} \!=\! 2 u_x^2 + 2 u u_{2x}, \quad
\frac{{\partial}^4 u}{\partial{x^4}} \!=\! u_{4x}, \quad
\frac{{\partial} ( u v )}{\partial{x}} = v u_x + u v_x, \quad etc..
\end{equation}
Gather the terms in the right hand sides of the equations in
(\ref{buildingblocksboussinesq}) to get
\begin{equation}
\label{boussinesqsetr}
{\cal R} = \{ \beta^2 u, \beta u^2, u^3, v^2, v u_x, u_x^2,
\beta v_x, u v_x, \beta u_{2x}, u u_{2x}, v_{3x}, u_{4x} \}.
\end{equation}
Using (\ref{zeroeulerscalarux}) and a similar formula for $v$,
for every term $t_i$ in ${\cal R}$ we compute
${\bf v}_i = {\cal L}_{{\bf u}(x)}^{(0)} t_i =
( {\cal L}_{u(x)}^{(0)} t_i, {\cal L}_{v(x)}^{(0)} t_i ).$
If ${\bf v}_i = (0,0)$ then $t_i$ is discarded and so is ${\bf v}_i.$
If ${\bf v}_i \ne (0,0)$ we verify whether or not ${\bf v}_i$ is linearly
independent of the non-zero vectors ${\bf v}_j,\,$ $j=1,2, \cdots, i-1.$
If independent, the term $t_i$ is kept, otherwise, $t_i$ is discarded and so
is ${\bf v}_i.$

Upon application of ${\cal L}_{{\bf u}(x)}^{(0)}$,
the first six terms in ${\cal R}$ lead to linearly independent vectors
${\bf v}_1$ through ${\bf v}_6.$
Therefore, $t_1$ through $t_6$ are kept
(and so are the corresponding vectors).
For $t_7 = \beta v_x$ we compute
${\bf v}_7 \!=\! {\cal L}_{{\bf u}(x)}^{(0)} (\beta v_x) \!=\!(0,0).$
So, $t_7$ is discarded and so is ${\bf v}_7.$
For $t_8 = u v_x$ we get
${\bf v}_8 \!=\! {\cal L}_{{\bf u}(x)}^{(0)} (u v_x)
\!=\!(v_x, -u_x) = -{\bf v}_5.$
So, $t_8$ is discarded and so is ${\bf v}_8.$

Proceeding in a similar fashion, $t_{9}, t_{10}, t_{11}$ and $t_{12}$
are discarded.
Thus, ${\cal R}$ is replaced by
\begin{equation}
\label{boussinesqsets}
{\cal S} = \{ \beta^2 u, \beta u^2, u^3, v^2, v u_x, u_x^2 \},
\end{equation}
which is free of divergences and divergence-equivalent terms.
Ignoring the highest-order terms (typically the last terms) in each of the
right hand sides of the equations in (\ref{buildingblocksboussinesq})
optimizes the procedure.
Indeed, ${\cal R}$ would have had six instead of twelve terms.
Coincidentally, in this example no further eliminations would be needed to
obtain ${\cal S}.$
Next, linearly combine the terms in ${\cal S}$ to get 
\begin{equation}
\label{candidaterhoboussinesq}
\rho =
c_1 \beta^2 u + c_2 \beta u^2 + c_3 u^3 + c_4 v^2 + c_5 v u_x + c_6 u_x^2.
\end{equation}
\vskip 0.0001pt
%
\noindent
{\bf Step 2}: {\bf Compute the undetermined coefficients $c_i$}
\vskip 4pt
\noindent
Compute $\D_t \rho.$
Here, $\rho$ is of order $M_1 \!=\!1$ in $u$ and order $M_2 \!=\! 0$ in $v.$
Hence, application of (\ref{operatordtforuandv}) gives
\begin{eqnarray}
\label{dtboussinesqrhoassumed}
\D_t \rho
\!&\!=&\! \frac{\partial \rho}{\partial u} \Id u_t
+ \frac{\partial \rho}{\partial u_{x}} \D_x u_t
+ \frac{\partial \rho}{\partial v} \Id v_t
\nonumber \\
\!&\!=&\! (c_1 \beta^2 + 2 c_2 \beta u + 3 c_3 u^2) u_t
+ (c_5 v + 2 c_6 u_x) u_{tx} + (2 c_4 v + c_5 u_x) v_t.
\end{eqnarray}
Use (\ref{boussinesqsystem}) to eliminate $u_t, u_{tx},$ and $v_t.$
Then, $E \!=\! - \D_t \, \rho $ evaluates to
\begin{eqnarray}
\label{Eboussinesq}
E \!&\!\!=\!&\! (c_1 \beta^2 + 2 c_2 \beta u + 3 c_3 u^2) v_x
+ (c_5 v + 2 c_6 u_x) v_{2x}
\nonumber \\
&&\!
+ (2 c_4 v + c_5 u_x) (\beta u_x - 3 u u_x - \alpha u_{3x}),
\end{eqnarray}
which must be exact.
Thus, require that
${\cal L}^{(0)}_{{\bf u}(x)} E =
({\cal L}^{(0)}_{u(x)} E, {\cal L}^{(0)}_{v(x)} E ) \equiv (0,0).$
Group like terms.
Set their coefficients equal to zero to obtain the parameterized system
\begin{equation}
\label{systemboussinesq}
\beta (c_2 - c_4) = 0, \;\; c_3 + c_4 = 0, \;\; c_5 = 0, \;\;
\alpha c_5 = 0, \;\; \beta c_5 = 0, \;\; \alpha c_4 - c_6 = 0,
\end{equation}
where $\alpha \ne 0$ and $\beta \ne 0.$
Investigate the eliminant of the system.
Set $c_1 = 1$ and obtain the solution
\begin{equation}
\label{solcboussinesq}
c_1 = 1, \quad c_2 = c_4, \quad
c_3 = - c_4, \quad c_5 = 0, \quad c_6 = \alpha c_4,
\end{equation}
which holds without condition on $\alpha$ and $\beta.$
Substitute (\ref{solcboussinesq}) into (\ref{candidaterhoboussinesq}) to get
\begin{equation}
\label{rho4bousall}
\rho = \beta^2 u + c_4 ( \beta u^2 - u^3 + v^2 + \alpha u_x^2 ).
\end{equation}
The density must be split into independent pieces.
Indeed, since $c_4$ is arbitrary, set $c_4 = 0$ or $c_4 = 1,$
thus splitting (\ref{rho4bousall}) into two independent densities,
$\rho = \beta^2 u \equiv u$ and
\begin{equation}
\label{rho4boussinesq}
\rho = \beta u^2 - u^3 + v^2 + \alpha u_x^2,
\end{equation}
which are $\rho^{(1)}$ and $\rho^{(4)}$ in 
(\ref{boussinesqconslaw1})-(\ref{boussinesqconslaw4}).
\vskip 4pt
\noindent
{\bf Step 3}: {\bf Compute the flux $J$}
\vskip 4pt
\noindent
Compute the flux corresponding to $\rho$ in (\ref{rho4boussinesq}).
Substitute (\ref{solcboussinesq}) into (\ref{Eboussinesq}).
Take the terms in $c_4$ and set $c_4 = 1.$
Thus,
\begin{equation}
\label{Eboussimp}
E = 2 \beta u v_x + 2 \beta v u_x  - 3 u^2 v_x - 6 u v u_x
+ 2 \alpha u_x v_{2x} - 2 \alpha v u_{3x},
\end{equation}
which is of order $M_1 = 3$ in $u$ and order $M_2 = 2$ in $v.$
Using (\ref{integrandhomotopyscalarux2}) and (\ref{integrandhomotopyscalarvx}),
one readily obtains
\begin{equation}
I_u E = 2 \beta u v - 6 u^2 v + 2 \alpha u_x v_x - 2 \alpha u_{2x} v,
\end{equation}
and
\begin{equation}
I_v E = 2 \beta u v - 3 u^2 v + 2 \alpha u_x v_x - 2 \alpha u_{2x} v.
\end{equation}
Hence, using (\ref{homotopyuvx}),
\begin{eqnarray}
\label{Jboussinesq}
J \!&\!=\!&\! {\cal H}_{u(x)} E
= \int_0^1\! (I_u E + I_v E)[\lambda {\bf u}] \,\frac{d\lambda}{\lambda}
\nonumber \\
\!&\!=\!&\!\int_0^1\!\left(
4 \beta \lambda u v - 9 \lambda^2 u^2 v
+ 4 \alpha \lambda u_x v_x - 4 \alpha \lambda u_{2x} v \right) d\lambda
\nonumber \\
\!&\!=\!&\!
2 \beta u v - 3 u^2 v + 2 \alpha u_x v_x - 2 \alpha u_{2x} v,
\end{eqnarray}
which is $J^{(4)}$ in (\ref{boussinesqconslaw4}).
One can set $\beta = 1$ at the end of the computations.
\vspace{-2mm}
\section{Conservation Laws of Systems of PDEs with Transcendental
Nonlinearities}
\label{transcendentalexamples}
We now turn to the symbolic computation of conservation laws of certain 
classes of PDEs with transcendental nonlinearities. 
We only consider PDEs where the transcendental functions act on one 
dependent variable $u$ (and not on $x-$derivatives of $u).$
In contrast to the examples in the previous sections, the candidate density 
will no longer have {\em constant} undetermined coefficients but 
{\em functional} coefficients which depend on the variable $u.$ 
Furthermore, we consider only PDEs which have one type of nonlinearity. 
For example, sine, or cosine, or exponential terms are fine but not a 
mixture of these functions.  
%
\subsection{The sine-Gordon Equation: Dilation Invariance and Conservation
Laws}
\label{sinegordonequation}
The {\rm sine-Gordon (sG) equation} appears in the literature
\cite{ABetalnc1971,GLrmp1971} in two different ways:
\vskip 3pt
\noindent
$\bullet$
In light-cone coordinates the sG equation, $u_{xt} = \sin u,$
has a mixed derivative term, which complicates matters.
We return to this type of equation in Section~\ref{applsGequation}.
\vskip 3pt
\noindent
$\bullet$
The sG equation in laboratory coordinates, $u_{2t} - u_{2x} = \sin u,$ can
be recast as
\begin{equation}
\label{sinegordon0}
u_t + v = 0, \quad v_t + u_{2x} + \sin u = 0,
\end{equation}
where $v(x,t)$ is an auxiliary variable.
System (\ref{sinegordon0}) is amenable to our approach, subject to
modifications to accommodate the transcendental nonlinearity.

The sG equation describes the propagation of crystal dislocations,
superconductivity in a Josephson junction, and ultra-short optical pulse
propagation in a resonant medium \cite{GLrmp1971}.
In mathematics, the sG equation is long known in the differential geometry of
surfaces of constant negative Gaussian curvature
\cite{PDandRJbook1989,ANbook1985}.

The sine-Gordon equation (\ref{sinegordon0}) is not uniform in rank unless
we replace it by
\begin{equation}
\label{sinegordon}
u_t + v = 0, \quad v_t + u_{2x} + \alpha \sin u = 0,
\end{equation}
where $\alpha$ is a real parameter with weight.
Indeed, substituting the Maclaurin series,
$\sin u = u - \frac{u^3}{3!} + \frac{u^5}{5!} - \cdots,$ and requiring
uniformity in rank yields
\begin{eqnarray}
\label{sinegordonweightequation1}
\!\!W(u) + W({\partial/\partial t}) \!\!&\!\!=\!\!&\!\! W(v),
\nonumber \\
\label{sinegordonweightequation2}
\!\!\!\!W(v) \!+\! W({\partial/\partial t}) \!\!&\!\!=\!\!&\!\! W(u) \!+\! 2
\!=\!W(\alpha) \!+\! W(u) \!=\!
W(\alpha) \!+\! 3 W(u) \!=\! W(\alpha) \!+\! 5 W(u) \!=\! \cdots .
\end{eqnarray}
This forces us to set $W(u) = 0$ and $W(\alpha) = 2.$
Consequently, (\ref{sinegordon}) is scaling invariant under the symmetry
\begin{equation}
\label{sinegordoncale}
(x, t, u, v, \alpha) \rightarrow
(\lambda^{-1} x, \lambda^{-1} t, \lambda^0 u,
\lambda^1 v, \lambda^2 \alpha),
\end{equation}
corresponding to
$W({\partial/\partial x}) = W({\partial/\partial t}) = 1, W(u) = 0, W(v) = 1,
W(\alpha) = 2.$
The first and second equations in (\ref{sinegordon}) are uniform of
ranks 1 and 2, respectively.

The first few (of infinitely many) density-flux pairs
\cite{PAthesis2003,RDandRBprsa1977} for the sG equation (\ref{sinegordon}) are
\begin{eqnarray}
\label{sinegordonconslaw1}
\rho^{(1)} \!&=&\! 2 \alpha \cos u + v^2 + u_x^2,
\quad\quad\quad\quad\;
J^{(1)} = 2 v u_x, \\
\label{sinegordonconslaw2}
\rho^{(2)} \!&=&\! 2 v u_x,
\quad\quad\quad\quad\quad\quad\quad\quad\quad\quad
J^{(2)} = - 2 \alpha \cos u + v^2 + u_x^2, \\
\label{sinegordonconslaw3}
\rho^{(3)} \!&=&\! 6 \alpha v u_x \cos u + v^3 u_x + v u_x^3 - 8 v_x u_{2x},\\
\label{sinegordonconslaw4}
\rho^{(4)} \!&=&\!
2 \alpha^2 \cos^2 u - 2 \alpha^2 \sin^2 u + 4 \alpha v^2 \cos u
+ v^4 + 20 \alpha u_x^2 \cos u + 6 v^2 u_x^2
\nonumber \\
\!\! && + u_x^4 - 16 v_x^2 - 16 u_{2x}^2,
\end{eqnarray}
$J^{(3)}$ and $J^{(4)}$ are not shown due to length.
Again, all densities and fluxes are uniform in rank
(before $\alpha$ is set equal to 1).
\subsection{Computation of a Conservation Law for the sine-Gordon System}
\label{applsGsystem}
We show how to compute densities ${\rho}^{(1)}$ and ${\rho}^{(2)},$ both
of rank 2, and their associated fluxes $J^{(1)}$ and $J^{(2)}.$
The candidate density will no longer have {\em constant} undetermined 
coefficients $c_i$ but {\em functional} coefficients $h_i(u)$ which depend 
on the variable with weight zero \cite{PAthesis2003}.
To avoid having to solve PDEs, we tacitly assume that there is only
{\em one} dependent variable with weight zero.
As before, the algorithm proceeds in three steps:
\vskip 5pt
\noindent
{\bf Step 1}: {\bf Construct the form of the density}
\vskip 4pt
\noindent
Augment the set of dependent variables with $\alpha$
(with non-zero weight) and replace $u$ by $u_x$ (since $W(u)=0).$
Hence, ${\cal V} = \{\alpha , u_x, v \}.$
Compute ${\cal R} = \{ \alpha, v^2, v^2, u_{2x}, v u_x, u_x^2 \}$ and
remove divergences and equivalent terms to get
${\cal S} = \{\alpha, v^2, u_x^2, v u_x \}.$
The candidate density
\begin{equation}
\label{candidaterhosg}
\rho = \alpha h_1(u) + h_2(u) v^2 + h_3(u) u_x^2 + h_4(u) v u_x,
\end{equation}
with undetermined functional coefficients $h_i(u).$
\vskip 5pt
\noindent
{\bf Step 2}: {\bf Compute the functions $h_i(u)$}
\vskip 4pt
\noindent
Compute
\begin{eqnarray}
\label{lastEsinegordon}
\D_t \rho \!&\!=\!&\!
\frac{\partial \rho}{\partial u} \Id u_t
+ \frac{\partial \rho}{\partial u_x} \D_x u_t
+ \frac{\partial \rho}{\partial v} \Id v_t
\nonumber \\
\!&\!=\!&\!
(\alpha h_1^{\prime} + v^2 h_2^{\prime} + u_x^2 h_3^{\prime}
+ v u_x h_4^{\prime} ) u_t + ( 2 u_x h_3 + v h_4 ) u_{tx}
+ ( 2 v h_2 + u_x h_4 ) v_t,
\end{eqnarray}
where $h_i^{\prime}$ denotes $\textstyle{\frac{dh_i}{du}}.$
After replacing $u_t$ and $v_t$ from (\ref{sinegordon}),
$E = - \D_t \rho$ becomes
\begin{equation}
\label{Esinegordon}
E = (\alpha h_1^{\prime} + v^2 h_2^{\prime} + u_x^2 h_3^{\prime}
+ v u_x h_4^{\prime} ) v + (2 u_x h_3 + v h_4 ) v_x
+ (2 v h_2 + u_x h_4 )(\alpha \sin u + u_{2x}).
\end{equation}
$E$ must be exact.
Therefore, require that ${\cal L}^{(0)}_{u(x)} E \equiv 0$ and
${\cal L}^{(0)}_{v(x)} E \equiv 0.$
Set the coefficients of like terms equal to zero to get a mixed linear
system of algebraic equations and ODEs:
\begin{eqnarray}
\label{odesystemsGequation}
&& h_2(u) - h_3(u) \!=\! 0, \quad h_2^{\prime}(u) \!=\! 0, \quad
h_3^{\prime}(u) \!=\! 0, \quad h_4^{\prime}(u) \!=\! 0,  \quad
h_2^{\prime\prime}(u) \!=\! 0, 
\nonumber \\
&& h_4^{\prime\prime}(u) = 0, \quad
2 h_2^{\prime}(u) - h_3^{\prime}(u) = 0, \quad
2 h_2^{\prime\prime}(u) - h_3^{\prime\prime}(u) = 0, 
\\
&& h_1^{\prime}(u)  + 2 h_2(u) \sin u = 0, \quad
h_1^{\prime\prime}(u) + 2 h_2^{\prime}(u) \sin u + 2  h_2(u) \cos u = 0.
\nonumber 
\end{eqnarray}
Solve the system \cite{PAthesis2003} and substitute the solution
\begin{equation}
\label{solhisinegordon}
{h_1}(u) = 2 c_1  \cos u + c_3, \quad {h_2}(u) = {h_3}(u) = c_1,
\quad {h_4}(u) = c_2,
\end{equation}
(with arbitrary constants $c_i)$ into (\ref{candidaterhosg}) to obtain
\begin{equation}
\label{rhosinegordon}
\rho = c_1 (2 \alpha \cos u + v^2 + u_x^2 ) + c_2 v u_x + c_3 \alpha.
\end{equation}
\vskip 5pt
\noindent
{\bf Step 3}: {\bf Compute the flux $J$}
\vskip 4pt
\noindent
Compute the flux corresponding to $\rho$ in (\ref{rhosinegordon}).
Substitute (\ref{solhisinegordon}) into (\ref{Esinegordon}), to get
\begin{equation}
\label{Eevalsinegordon}
E = c_1 (2 u_x v_x + 2 v u_{2x} )
    + c_2 ( \alpha u_x \sin u + v v_x + u_x u_{2x}  ).
\end{equation}
Since $E \!=\!\D_x J,\;$ one must integrate to obtain $J.$
Using (\ref{integrandhomotopyscalarux}) and (\ref{integrandhomotopyscalarvx})
one gets
$I_u E \!=\! 2 c_1 v u_x + c_2 ( \alpha u \sin u + u_x^2 )$ and
$I_v E \!=\! 2 c_1 v u_x + c_2 v^2.$
Using (\ref{homotopyuvx}),
\begin{eqnarray}
J &=& {\cal H}_{{\bf u}(x)} E =
\int_0^1 ( I_u E + I_v E )[\lambda {\bf u}] \, \frac{d\lambda}{\lambda}
\nonumber \\
&=& \int_0^1 \left(
4 c_1 \lambda v u_x
+ c_2 ( \alpha u \sin(\lambda u) + \lambda v^2 + \lambda u_x^2 ) \right)
\, d\lambda
\nonumber \\
&=& c_1 (2 v u_x)
+ c_2 \left(- \alpha \cos u + \frac{1}{2} v^2 + \frac{1}{2} u_x^2 \right).
\label{fluxforEevalsinegordon}
\end{eqnarray}
Finally, split density (\ref{rhosinegordon}) and flux
(\ref{fluxforEevalsinegordon}) into independent pieces (for $c_1$ and $c_2)$
to get
\begin{eqnarray}
\label{rho2J2singgordon}
\rho^{(1)} \!&\!=\!&\! 2 \alpha  \cos u + v^2 + u_x^2,
\quad\quad\quad J^{(1)} = 2 v u_x, \\
\label{rho3J3singgordon}
\rho^{(2)} \!&\!=\!&\! v u_x, \quad\quad\quad\quad\quad\quad\quad\quad\quad\;
J^{(2)} = - \alpha \cos u + \frac{1}{2} v^2 + \frac{1}{2} u_x^2.
\end{eqnarray}
For $E$ in (\ref{Eevalsinegordon}), $J$ in (\ref{fluxforEevalsinegordon})
can easily be computed by hand \cite{PAthesis2003}.
However, the computation of fluxes corresponding to densities of ranks $\ge 2$
is cumbersome and requires integration with the homotopy operator.
\vspace{-2mm}
\section{Conservation Laws of Scalar Equations with Transcendental and Mixed
Derivative Terms}
\label{mixedderivativePDE}
Our method to compute densities and fluxes of scalar equations with
transcendental terms and a mixed derivative term (i.e.\ $u_{xt})$ is an 
adaptation of the technique shown in Section~\ref{continuousexamples}.
We only consider single PDEs with one type of transcendental nonlinearity. 
Since we are no longer dealing with evolution equations, 
densities and fluxes could dependent on $u_t, u_{2t},$ etc.\ 
We do not cover such cases; instead, we refer the reader to \cite{TW2002}. 
%
\subsection{The sine-Gordon Equation in Light-Cone Coordinates}
\label{applsGequation}
In light-cone coordinates (or characteristic coordinates) the sG equation,
\begin{equation}
\label{sGlightcone}
u_{xt} = \sin u,
\end{equation}
has a mixed derivative as well as a transcendental term.
A change of variables, $\Phi = u_x, \Psi = -1 + \cos u,$ allows one to
replace (\ref{sGlightcone}) by
\begin{equation}
\label{sGlightconepoly}
\Phi_{xt} - \Phi - \Phi \Psi = 0, \quad 2 \Psi + \Psi^2 + \Phi_{t}^2 = 0,
\end{equation}
without transcendental terms.
Unfortunately neither (\ref{sGlightcone}) nor (\ref{sGlightconepoly}) can be
written as a system of evolution equations.
As shown in Section~\ref{sinegordonequation}, to deal with the transcendental
nonlinearity, which imposes $W(u) = 0,$ one has to replace
(\ref{sGlightcone}) by
\begin{equation}
\label{sGlightconealpha}
u_{xt} = \alpha \sin u,
\end{equation}
where $\alpha$ is an auxiliary parameter with weight.
Indeed, (\ref{sGlightconealpha}) is dilation invariant under the
scaling symmetry
\begin{equation}
\label{sGequationscale}
(x, t, u, \alpha)
\rightarrow
( {\lambda}^{-1} x, \lambda^{-1} t, {\lambda}^{0} u, {\lambda}^{2} \alpha ),
\end{equation}
corresponding to $W(\partial/\partial x) = W(\partial/\partial t) = 1,
W(u) = 0,$ and $W(\alpha) = 2.$
The density-flux pairs \cite{PAthesis2003,RDandRBprsa1977} of ranks
$2, 4, 6,$ and $8$ (which are independent of $u_t, u_{2t},$ etc.), are
\begin{eqnarray}
\label{sinegordonlightconeconslaw1}
\rho^{(1)} \!&=&\! u_x^2,
\quad\quad\quad\quad\quad\quad\quad\quad\quad\;\;\;
J^{(1)} = 2 \alpha \cos u, \\
\label{sinegordonlightconeconslaw2}
\rho^{(2)} \!&=&\! u_x^4 - 4 u_{2x}^2
\quad\quad\quad\quad\quad\quad\quad\,
J^{(2)} = 4 \alpha u_x^2 \cos u, \\
\label{sinegordonlightconeconslaw3}
\rho^{(3)} \!&=&\! u_x^6 - 20 u_x^2 u_{2x}^2 + 8 u_{3x}^2,
\quad\quad
J^{(3)} =
2 \alpha ( 3 u_x^4 \cos u + 8 u_x^2 u_{2x} \sin u - 4 u_{2x}^2 \cos u ), \\
\label{sinegordonlightconeconslaw4}
\rho^{(4)} \!&=&\!
5 u_x^8 - 280 u_x^4 u_{2x}^2 - 112 u_{2x}^4 + 224 u_x^2 u_{3x}^2
- 64 u_{4x}^2, \\
\label{sinegordonlightconeconslaw4bis}
J^{(4)} \!&=&\!
8 \alpha \left(
5 u_x^6 \cos u + 40 u_x^4 u_{2x} \sin u + 20 u_x^2 u_{2x}^2 \cos u
+ 16 u_{2x}^3 \sin u - 16 u_x^3 u_{3x} \cos u \right.
\nonumber \\
&& \left. - 48 u_x u_{2x} u_{3x} \sin u + 8 u_{3x}^2 \cos u \right).
\end{eqnarray}
There are infinitely many density-flux pairs (all of even rank).
Since $u_{xt} = (u_x)_t,$ one can view (\ref{sGlightconealpha}) 
as an evolution equation in a new variable, $U = u_x,$ and construct densities
as linear combinations with constant coefficients of monomials in $U$
and its $x-$derivatives.
As before, each monomial has a (pre-selected) rank.
To accommodate the transcendental term(s) one might be incorrectly tempted 
to linearly combine such monomials with functional coefficients $h_i(u)$ 
instead of constant coefficients $c_i.$
For example, however, for rank $R = 2,$ one should take $\rho = c_1 u_x^2$ 
instead of $\rho = h_1(u) u_x^2,$ because the latter would lead to
$\D_t \rho = h_1^{\prime} u_t u_x^2 + 2 h_1 u_x u_{2x}$ and
$u_t$ cannot be replaced from (\ref{sGlightconealpha}).
\subsection{Examples of Equations with Transcendental Nonlinearities}
\label{othertranscendentalequations}
In this section we consider additional PDEs of the form $u_{xt} = f(u),$
where $f(u)$ has transcendental terms.
Using the Painlev\'e integrability test, researchers \cite{SAandGB2002a}
have concluded that the only PDEs of that type that are completely
integrable are equivalent to one of the standard forms of the nonlinear
Klein-Gordon equation \cite{MAandPCbook1991,SAandGB2002a}.
These standard forms (in light-cone coordinates) include the sine-Gordon
equation, $u_{xt} = \sin u,$ discussed in Section~\ref{sinegordonequation},
the sinh-Gordon equation $u_{xt} = \sinh u,$ the Liouville equation
$u_{xt} = {\rm e}^u,$ and the double Liouville equations,
$u_{xt} = {\rm e}^u \pm {\rm e}^{-2u}.$
The latter is also referred to in the literature as the Tzetzeica and
Mikhailov equations.
For each of these equations one can compute conservation laws with the
method discussed in Section~\ref{applsGequation}.
Alternatively, if these equations were transformed into laboratory
coordinates, one would apply the method of Section~\ref{applsGsystem}.
The multiple sine-Gordon equations, e.g.\ $u_{xt} = \sin u + \sin 2u,$
have only a finite number of conservation laws and are not completely
integrable, as supported by other evidence \cite{MAandPCbook1991}.

The sinh-Gordon equation, $u_{xt} = \sinh u$, arises as a special case of the
Toda lattice discussed in Section~\ref{todasection}.
It also describes the dynamics of strings in constant curvature
space-times \cite{ALandNprdS1996}.
In thermodynamics, the sinh-Gordon equation can be used to calculate
partition and correlation functions, and thus support Langevin simulations
\cite{AKandSHprl1997}.
In Table~\ref{tablesinhGordon}, we show a few density-flux pairs for the
sinh-Gordon equation in light-cone coordinates,  $u_{xt} = \alpha \sinh u.$
As with the sG equation (\ref{sGlightcone}),
$W(\partial/\partial x) = W(\partial/\partial t) = 1,$
$W(u) = 0,$ and $W(\alpha) = 2.$
The ranks in the first column of Table~\ref{tablesinhGordon} correspond
to the ranks of the densities, which are polynomial in $U = u_x$ and its
$x-$derivatives.
The sinh-Gordon equation has infinitely many conservation laws and is known 
to be completely integrable \cite{MAandPCbook1991}.
%
\begin{table}[hbt]
\caption{Conservation Laws of the sinh-Gordon equation,
$u_{xt} = \alpha \sinh u$}
\label{tablesinhGordon}
\begin{center}
\begin{tabular}{||c|l|l||}
\hline \hline
\rule[-7pt]{0pt}{21pt}Rank
& $ \hskip 2pc $\;Density $(\rho )$ & $ \hskip 7pc $\;\;Flux $(J)$
\\ \hline
\rule[-7pt]{0pt}{21pt}$2$ & $ u_x^2 $
& $ - 2 \alpha \cosh u $
\\ \hline
\rule[-7pt]{0pt}{21pt}$4$ & $ u_x^4 + 4 u_{2x}^2 $
& $ - 4 \alpha u_x^2 \cosh u $
\\ \hline
\rule{0pt}{14pt}$6$ & $ u_x^6 + 20 u_x^2 u_{2x}^2 + 8 u_{3x}^2 $
& $- 2 \alpha [ (3 u_x^4 + 4 u_{2x}^2 ) \cosh u + 8 u_x^2 u_{2x} \sinh u ]$
\\ \hline
\rule{0pt}{14pt}$8$ & $ 5 u_x^8 + 280 u_x^4 u_{2x}^2 - 112 u_{2x}^4 $
& $ - 8 \alpha \left[
(5 u_x^6 - 20 u_x^2 u_{2x}^2 + 16 u_x^3 u_{3x} +8 u_{3x}^2) \cosh u \right.$
\\
& $ \quad + 224 u_x^2 u_{3x}^2 + 64 u_{4x}^2 $
& $ \left.
\quad\;\;\; + (40 u_x^4 u_{2x} - 16 u_{2x}^3 + 48 u_x u_{2x} u_{3x}) \sinh u
\right]. $
\\
\hline \hline
\end{tabular}
\end{center}
\end{table}
\vspace{-2mm}

The Liouville equation, $u_{xt} = {\rm e}^u,$ plays an important role in
modern field theory \cite{LLbook1997}, e.g.\ in the theory of strings, 
where the quantum Liouville field appears as a conformal anomaly 
\cite{AKaam2002}.
The first few (of infinitely many) density-flux pairs for the Liouville
equation in light-cone coordinates,  $u_{xt} = \alpha {\rm e}^u,$ are given
in Table~\ref{tableLiouville}.
As before, $W(\partial/\partial x) = W(\partial/\partial t) = 1,$
$W(u) = 0,$ and $W(\alpha) = 2.$
The ranks in the table refer to the ranks of the densities.
Dodd and Bullough \cite{RDandRBprsa1977} have shown that the Liouville 
equation has no densities of ranks $3,5,$ and $7.$
As shown in Table~\ref{tableLiouville}, there are two densities of rank $6,$
and three densities of rank $8.$
Our results agree with those in \cite{RDandRBprsa1977}, where one can also
find the unique density of rank $9$ and the four independent densities of
rank $10.$

\begin{table}[t]
\caption{Conservation Laws of the Liouville equation,
$u_{xt} = \alpha {\rm e}^u$}
\label{tableLiouville}
\begin{center}
\begin{tabular}{||c|l|l||}
\hline \hline
\rule[-7pt]{0pt}{21pt}R
& $ \hskip 5pc $Density $(\rho )$ & $ \hskip 7pc $ Flux $(J)$
\\ \hline
\rule[-7pt]{0pt}{21pt}$2$ & $ u_x^2 $ & $ - 2 \alpha {\rm e}^u $
\\ \hline
\rule[-7pt]{0pt}{21pt}$4$ & $ u_x^4 + 4 u_{2x}^2 $
& $ - 4 \alpha u_x^2 {\rm e}^u $
\\ \hline
\rule{0pt}{14pt}$6$ & $ c_1 ( u_x^6 - 20 u_{2x}^3 - 12 u_{3x}^2 ) $
& $ - \alpha \left[ 6 c_1 (u_x^4 - 4 u_x^2 u_{2x} - 2 u_{2x}^2 ) \right.$ \\
\rule{0pt}{14pt} & $ + c_2 ( u_x^2 u_{2x}^2 + u_{2x}^3 + u_{3x}^2 ) $
& $ \left. \quad\;\; + c_2 u_{2x} ( 2 u_x^2 + u_{2x} ) \right] {\rm e}^u $
\\ \hline
\rule{0pt}{14pt}$8$ & $ c_1 ( u_x^8 - 56 u_x^2 u_{2x}^3 - 168 u_x^2 u_{3x}^2 $
& $- \alpha \left[
   8 c_1 ( u_x^6 - 6 u_x^4 u_{2x} + 3 u_x^2 u_{2x}^2 - 20 u_{2x}^3 \right.$ \\
\rule{0pt}{14pt} & $ \quad - 672 u_{2x} u_{3x}^2 - 144 u_{4x}^2 ) $
& $ \left. \quad\;\; - 36 u_x^3 u_{3x} - 108 u_x u_{2x} u_{3x} - 18 u_{3x}^2 )
\right. $ \\
\rule{0pt}{14pt} & $ + c_2(u_x^4 u_{2x}^2 + u_x^2 u_{2x}^3 +5 u_x^2 u_{3x}^2 $
& $ \left. \; + c_2(2 u_x^4 u_{2x} -u_x^2 u_{2x}^2 + 4 u_{2x}^3
  +8 u_x^3 u_{3x} \right.$ \\
\rule{0pt}{14pt} & $ \quad\;\; + 18 u_{2x} u_{3x}^2 + 4 u_{4x}^2 ) $
& $ \left. \quad\;\; + 24 u_x u_{2x} u_{3x} + 4 u_{3x}^2 ) \right. $ \\
\rule{0pt}{14pt} & $ + c_3 ( u_{2x}^4 + 3 u_x^2 u_{3x}^2 + 15 u_{2x} u_{3x}^2
+ 3 u_{4x}^2 ) $
& $ \left. \; + c_3 ( 4 u_{2x}^3 + 6 u_x^3 u_{3x} + 18 u_x u_{2x} u_{3x}
+ 3 u_{3x}^2 ) \right] {\rm e}^u $
\\ \hline \hline
\end{tabular}
\end{center}
\end{table}

\begin{table}[h]
\caption{Conservation Laws of the double Liouville equation,
$u_{xt} = \alpha ({\rm e}^u - {\rm e}^{-2u})$ }
\label{tabledoubleLiouville}
\begin{center}
\begin{tabular}{||c|l|l||}  \hline \hline
\rule[-7pt]{0pt}{21pt}Rank
& $ \hskip 3pc $Density $(\rho )$ & $ \hskip 6pc $ Flux $(J)$
\\ \hline
\rule[-7pt]{0pt}{21pt}$2$ & $ u_x^2 $
& $ - \alpha ( 2 {\rm e}^u + {\rm e}^{-2u} )$
\\ \hline
\rule[-7pt]{0pt}{21pt}$4$ & ------ & ------
\\ \hline
\rule{0pt}{14pt}$6$ & $u_x^6 + 15 u_x^2 u_{2x}^2 - 5 u_{2x}^3 + 3 u_{3x}^2 $
&$-3 \alpha \left[ (2 u_x^4 + 2 u_x^2 u_{2x} + u_{2x}^2 ) {\rm e}^u \right.$ \\
& & $ \left. \quad\;\; + ( u_x^4 - 8 u_x^2 u_{2x} + 2 u_{2x}^2 ) {\rm e}^{-2u}
\right] $ \\ \hline
\rule{0pt}{14pt}$8$ &$u_x^8 +42 u_x^4 u_{2x}^2 -14 u_x^2 u_{2x}^3 -7 u_{2x}^4$
&$ -\alpha \left[ ( 8 u_x^6 + 36 u_x^4 u_{2x} -18 u_x^2 u_{2x}^2 - 20 u_{2x}^3
    \right. $ \\
\rule{0pt}{14pt}& $\;\; + 21 u_x^2 u_{3x}^2 - 21 u_{2x} u_{3x}^2 + 3 u_{4x}^2$
& $ \left. \quad\;\; + 6 u_x^3 u_{3x} + 18 u_x u_{2x} u_{3x} + 3 u_{3x}^2)
{\rm e}^{u} \right. $ \\
& & $ \left. \quad\;\; + (4 u_x^6 - 72 u_x^4 u_{2x} - 18 u_x^2 u_{2x}^2
- 4 u_{2x}^3 \right. $ \\
\rule{0pt}{14pt} &
&$ \left. \quad\;\; + 48 u_x^3 u_{3x} - 72 u_x u_{2x} u_{3x} + 6 u_{3x}^2 )
{\rm e}^{-2\,u} \right] $ \\ \hline
\rule[-7pt]{0pt}{21pt}$10$ & ------ & ------ \\
\hline \hline
\end{tabular}
\end{center}
\end{table}

The double Liouville equations,
\begin{equation}
\label{doubleLiouville}
u_{xt} = {\rm e}^u \pm {\rm e}^{-2u},
\end{equation}
arise in the field of ``laser-induced vibrational predesorption of
molecules physisorbed on insulating substrates.''
More precisely, (\ref{doubleLiouville}) is used to investigate the dynamics 
of energy flow of excited admolecules on insulating substrates
\cite{YOetaljcp1997}.
Double Liouville equations are also relevant in studies of global properties
of scalar-vacuum configurations in general relativity and similarly 
systems in alternative theories of gravity \cite{KBapp2001}.

In Table~\ref{tabledoubleLiouville}, we show some density-flux pairs of
$u_{xt} = \alpha ( {\rm e}^u - {\rm e}^{-2u} ).$
There are no density-flux pairs for ranks $4$ and $10.$
We computed a density-flux pair for rank $12$ (not shown due to length).
The results for (\ref{doubleLiouville}) with the plus sign are similar.
\vskip 15pt
\noindent
{\Large \bf Part II: Nonlinear Differential-Difference Equations}
\vskip 8pt
\noindent
In the second part of this chapter we discuss two distinct methods to 
construct conservation laws of nonlinear DDEs.
The first method follows closely the technique for PDEs discussed in Part I.
It is quite effective for certain classes of DDEs, including the
Kac-van Moerbeke and Toda lattices, but far less effective for more
complicated lattices, such as the Bogoyavlenskii and the Gardner lattices.
The latter examples are treated with a new method 
based on a leading order analysis proposed by Hickman \cite{HickmanJNMP2008}.
\vspace{-2mm}
\section{Nonlinear DDEs and Conservation Laws}
\label{notationexamples}
We consider autonomous nonlinear systems of DDEs of the form
\begin{equation}
\label{DDEsystem}
{\dot{\bf u}}_n = {\bf F} ( {\bf u}_{n-l}, ...,{\bf u}_{n-1}, {\bf u}_{n},
                            {\bf u}_{n+1},...,{\bf u}_{n+m}),
\end{equation}
where ${\bf u}_{n}$ and ${\bf F}$ are vector-valued functions with $N$
components.
We only consider DDEs with one discrete variable, denoted by integer $n.$
In many applications, $n$ comes from a discretization of a space variable.
The dot stands for differentiation with respect to the continuous
variable which frequently is time, $t.$
We assume that ${\bf F}$ is polynomial with constant coefficients,
although this restriction can be waived for the method presented in
Section~\ref{newmethod}.
No restrictions are imposed on the degree of nonlinearity of ${\bf F}.$
If parameters are present in (\ref{DDEsystem}), they will be denoted by
lower-case Greek letters.

${\bf F}$ depends on ${\bf u}_n$ and a finite number of forward and
backward shifts of ${\bf u}_n.$
We identify $l$ with the furthest negative shift of any variable in the
system, and $m$ with the furthest positive shift of any variable in
the system.
No restrictions are imposed on the integers $l$ and $m,$ which measure the
degree of non-locality in (\ref{DDEsystem}).

By analogy with $\D_x,$ we define the shift operator $\D$ by
$\D {\bf u}_n = {\bf u}_{n+1}.$
The operator $\D$ is often called the {\em up-shift operator} or
forward- or right-shift operator.
Its inverse, $\D^{-1},$ is the {\em down-shift operator} or
backward- or left-shift operator, $\D^{-1} {\bf u}_n = {\bf u}_{n-1}.$
The action of the shift operators is extended to functions by acting on
their arguments.
For example,
\begin{eqnarray}
\label{Dfunction}
\D {\bf F} ({\bf u}_{n-l}, ...,{\bf u}_{n-1}, {\bf u}_{n},
{\bf u}_{n+1},...,{\bf u}_{n+m})
\!&\!=\!&\! {\bf F} (\D {\bf u}_{n-l}, ..., \D {\bf u}_{n-1},
\D {\bf u}_{n}, \D {\bf u}_{n+1},..., \D {\bf u}_{n+m})
\nonumber \\
\!&\!=\!&\! {\bf F} ({\bf u}_{n-l+1}, ...,{\bf u}_{n}, {\bf u}_{n+1},
{\bf u}_{n+2},...,{\bf u}_{n+m+1}).
\end{eqnarray}
Following \cite{MHandWHprsa2003}, we generate (\ref{DDEsystem}) from
\begin{equation}
\label{DDEsystemshifted}
{\dot{\bf u}}_0 = {\bf F}
( {\bf u}_{-l}, {\bf u}_{-l+1}, ...,{\bf u}_{-1}, {\bf u}_{0},
{\bf u}_{1},...,{\bf u}_{m-1},{\bf u}_{m}),
\end{equation}
where ${\dot{\bf u}}_n = \D^n {\dot{\bf u}}_0 = \D^n {\bf F}.$
To further simplify the notation, we denote the zero-shifted dependent
variable, ${\bf u}_{0},$ by ${\bf u}.$
Shifts of ${\bf u}$ are generated by repeated application of $\D.$
For instance, ${\bf u}_k = \D^k {\bf u}.$
The {\rm identity operator} is denoted by $\Id,$ where
$\D^0 {\bf u} = \Id {\bf u} = {\bf u}.$

A {\em conservation law} of (\ref{DDEsystemshifted}),
\begin{equation}
\label{conslawdde}
\D_t \, \rho + \Delta \, J = 0,
\end{equation}
links a {\em conserved density} $\rho$ to an {\em associated flux} $J,$
where both are scalar functions depending on ${\bf u}$ and its shifts.
In (\ref{conslawdde}), which holds on solutions of (\ref{DDEsystemshifted}),
$\D_t$ is the total derivative with respect to time and
$\Delta = \D - \Id$ is the {\rm forward difference operator}.

For readability (in particular, in the examples), the components of
${\bf u}$ will be denoted by $u, v, w, $ etc.\
In what follows we consider only autonomous functions,
i.e.\ ${\bf F}, \rho,$ and $J$ do not explicitly depend on $t.$

The time derivatives are defined in a similar way as in the continuous case,
see (\ref{kdvDtrho3}) and (\ref{operatordtforuandv}).
We show the discrete analog of (\ref{operatordtforuandv}).
For a density
$\rho(u_{p}, u_{p+1}, \cdots, u_q, v_r, v_{r+1}, \cdots, v_s)$, involving
two dependent variables $(u, v)$ and their shifts, the time derivative is
computed as
\begin{eqnarray}
\label{Dtdiscreteuv}
\D_t \rho \!&\!=\!&\!
\sum_{k=p}^{q} \frac{\partial \rho}{\partial u_k} {\dot u}_k
+ \sum_{k=r}^{s} \frac{\partial \rho}{\partial v_k} {\dot v}_k
\nonumber \\
\!&\!=\!&\! \left(
\sum_{k=p}^{q} \frac{\partial \rho}{\partial u_k} \D^k \right) {\dot u}
+ \left(
\sum_{k=r}^{s} \frac{\partial \rho}{\partial v_k} \D^k \right) {\dot v},
\end{eqnarray}
since $\D$ and ${\rm d}/{\rm d}t$ commute.
Obviously, the difference operator extends to functions.
For example, $\Delta J = \D \, J - J$ for a flux, $J.$

A density is {\em trivial} if there exists a function $\psi$ so that
$\rho = \Delta \psi.$
Similar to the continuous case, we say that two densities, $\rho^{(1)}$
and $\rho^{(2)},$ are {\em equivalent} if and only if
$\rho^{(1)} + k \rho^{(2)} = \Delta \psi,$ for some $\psi$ and some
non-zero scalar $k.$

It is paramount that the density is free of equivalent terms for if such
terms were present, they could be moved into the flux $J.$
Instead of working with different densities, we will use the equivalence
of monomial terms $t_i$ in the same density (of a fixed rank).
Compositions of $\D$ or $\D^{-1}$ define an {\em equivalence relation\/}
$(\equiv)$ on monomial terms.
Simply stated, all shifted terms are {\em equivalent}, e.g.\
$u_{-1} v_{1} \equiv u v_{2} \equiv u_{2} v_{4} \equiv u_{-3} v_{-1}$
since
\begin{equation}
\label{equivalenceexample}
u_{-1} v_1 = u v_2 - \Delta \, ( u_{-1} v_1 )
= u_2 v_4 - \Delta \, ( u_1 v_3 + u v_2 + u_{-1} v_1 )
= u_{-3} v_{-1} + \Delta \, ( u_{-2} v + u_{-3} v_{-1}).
\end{equation}
This equivalence relation holds for any function of the dependent variables,
but for the construction of conserved densities we will apply it only to
monomials.

In the algorithm used in Sections~\ref{applkvmlattice},~\ref{appltodalattice}, 
and~\ref{applALlattice}, we will use the following {\em equivalence criterion}:
two monomial terms, $t_1$ and $t_2$, are equivalent, $t_1 \equiv t_2,$
if and only if $t_1 = \D^r \, t_2$ for some integer $r.$
Obviously, if $t_1 \equiv t_2,$ then $t_1 = t_2 + \Delta J$
for some polynomial $J,$ which depends on ${\bf u}$ and its shifts.
For example, $u_{-2} u \equiv u_{-1} u_1$ because
$u_{-2} u = \D^{-1} u_{-1} u_1.$
Hence, $u_{-2} u = u_{-1} u_1 + [- u_{-1} u_1 + u_{-2} u ]
= u_{-1} u_1 + \Delta J,$ with $J = -u_{-2} u.$

For efficiency we need a criterion to choose a unique representative from
each equivalence class.
There are a number of ways to do this.
We define the {\em canonical} representative as that member that has
(i) no negative shifts and (ii) a non-trivial dependence on the {\em local}
(that is, zero-shifted) variable.
For example, $u u_2$ is the canonical representative of the class
$\{ \cdots, u_{-2} u, u_{-1} u_1, u u_2, u_1 u_3, \cdots \}.$
In the case of e.g.\ two variables $(u$ and $v)$,
$u_2 v$ is the canonical representative of the class
$\{ \cdots, u_{-1} v_{-3}, u v_{-2}, u_1 v_{-1}, u_2 v, u_3 v_1, \cdots \}.$

Alternatively, one could choose a variable ordering and then choose the
member that depends on the zero-shifted variable of lowest lexicographical
order.
The code in \cite{WHwebsite2004} uses lexicographical ordering
of the variables, i.e.\ $u \prec v \prec w,$ etc.
Thus, $u v_{-2}$ (instead of $u_2 v)$ is chosen as the canonical
representative of
$\{ \cdots, u_{-1} v_{-3}, u v_{-2}, u_1 v_{-1}, u_2 v, u_3 v_1, \cdots \}.$

It is easy to show \cite{HickmanJNMP2008} that if $\rho$ is a density then
$\D^k \rho$ is also a density.
Hence, using an appropriate ``up-shift" all negative shifts in a density can
been removed.
Thus, without loss of generality, we may assume that a density that depends
on $q$ shifts has {\em canonical} form
$\rho({\bf u}, {\bf u}_1, \cdots, {\bf u}_q).$
\vspace{-2mm}
\section{The Method of Undetermined Coefficients for DDEs}
\label{discretemuc}
In this section we show how polynomial conservation laws can be computed 
for a scalar DDE,
\begin{equation}
\label{scalardde}
{\dot u} = F( u_{-l}, u_{-l+1}, \cdots, u, \cdots, u_{m-1}, u_m ).
\end{equation}
The Kac-van Moerbeke example is used to illustrate the steps.
%
\subsection{A Classic Example: The Kac-van Moerbeke Lattice}
\label{kacvanmoerbeke}
The {\rm Kac-van Moerbeke (KvM) lattice} \cite{RHandJS1976,MKandPvMam1975},
also known as the Volterra lattice,
\begin{equation}
\label{kvmlatticeoriginal}
{\dot {u}}_n = u_n (u_{n+1} - u_{n-1}),
\end{equation}
arises in the study of Langmuir oscillations in plasmas, population dynamics,
etc.\
Eq.\ (\ref{kvmlatticeoriginal}) appears in the literature in various forms,
including
${\dot R}_n = \frac{1}{2} ( {\rm exp}(-R_{n-1}) - {\rm exp}(-R_{n+1} ) ),$
and ${\dot w}_n = w_n ( w_{n+1}^2 - w_{n-1}^2),$ which relate to
(\ref{kvmlatticeoriginal}) by simple transformations \cite{GTbook2000}.
We continue with (\ref{kvmlatticeoriginal}) and, adhering to the simplified
notation, write it as
\begin{equation}
\label{kvmlattice}
{\dot u} = u (u_1 - u_{-1}),
\end{equation}
or, with the conventions adopted above, $ {\dot u} = u (\D u - \D^{-1} u).$

Lattice (\ref{kvmlattice}) is invariant under the scaling symmetry
$(t, u) \rightarrow ({\lambda}^{-1} t, \lambda u).$
Hence, $u$ corresponds to one derivative with respect to $t,$
i.e.\ $u \sim \frac{\rm d}{\rm dt}.$
In analogy to the continuous case, we define the {\em weight} $W$ of a
variable as the exponent of $\lambda$ that multiplies the variable
\cite{UGandWHpd1998,UGandWHacm1999}.
We assume that shifts of a variable have the same weights,
that is, $W(u_{-1}) = W(u) = W(u_1).$
Weights of dependent variables are nonnegative and rational.
The {\em rank} of a monomial equals the total weight of the monomial.
An expression (or equation) is {\em uniform in rank} if all its monomial
terms have equal rank.

Applied to (\ref{kvmlattice}), $W({\rm d}/{\rm d} t) = W(\D_t) = 1$
and $W(u) = 1.$
Conversely, the scaling symmetry can be computed with linear algebra
as follows.
Setting $W({\rm d}/{\rm d}t) = 1$ and requiring that (\ref{kvmlattice})
is uniform in rank yields $W(u) + 1 = 2 W(u).$
Thus, $W(u) = 1,$ which agrees with the scaling symmetry.

Many integrable nonlinear DDEs are scaling invariant.
If not, they can be made so by extending the set of dependent variables
with parameters with weights.
Examples of such cases are given in Sections~\ref{ALlattice} 
and~\ref{gardnerlatticeinvariance}.

The KvM lattice has infinitely many polynomial density-flux pairs.
We give the conserved densities of rank $R \leq 4$ with associated fluxes
($J^{(4)}$ is omitted due to length):
%
\begin{eqnarray}
\label{kvmrho1J1}
\!\!\!\!\rho^{(1)} \!&\!=\!&\! u,
\quad \quad \quad \quad\quad\quad\quad\quad\quad\quad\,
J^{(1)} \!=\! - u u_{-1} ,  \\
\label{kvmrho1J2}
\!\!\!\!\rho^{(2)} \!&\!=\!&\! \frac{1}{2} u^2 + u u_1,
\quad\quad\quad\quad\quad\quad\;
J^{(2)} \!=\! -( u_{-1} u^2 - u_{-1} u u_1 ), \\
\label{kvmrho3J3}
\!\!\!\!\rho^{(3)} \!&\!=\!&\! \frac{1}{3} u^3 + u u_1 (u + u_1 + u_2), \;\,
J^{(3)} \!=\!
-( u_{-1} u^3 + 2 u_{-1} u^2 u_1 + u_{-1} u u_1^2 + u_{-1} u u_1 u_2 ), \\
\label{kvmrho4}
\!\!\!\!\rho^{(4)} \!&\!=\!&\!
\frac{1}{4} u^4 + u^3 u_1 + \frac{3}{2} u^2 u_1^2
+ u u_1^2 (u_1 + u_2) + u u_1 u_2 (u + u_1 + u_2 + u_3).
\end{eqnarray}
In addition to infinitely many polynomial conserved densities,
(\ref{kvmlattice}) has a non-polynomial density,
$\rho^{(0)} = \ln u$ with flux $J^{(0)} = - (u + u_{-1}).$
We discuss the computation of non-polynomial densities in 
Section~\ref{newmethod}.
%
\subsection{The Method of Undetermined Coefficients Applied to a Scalar 
Nonlinear DDE}
\label{applkvmlattice}
We outline how densities and fluxes can be constructed for a scalar DDE 
(\ref{scalardde}).
Using (\ref{kvmlattice}) as an example, we compute $\rho^{(3)}$ of rank
$R = 3$ and associated flux $J^{(3)}$ of rank $R = 4,$ both listed in 
(\ref{kvmrho3J3}).
\vskip 3pt
\noindent
$\bullet$
Select the rank $R$ of $\rho.$
Start from the set ${\cal V}$ of dependent variables
(including parameters with weight, when applicable), and form a set ${\cal M}$
of all non-constant monomials of rank $R$ or less (without shifts).
For each monomial in ${\cal M}$ introduce the right number of $t-$derivatives
to adjust the rank of that term.
Using the DDE, evaluate the $t-$derivatives, strip off the numerical
coefficients, and gather the resulting terms in a set ${\cal R}.$
For the KvM lattice (\ref{kvmlattice}),
${\cal V} = \{ u \}$ and ${\cal M} = \{ u^3, u^2, u \}.$
Since $u^3, u^2,$ and $u$ have ranks $3, 2,$ and $1,$ respectively,
one computes
\begin{equation}
\label{buildingblockskvm}
\frac{{\rm d}^0 u^3}{{\rm d}t^0} = u^3, \quad
\frac{{\rm d} u^2}{{\rm d}t}
= 2 u {\dot u} = 2 u^2 ( u_1 - u_{-1} )
= 2 u^2 u_1 - 2 u_{-1} u^2,
\end{equation}
and
\begin{eqnarray}
\frac{ {\rm d}^2 u }{ {\rm d} t^2 }
&\!=\!& \frac{ {\rm d} {\dot u} }{ {\rm d} t }
= \frac{ {\rm d} \left( u (u_1 - u_{-1}) \right) }{ {\rm d} t }
= {\dot u} (u_1 - u_{-1}) + u ({\dot u}_1 - {\dot u}_{-1})
\nonumber \\
&\!=\!& u (u_1 - u_{-1})^2 + u ( u_1 (u_2 - u) - u_{-1} (u - u_{-2}) )
\nonumber \\
&\!=\!&
u u_1^2 - 2 u_{-1} u u_1 + u_{-1}^2 u + u u_1 u_2 - u^2 u_1 - u_{-1} u^2 +
u_{-2} u_{-1} u,
\end{eqnarray}
where (\ref{kvmlattice}) (and its shifts) has been used to remove the time
derivatives.
Build ${\cal R}$ using the terms from the right hand sides of the equations
in (\ref{buildingblockskvm}) and ignoring numerical coefficients,
\begin{equation}
\label{listrkvmlattice}
{\cal R} \!=\! \{ u^3, u^2 u_1, u_{-1} u^2, u u_1^2, u_{-1} u u_1, u_{-1}^2 u,
                  u u_1 u_2, u_{-2} u_{-1} u \}.
\end{equation}
\vskip 3pt
\noindent
$\bullet$
Identify the elements in ${\cal R}$ that belong to the same equivalence
classes, replace them by their canonical representatives, and remove
all duplicates.
Call the resulting set ${\cal S},$ which has the building blocks of a
candidate density.
Continuing with (\ref{listrkvmlattice}),
$u_{-2} u_{-1} u \equiv u_{-1} u u_1 \equiv u u_1 u_2.$
Likewise, $u_{-1} u^2 \equiv u u_1^2$ and $ u_{-1}^2 u \equiv u^2 u_1.$
Thus, $ {\cal S} = \{ u^3, u^2 u_1, u u_1^2, u u_1 u_2 \} .$
\vskip 5pt
\noindent
$\bullet$
Form an arbitrary linear combination of the elements in ${\cal S}.$
This is the candidate $\rho.$
Continuing with the example,
\begin{equation}
\label{kvmrho3candidate}
\rho = c_1 \, u^3 + c_2 \, u^2 u_1 + c_3 \, u u_1^2 + c_4 \, u u_1 u_2.
\end{equation}
\vskip 3pt
\noindent
$\bullet$
Compute
\begin{equation}
\label{Dtdiscreteu}
\D_t \rho
= \sum_{k=0}^{q} \frac{\partial \rho}{\partial u_k} {\dot u}_k
= \left( \sum_{k=0}^{q} \frac{\partial \rho}{\partial u_k} \D^k
\right) {\dot u},
\end{equation}
where $q$ is the highest shift in $\rho.$
Using (\ref{kvmrho3candidate}) where $q = 2,$
\begin{eqnarray}
\label{kvmDtrho3explicit}
\!\!\!\!\!\!\!\!\!\!\D_t \rho \!&\!=\!&\!
\left(
 \frac{\partial \rho}{\partial u} \Id
 + \frac{\partial \rho}{\partial u_1} \D
 + \frac{\partial \rho}{\partial u_2} \D^2
 \right) {\dot u} \nonumber \\
\!\!&\!=\!&\! \left(
( 3 c_1 u^2 + 2 c_2 u u_1 + c_3 u_1^2 + c_4 u_1 u_2 ) \Id
\!+\! ( c_2 u^2 + 3 c_3 u u_1 + c_4 u u_2 ) \D
\!+\! c_4 u u_1 \D^2 \right) {\dot u}.
\end{eqnarray}
\vskip 2pt
\noindent
$\bullet$
Evaluate $\D_t \rho$ on the DDE (\ref{scalardde}) by replacing ${\dot u}$
by $F.$
Call the result $E.$
In (\ref{kvmlattice}), $F = u (u_1 - u_{-1}).$
The evaluated form of (\ref{kvmDtrho3explicit}) is
\begin{eqnarray}
\label{kvmDtrho3evaluated}
\!\!\!E \!&\!=\!&\!
( 3 c_1 u^2 + 2 c_2 u u_1 + c_3 u_1^2 + c_4 u_1 u_2 ) u (u_1 - u_{-1})
+ ( c_2 u^2 + 2 c_3 u u_1 + c_4 u u_2 ) u_1 (u_2 - u)
\nonumber \\
&& + ( c_4 u u_1 ) u_2 (u_3 - u_1)
\nonumber \\
\!&\!=\!&\!
(3 c_1 - c_2) u^3 u_1 - 3 c_1 u_{-1} u^3 + 2 (c_2 - c_3) u^2 u_1^2
- 2 c_2 u_{-1} u^2 u_1 + c_3 u u_1^3 - c_3 u_{-1} u u_1^2
\nonumber \\
\!&\!&\! - c_4 u_{-1} u u_1 u_2 + (c_2 - c_4) u^2 u_1 u_2 + 2 c_3 u u_1^2 u_2
+ c_4 u u_1 u_2^2 + c_4 u u_1 u_2 u_3.
\end{eqnarray}
\vskip 3pt
\noindent
$\bullet$
Set $J=0.$
Transform $E$ into its canonical form.
In doing so modify $J$ so that $E + \Delta \, J$ remains unchanged.
For example in (\ref{kvmDtrho3evaluated}),
replace $-3 c_1 u_{-1} u^3$ in $E$ by $-3 c_1 u u_1^3$ and
add $-3 c_1 u_1 u^3$ to $J$ since $u u_1^3 - [u u_1^3 - u_{-1} u^3 ].$
Do the same for all the other terms which are not in canonical form.
After grouping like terms, (\ref{kvmDtrho3evaluated}) becomes
\vskip 3pt
\noindent
\begin{eqnarray}
\label{patternmatchedEkvmlattice}
E &=&
(3 c_1 - c_2) u^3 u_1 + (c_3 - 3 c_1) u u_1^3 + 2 (c_2 - c_3) u^2 u_1^2
+ 2 (c_3 - c_2) u u_1^2 u_2
\nonumber \\
&& + (c_4 - c_3) u u_1 u_2^2 + (c_2 - c_4) u^2 u_1 u_2,
\end{eqnarray}
with
\begin{equation}
\label{candidateJ3kvmlattice}
J = - ( 3 c_1 u_{-1} u^3 + 2 c_2 u_{-1} u^2 u_1 + c_3 u_{-1} u u_1^2
    + c_4 u_{-1} u u_1 u_2 ).
\end{equation}
\vskip 1pt
\noindent
$\bullet$ $E$ is now the obstruction to $\rho$ being a density.
Set $E=0$ and solve for the undetermined coefficients $c_i.$
Thus,
\begin{equation}
3 c_1 - c_2 = 0,\; 3 c_1 - c_3 = 0,\; c_2 - c_3 = 0,\; c_3 - c_4 = 0,\;
c_2 - c_4 = 0,
\end{equation}
which yields $c_2 = c_3 = c_4 = 3 c_1,$ where $c_1$ is arbitrary.
\vskip 3pt
\noindent
$\bullet$
Substitute the solution for the $c_i$ into the candidates for $\rho$ and $J$
to obtain the final density and associated flux
(up to a common arbitrary factor which can be set to 1 or any other
nonzero value).
For the example, setting $c_1 = \frac{1}{3}$ and substituting
$c_2 = c_3 = c_4 = 1$ into (\ref{kvmrho3candidate}) and
(\ref{candidateJ3kvmlattice}) yields $\rho^{(3)}$ and $J^{(3)}$
as given in (\ref{kvmrho3J3}).
%
%
\vspace{-2mm}
\section{Discrete Euler and Homotopy Operators}
\label{toolsDDEs}
For simplicity, we will consider the case of only one discrete (dependent)
variable $u.$
First, we remove negative shifts from $E.$
Thus, $\tilde{E}=\D^l E$ where $l$ corresponds to the lowest shift in $E.$
\vskip 6pt
\noindent
{\bf The discrete variational derivative (discrete Euler operator)}
\vskip 5pt
\noindent
An expression $E$ is {\em exact} if and only if it is a total difference.
The following exactness test is well-known
\cite{VAetaltmp2000,MHandWHprsa2003}:
A necessary and sufficient condition for a function $E,$ with positive shifts,
to be exact is that ${\cal L}^{(0)}_u E \equiv 0,$
where ${\cal L}^{(0)}_u$ is the {\em discrete variational derivative}
(discrete Euler operator of order zero) \cite{VAetaltmp2000} defined by
\begin{eqnarray}
\label{discreteeuleroperatoru}
{\cal L}^{(0)}_u E
\!&\!=\!&\!
\frac{\partial }{\partial u} \left( \sum_{k=0}^{m} \D^{-k} \right) E
= \frac{\partial }{\partial u} \left( \Id + \D^{-1}
  + \D^{-2} + \D^{-3} + \cdots + \D^{-m} \right) E,
\nonumber \\
\!&\!=\!&\!
\Id \frac{\partial E}{\partial u}
+ \D^{-1} \frac{\partial E}{\partial u_1}
+ \D^{-2} \frac{\partial E}{\partial u_2}
+ \D^{-3} \frac{\partial E}{\partial u_3}
+ \cdots + \D^{-m} \frac{\partial E}{\partial u_m},
\end{eqnarray}
where $m$ is highest shift occurring in $E.$
\vskip 5pt
\noindent
{\bf Application}.
We return to (\ref{kvmDtrho3evaluated}) where $l = 1$.
Therefore,
\begin{eqnarray}
\label{tildeEkvmlattice}
{\tilde E} = \D E
\!&\!=\!&\! (3 c_1 - c_2) u_1^3 u_2 - 3 c_1 u u_1^3
+ 2 (c_2 - c_3) u_1^2 u_2^2 - 2 c_2 u u_1^2 u_2 + c_3 u_1 u_2^3
- c_3 u u_1 u_2^2
\nonumber \\
\!&\!&\! - c_4 u u_1 u_2 u_3 + (c_2 - c_4) u_1^2 u_2 u_3
+ 2 c_3 u_1 u_2^2 u_3 + c_4 u_1 u_2 u_3^2 + c_4 u_1 u_2 u_3 u_4,
\end{eqnarray}
which is free of negative shifts.
Applying (\ref{discreteeuleroperatoru}) to (\ref{tildeEkvmlattice}),
where $m = 4,$ gives
\begin{eqnarray}
\label{eulerzerouE}
{\cal L}^{(0)}_u {\tilde E}
&\!=\!& \frac{\partial }{\partial u} \left(
\Id + \D^{-1} + \D^{-2} + \D^{-3} +\D^{-4}
\right) {\tilde E}
\nonumber \\
\!&\!=\!&\! 3 (3 c_1 - c_2) u^2 u_{1} + 3 (c_3 - 3 c_1) u_{-1} u^2
+ 2 (c_2 - c_4) u u_1 u_2 + 4 (c_2 - c_3) u u_1^2
\nonumber\\
\!&\!&\! + 4 (c_3 - c_2) u_{-1} u u_1 + 2 (c_3 - c_2) u_1^2 u_2
+ (c_3 - 3 c_1) u_1^3 + (c_4 - c_3) u_{-1} u_1^2
\nonumber\\
\!&\!&\! + (c_4 - c_3) u_1 u_2^2 + (3 c_1 - c_2) u_{-1}^3
+ (c_2 - c_4) u_{-1}^2 u_1 + 4 (c_2 - c_3) u_{-1}^2 u
\nonumber\\
\!&\!&\! + 2 (c_3 -c_2) u_{-2} u_{-1}^2 + 2 (c_4 - c_3 ) u_{-2} u_{-1} u
+ (c_2 - c_4) u_{-2}^2 u_{-1},
\end{eqnarray}
which, as expected, vanishes identically when
$c_1 = \frac{1}{3}, c_2 = c_3 = c_4 = 1.$

Due to the large amount of terms generated by the Euler operator,
this method for finding the undetermined coefficients is much less efficient
than the ``splitting and shifting" approach illustrated on the same example
in Section~\ref{applkvmlattice}.
\vskip 6pt
\noindent
{\bf The discrete homotopy operator}
\vskip 5pt
\noindent
As in the continuous case, the discrete homotopy operator reduces the
inversion of the difference operator, $\Delta = \D - \Id,$ to a problem
of single-variable calculus.
Indeed, assuming that $E$ is exact and free of negative shifts,
the flux $J = \Delta^{-1} E$ can be computed without ``summation by parts."
Instead, one computes a single integral with respect to an auxiliary
variable denoted by $\lambda$ (not to be confused with $\lambda$ in
Section~\ref{kacvanmoerbeke}).

Consider an exact expression $E$ (of one variable $u),$ free of negative
shifts, and with highest shift $m.$
The {\em discrete homotopy operator} is defined
\cite{PHandEMfcm2004,EMandPHams2002,EMandRQcrm2004} by
\begin{equation}
\label{discretehomotopyscalaru}
{\cal H}_u E =
\int_{0}^{1} \left( I_u E \right) [\lambda u] \, \frac{d \lambda}{\lambda},
\end{equation}
with
\begin{equation}
\label{integranddiscretehomotopyscalaru}
I_u E = \sum_{k=1}^{m} \left(
  \sum_{i=0}^{m-k} u_i \frac{\partial }{\partial u_i} \right) \D^{-k} E
= \sum_{k=1}^{m} \left( \sum_{i=1}^{k} \D^{-i} \right)
  u_k \frac{\partial E}{\partial u_k},
%
\end{equation}
where $(I_u E) [\lambda u]$ means that in $I_u E$ one replaces
$u \rightarrow \lambda u, \, u_1 \rightarrow \lambda u_1,\,
u_2 \rightarrow \lambda u_2,\, {\rm etc.}$
The formulas in (\ref{integranddiscretehomotopyscalaru}) are
equivalent to the one in \cite{WHandBDandDPmcs2007}, which in
turn is equivalent to the formula in terms of discrete higher Euler operators
\cite{WHetalbirkhauser2005,WHetalcrm2004}.
Given an exact function $E$ without negative shifts one has
$J = \Delta^{-1} E = {\cal H}_u E.$
A proof can be found in \cite{PHandEMfcm2004,EMandRQcrm2004}.
\vskip 5pt
\noindent
{\bf Application.}
Upon substitution of $c_1 = \frac{1}{3}, c_2 = c_3 = c_4 = 1$ into
(\ref{tildeEkvmlattice}), we obtain
\begin{equation}
\label{tileEexactkvm}
{\tilde E} \!=\! \D E \!=\!
 - u u_1^3 - 2 u u_1^2 u_2 + u_1 u_2^3 - u u_1 u_2^2 - u u_1 u_2 u_3
 + 2 u_1 u_2^2 u_3 + u_1 u_2 u_3^2 + u_1 u_2 u_3 u_4,
\end{equation}
where the highest shift is $m = 4.$
Using (\ref{integranddiscretehomotopyscalaru}),
\begin{eqnarray}
\label{kvmIufortildeE}
I_u {\tilde E} \!\!&\!=\!&\!\!
\sum_{k=1}^{4} \left( \sum_{i=1}^{k} \D^{-i} \right)
 u_k \frac{\partial {\tilde E}}{\partial u_k}
= ( \D^{-1} ) u_1 \frac{\partial {\tilde E}}{\partial u_1}
+ ( \D^{-1} + \D^{-2} ) u_2 \frac{\partial {\tilde E}}{\partial u_2}
\nonumber \\
\!&\!&\!
+ ( \D^{-1} + \D^{-2} + \D^{-3} ) u_3 \frac{\partial {\tilde E}}{\partial u_3}
+ ( \D^{-1} + \D^{-2} + \D^{-3} + \D^{-4} )
u_4 \frac{\partial {\tilde E}}{\partial u_4}
\nonumber \\
\!&\!=\!&\!
\D^{-1} u_1
  ( - 3 u u_1^2 - 4 u u_1 u_2 + u_2^3 - u u_2^2 - u u_2 u_3 + 2 u_2^2 u_3
   + u_2 u_3^2 + u_2 u_3 u_4 )
\nonumber \\
\!&\!&\! + ( \D^{-1} + \D^{-2} ) u_2
 ( - 2 u u_1^2 + 3 u_1 u_2^2 - 2 u u_1 u_2 - u u_1 u_3 + 4 u_1 u_2 u_3
  + u_1 u_3^2 + u_1 u_3 u_4 )
\nonumber \\
\!&\!&\! + ( \D^{-1} + \D^{-2} + \D^{-3} ) u_3
 ( - u u_1 u_2 + 2 u_1 u_2^2 + 2 u_1 u_2 u_3 + u_1 u_2 u_4 )
\nonumber \\
\!&\!&\! + ( \D^{-1} + \D^{-2} +  \D^{-3} + \D^{-4} ) u_4 ( u_1 u_2 u_3 )
\nonumber \\
\!&\!=\!&\! 4 ( u u_1^3 + 2 u u_1^2 u_2 + u u_1 u_2^2 + u u_1 u_2 u_3 ),
\end{eqnarray}
which has the correct terms of $J^{(3)}$ but incorrect coefficients.
Finally, using (\ref{discretehomotopyscalaru})
\begin{eqnarray}
\label{tildeJkvm}
{\tilde J} \!&\!=\!&\! {\cal H}_u (- {\tilde E} )
= - \int_0^1\! (I_u {\tilde E})[\lambda u] \,\frac{d\lambda}{\lambda}
\nonumber \\
\!&\!=\!&\! - 4 \int_0^1\! \left(
u u_1^3 + 2 u u_1^2 u_2 + u u_1 u_2^2 + u u_1 u_2 u_3 \right) \lambda^3
\, d\lambda
\nonumber \\
\!&\!=\!&\! - (u u_1^3 + 2 u u_1^2 u_2 + u u_1 u_2^2 + u u_1 u_2 u_3).
\end{eqnarray}
Recall that ${\tilde E} = \D E.$
Hence,
\begin{equation}
\label{Jkvm}
J = \D^{-1} {\tilde J} =
- (u_{-1} u^3 + 2 u_{-1} u^2 u_1 + u_{-1} u u_1^2 + u_{-1} u u_1 u_2),
\end{equation}
which corresponds to $J^{(3)}$ in (\ref{kdvrho3J3}).

The homotopy method is computationally inefficient.
Even for a simple example, like (\ref{kvmIufortildeE}), the integrand has a
large number of terms, most of which eventually cancel.
To compute the flux we recommend the ``splitting and shifting" approach
which was illustrated (on the same example) in Section~\ref{applkvmlattice}.

The generalization of the exactness test to an expression $E$ with multiple
dependent variables $(u,v,\cdots)$ is straightforward.
For example, an expression $E$ of discrete variables $u,v$ and their
forward shifts will be exact if and only if
${\cal L}^{(0)}_{\bf u} E =
( {\cal L}^{(0)}_u E, {\cal L}^{(0)}_v E ) \equiv (0,0),$
where ${\cal L}^{(0)}_v$ is defined analogously to
(\ref{discreteeuleroperatoru}).
Similar to the continuous case, the homotopy operator formulas
(\ref{discretehomotopyscalaru}) and (\ref{integranddiscretehomotopyscalaru})
straightforwardly generalize to multiple dependent variables.
The reader is referred to
\cite{WHetalbirkhauser2005,WHandBDandDPmcs2007,WHetalcrm2004}
for details.
%
%
\vspace{-2mm}
\section{Conservation Laws of Nonlinear Systems of DDEs}
\label{systemsDDEs}
We use the method discussed in Section~\ref{applkvmlattice} 
to compute conservation laws for the Toda and Ablowitz-Ladik lattices. 
Using the latter lattice, we illustrate a ``divide and conquer" strategy, 
based on multiple scales, which allows on to circumvent difficulties in 
the computation of densities of DDEs that are not dilation invariant.
%
\subsection{The Toda Lattice}
\label{todasection}
One of the earliest and most famous examples of completely integrable DDEs 
is the Toda lattice \cite{MTjpsj1967,MTbook1981},
\begin{equation}
\label{todalatticeexponential}
{\ddot{y}}_n = \exp{(y_{n-1} - y_n)} - \exp{(y_n - y_{n+1})},
\end{equation}
where $y_n$ is the displacement from equilibrium of the $n\/$th particle
with unit mass under an exponential decaying interaction force between
nearest neighbors.
With the change of variables,
$u_n = {\dot{y}}_n, v_n = \exp{(y_{n} - y_{n+1})},$
lattice (\ref{todalatticeexponential}) can be written in algebraic form
\begin{equation}
\label{todalatticeoriginal}
{\dot{u}}_n = v_{n-1} - v_n, \quad {\dot{v}}_n = v_n (u_n - u_{n+1}).
\end{equation}
Adhering to the simplified notation, we continue with
\begin{equation}
\label{todalattice}
{\dot u} = v_{-1} - v, \quad {\dot v} = v (u - u_{1}).
\end{equation}
As before, we set $W({\rm d}/{\rm d} t) = 1,$ assign unknown weights,
$W(u), W(v),$ to the dependent variables, and require that each equation
in (\ref{todalattice}) is uniform in rank.
This yields
\begin{equation}
W(u) + 1 = W(v), \quad\quad W(v) + 1 = W(u) + W(v).
\end{equation}
The solution $W(u) = 1, W(v) = 2$ reveals that (\ref{todalattice})
is invariant under the scaling symmetry
\begin{equation}
\label{scalingtoda}
(t, u, v) \rightarrow (\lambda^{-1} t, \lambda u, {\lambda}^{2} v),
\end{equation}
where $\lambda$ is an arbitrary parameter.

The Toda lattice has infinitely many conservation laws \cite{MHphysrev1974}.
The first two density-flux pairs are easy to compute by hand.
Here we give the densities of rank $R \leq 4$ with associated fluxes,
$J^{(4)}$ being omitted due to length:
\begin{eqnarray}
\label{todarho1J1}
\rho^{(1)} \!&\!=\!&\! u,
\quad \quad\quad\quad\quad\quad\quad\quad\quad\,
J^{(1)} \!=\! v_{-1},  \\
\label{todarho1J2}
\rho^{(2)} \!&\!=\!&\! \frac{1}{2} u^2 + v,
\quad\quad\quad\quad\quad\quad\;
J^{(2)} \!=\! u v_{-1}, \\
\label{todarho3J3}
\rho^{(3)} \!&\!=\!&\! \frac{1}{3} u^3 + u (v_{-1} + v), \quad\quad\,
J^{(3)} \!=\! u_{-1} u v_{-1} + v_{-1}^2, \\
\label{todarho4}
\rho^{(4)} \!&\!=\!&\!
\frac{1}{4} u^4 + u^2 (v_{-1} + v) + u u_1 v + \frac{1}{2} v^2 + v v_1.
\end{eqnarray}
%
\subsection{Computation of a Conservation Law of the Toda Lattice}
\label{appltodalattice}
As an example, we compute density $\rho^{(3)}$ (of rank $R = 3)$ and 
associated flux $J^{(3)}$ (of rank $4)$ in (\ref{todarho3J3}).
\vskip 5pt
\noindent
{\bf Step 1: Construct the form of the density}
\vskip 4pt
\noindent
Start from ${\cal V} = \{ u, v \},$ i.e.\ the set of dependent variables
with weight.
List all monomials in $u$ and $v$ of rank $R = 3$ or less:
${\cal M} \!=\! \{ u^3, u^2, u v, u, v \}.$
Next, for each monomial in ${\cal M}$, introduce the correct number of
$t$-derivatives so that each term has rank $3.$
Using (\ref{todalattice}), compute
\begin{eqnarray}
\label{todaweightadjust}
&& \frac{{\rm d}^0 u^3}{ {\rm d}t^0} = u^3,
\quad\quad\quad\quad\quad\quad\quad\quad\quad
\frac{{\rm d}^0 u v}{{\rm d}t^0} = u v,
\nonumber \\
&& \frac{{\rm d} u^2}{{\rm d}t} = 2 u {\dot u} = 2 u v_{-1} - 2 u v,
\quad\quad
\frac{{\rm d} v}{{\rm d}t} = {\dot v} =  u v - u_1 v,
\nonumber \\
&& \frac{{\rm d}^2 u}{{\rm d}t^2} = \frac{{\rm d}{\dot u}}{{\rm d}t}
   = \frac{{\rm d} (v_{-1} - v)}{{\rm d}t}
   = u_{-1} v_{-1} - u v_{-1} - u v + u_1 v.
\end{eqnarray}
Gather the terms in the right hand sides in (\ref{todaweightadjust}) to get
${\cal R} = \{ u^3, u v_{-1}, u v, u_{-1} v_{-1}, u_1 v \}.$

Identify members belonging to the same equivalence classes and replace them
by their canonical representatives.
For example, $u v_{-1} \equiv u_1 v.$
Adhering to lexicographical ordering, we will use $u v_{-1}$ instead of
$u_1 v.$
Doing so, replace ${\cal R}$ by ${\cal S} = \{ u^3, u v_{-1}, u v \}, $
which has the building blocks of the density.
Linearly combine the monomials in ${\cal S}$ with undetermined
coefficients $c_i$ to get the candidate density of rank $3:$
\begin{equation}
\label{formrho3toda}
\rho = c_1 \, u^3 + c_2 \, u v_{-1} + c_3 \, u v.
\end{equation}
\vskip 2pt
\noindent
{\bf Step 2: Compute the undetermined coefficients $c_i$}
\vskip 2pt
\noindent
Compute $\D_t \rho$ and use (\ref{todalattice}) to eliminate ${\dot u}$ and
${\dot v}$ and their shifts.
Thus,
\begin{eqnarray}
\label{Etodalattice}
E \!&\!=\!&\!
  ( 3 c_1 - c_2 ) u^2 v_{-1} + (c_3 - 3 c_1 ) u^2 v + (c_3 - c_2) v_{-1} v
  + c_2 u_{-1} u v_{-1}  + c_2 v_{-1}^2
\nonumber \\
\!&\!&\! - c_3 u u_1 v - c_3 v^2.
\end{eqnarray}
\vskip 2pt
\noindent
%
%
{\bf Step 3: Find the associated flux $J$}
\vskip 2pt
\noindent
Transform (\ref{Etodalattice}) into canonical form to obtain
\begin{equation}
\label{Etodacanonicalform}
E = ( 3 c_1 - c_2 ) u_1^2 v + (c_3 - 3 c_1 ) u^2 v + (c_3 - c_2) v v_1
    + c_2 u u_1 v + c_2 v^2 - c_3 u u_1 v - c_3 v^2
\end{equation}
with
\begin{equation}
\label{formJ3todacanonicalform}
J = (3 c_1 - c_2) u^2 v_{-1} + (c_3 - c_2) v_{-1} v + c_2 u_{-1} u v_{-1}
    + c_2 v_{-1}^2.
\end{equation}
Set $E = 0$ to get the linear system
\begin{equation}
\label{systemtodacanonicalform}
3 c_1 - c_2 = 0, \quad c_3 - 3 c_1 = 0, \quad c_2 - c_3 = 0.
\end{equation}
Set $c_1 = \frac{1}{3}$ and substitute the solution
$c_1 = \frac{1}{3}, c_2 = c_3 = 1,$ into (\ref{formrho3toda}) and
(\ref{formJ3todacanonicalform}) to obtain $\rho^{(3)}$ and $J^{(3)}$
in (\ref{todarho3J3}).
%
%
\subsection{The Ablowitz-Ladik Lattice}
\label{ALlattice}
In \cite{MAandJLjmp1975,MAandJLsam1976,MAandJLjmp1976},
Ablowitz and Ladik derived and studied the following completely 
integrable discretization of the nonlinear Schr\"odinger equation:
\begin{equation}
\label{ablowitzladikcomplex}
i \, {\dot u}_n =
u_{n+1} - 2 u_n + u_{n-1} \pm u_n^{*} u_n ( u_{n+1} + u_{n-1} ),
\end{equation}
where $u_n^{*}$ is the complex conjugate of $u_n.$
We continue with (\ref{ablowitzladikcomplex}) with the plus sign;
the case with the negative sign is analogous.
Instead of splitting $u_n$ into its real and imaginary parts, we treat
$u_n$ and $v_n = u_n^{*}$ as independent variables and augment
(\ref{ablowitzladikcomplex}) with its complex conjugate equation,
\begin{eqnarray}
\label{ablowitzladikoriginal}
{\dot u}_n &=& (u_{n+1} - 2 u_n + u_{n-1}) + u_n v_n (u_{n+1} + u_{n-1}),
\nonumber \\
{\dot v}_n &=& -( v_{n+1} - 2 v_n + v_{n-1} ) -  u_n v_n (v_{n+1} + v_{n-1}),
\end{eqnarray}
where $i$ has been absorbed into a scale on $t.$
Since $v_n = u_n^{*}$ we have $W(v_n) = W(u_n).$
Neither equation in (\ref{ablowitzladikoriginal}) is dilation invariant.
To circumvent this problem we introduce an auxiliary parameter $\alpha$
with weight, and replace (\ref{ablowitzladikoriginal}) by
\begin{eqnarray}
\label{ablowitzladiklattice}
{\dot u} &=& \alpha ( u_1 - 2 u + u_{-1} ) + u v ( u_1 + u_{-1} ),
\nonumber \\
{\dot v} &=& - \alpha ( v_1 - 2 v + v_{-1} ) - u v ( v_1 + v_{-1} ),
\end{eqnarray}
presented in the simplified notation.
Both equations in (\ref{ablowitzladiklattice}) are uniform in rank provided
\begin{eqnarray}
\label{ablowitzladikweightequations}
W(u) + 1 &\!=\!& W(\alpha) + W(u) = 2 W(u) + W(v),
\nonumber \\
W(v) + 1 &\!=\!& W(\alpha) + W(v) = 2 W(v) + W(u),
\end{eqnarray}
which holds when $W(u) + W(v) = W(\alpha) = 1.$
Since $W(u) = W(v)$ we have $W(u) = W(v) = \frac{1}{2},$ and $W(\alpha) = 1,$
which expresses that (\ref{ablowitzladiklattice}) is invariant under
the scaling symmetry
\begin{equation}
\label{ablowitzladikscale}
(t, u, v, \alpha) \rightarrow (\lambda^{-1} t, \lambda^{\frac{1}{2}} u,
{\lambda}^{\frac{1}{2}} v, \lambda \alpha ).
\end{equation}
We give the conserved densities of (\ref{ablowitzladiklattice}) of ranks $2$
through $4.$
Only the flux of rank $3$ associated to $\rho^{(1)}$ is shown.
The others are omitted due to length.
\begin{eqnarray}
\label{rho1ablowitzladiklattice}
\rho^{(1)} \!&\!=\!&\! \alpha ( c_1 u v_{-1} + c_2 u v_1 ), \\
\label{J1ablowitzladiklattice}
J^{(1)} \!&\!=\!&\! - \alpha {\left(
c_1 ( \alpha u v_{-2} - \alpha u_{-1} v_{-1} + u_{-1} u v_{-2} v_{-1} )
+ c_2 ( \alpha u v - \alpha u_{-1} v_1 - u_{-1} u v v_1 ) \right)}, \\
\label{rho2ablowitzladiklattice}
\rho^{(2)} \!&\!=\!&\!
\alpha \left( c_1 ( \tfrac{1}{2} u^2 v_{-1}^2 + u u_1 v_{-1} v
+ \alpha u v_{-2} )
+ c_2 ( \tfrac{1}{2} u^2 v_1^2 + u u_1 v_1 v_2 + \alpha u v_2 ) \right), \\
\label{rho3ablowitzladiklattice}
\rho^{(3)} \!&\!=\!&\! \alpha \left(
c_1 \left[ \tfrac{1}{3} u^3 v_{-1}^3
+ u u_1 v_{-1} v ( u v_{-1} + u_1 v + u_2 v_1 )
+ \alpha u v_{-1} ( u v_{-2} + u_1 v_{-1} ) \right. \right. \nonumber \\
\!&\!&\! \left. \left.
+ \alpha u v (u_1 v_{-2} + u_2 v_{-1} ) + \alpha^2 u v_{-3} \right]
+ c_2 \left[ \tfrac{1}{3} u^3 v_1^3
+ u u_1 v_1 v_2 ( u v_1 + u_1 v_2 + u_2 v_3 ) \right. \right. \nonumber \\
\!&\!&\! \left. \left. + \alpha u v_2 ( u v_1 + u_1 v_2 )
+ \alpha u v_3 ( u_1 v_1 + u_2 v_2 ) + \alpha^2 u v_3 \right] \right),
\end{eqnarray}
where $c_1$ and $c_2$ are arbitrary constants.
Our results confirm those in \cite{MAandJLjmp1976}.
Since our method is restricted to polynomial densities we cannot compute the
density with a logarithmic term,
\begin{equation}
\label{hamiltonian}
\rho_n =
\left( u_n^{*} ( u_{n-1} + u_{n+1} ) - 2 \ln (1 + u_n u_n^{*}) \right),
\end{equation}
which corresponds \cite{MAandBHSIAMjam1990,MAandJLjmp1976} to the 
Hamiltonian of (\ref{ablowitzladikcomplex}), viz.\ $H = -i \sum_n \rho_n.$ 
%
\subsection{Computation of a Conservation Law of the Ablowitz-Ladik Lattice}
\label{applALlattice}
To make (\ref{ablowitzladikoriginal}) scaling invariant we had to introduce
an auxiliary parameter $\alpha.$
This complicates matters in two ways as we will show in this section.
First, to compute rather simple conserved densities, like $\rho = u v_{-1}$
and $\rho = u v_1,$ we will have to select rank $R = 2$ for which the
candidate density has twenty terms.
However, eighteen of these terms eventually drop out.
Second, conserved densities corresponding to lower ranks might appear in
the result.
These lower-rank densities are easy to recognize for they are multiplied with
arbitrary coefficients $c_i.$
Consequently, when parameters with weight are introduced, the densities
corresponding to distinct ranks are no longer linearly independent.

We compute $\rho^{(1)}$ of rank $R = 2$ in (\ref{rho1ablowitzladiklattice})
with associated flux $J^{(1)}$ in ({\ref{J1ablowitzladiklattice}).
Note that $\rho^{(1)}$ and $J^{(1)}$ cannot be computed with the steps below
when $R = 1.$
\vskip 5pt
\noindent
{\bf Step 1}: {\bf Construct the form of the density}
\vskip 4pt
\noindent
Augment the set of dependent variables with the parameter $\alpha$
(with non-zero weight).
Hence, ${\cal V} = \{ u, v, \alpha \}.$
Construct
\begin{equation}
\label{MlistALlattice}
{\cal M} = \{ u, v, u^2, u v, v^2, \alpha u, u^3, \alpha v, u^2 v, u v^2, v^3,
\alpha u^2, u^4, \alpha u v, u^3 v, \alpha v^2, u^2 v^2, u v^3, v^4 \},
\end{equation}
which contains all non-constant monomials of (chosen) rank $2$ or less
(without shifts).
As with the previous examples, for each element in ${\cal M}$ add the right
number of $t-$derivatives.
Use (\ref{ablowitzladiklattice}) to evaluate the $t-$derivatives, gather
the terms in the right hand sides, introduce the canonical representatives
(based on lexicographical ordering), and remove duplicates to get
\begin{eqnarray}
\label{Slistablowitzladiklattice}
{\cal S} \!&\!=\!&\!
\{ \alpha u^2, u^4, \alpha u u_1, \alpha u v_{-1}, \alpha u v,
u^3 v, u^2 u_1 v, u^2 v_{-1} v, \alpha v^2, u^2 v^2,
\nonumber \\
&& u u_1 v^2, u v_{-1} v^2, u v^3, v^4, \alpha u v_1,
u u_1^2 v_1, \alpha v v_1, u^2 v v_1, u v^2 v_1, u u_1 v_1^2 \}.
\end{eqnarray}
Linearly combine the monomials in ${\cal S}$ with undetermined coefficients
$c_i$ to get the candidate density of rank $2$:
\begin{eqnarray}
\label{formrho3ablowitzladiklattice}
\rho \!&\!=\!&\!
c_1 \, \alpha u^2 + c_2 \, u^4 + c_3 \, \alpha u u_1 + c_4 \, \alpha u v_{-1}
+ c_5 \, \alpha u v + c_6 \, u^3 v + c_7 \, u^2 u_1 v
\nonumber \\
\!&\!&\! + c_8 \, u^2 v_{-1} v + c_9 \, \alpha v^2 +
c_{10} \, u^2 v^2 + c_{11} \, u u_1 v^2 +
c_{12} \, u v_{-1} v^2 + c_{13} \, u v^3 + c_{14} \, v^4
\nonumber \\
\!&\!&\! + c_{15} \, \alpha u v_1 + c_{16} \, u u_1^2 v_1 +
c_{17} \, \alpha v v_1 + c_{18} \, u^2 v v_1 +
+ c_{19} \, u v^2 v_1 + c_{20} \, u u_1 v_1^2.
\end{eqnarray}
\vskip 3pt
\noindent
{\bf Step 2: Compute the undetermined coefficients $c_i$}
\vskip 2pt
\noindent
The computations proceed as in the examples in Sections~\ref{applkvmlattice}
and~\ref{appltodalattice}.
Thus, compute $\D_t \rho$ and use (\ref{ablowitzladiklattice}) to eliminate
${\dot u}$ and ${\dot v}$ and their shifts.
Next, bring the expression $E$ into canonical form to obtain the
linear system for the undetermined coefficients $c_i.$

Without showing the lengthy computations, one finds that all constants
$c_i = 0,$ except $c_4$ and $c_{15}$ which are arbitrary.
Substitute the coefficients into (\ref{formrho3ablowitzladiklattice})
to get $\rho^{(1)}$ in (\ref{rho1ablowitzladiklattice}).
\vskip 5pt
\noindent
{\bf Step 3}: {\bf Find the associated flux $J$}
\vskip 2pt
\noindent
The associated flux comes for free when $E$ is transformed into
canonical form.
Alternatively, one could apply the homotopy approach for multiple dependent
variables
\cite{WHetalbirkhauser2005,WHetalcrm2004} to compute the flux.
In either case, one gets $J^{(1)}$ in (\ref{J1ablowitzladiklattice}).
%
\subsection{A ``Divide and Conquer" Strategy}
\label{applALlatticedivideconquer}
It should be clear from the example in Section~\ref{applALlattice} 
that our method is not efficient if the densities are not of the form
(\ref{kvmrho1J1})-(\ref{kvmrho4}) for the KvM lattice and
(\ref{todarho1J1})-(\ref{todarho4}) for the Toda lattice.
Indeed, the densities for the AL lattice in
(\ref{rho1ablowitzladiklattice})-(\ref{rho3ablowitzladiklattice}) are
quite different in structure.
Therefore, in \cite{HEthesis2003}, Eklund presented alternate strategies
to deal more efficiently with DDEs, in particular with those involving
weighted parameters such as (\ref{ablowitzladiklattice}).

A first alternative is to work with {\em multiple scales} by setting either
$W(\D_t) = 0$ or $W(\D_t) = 1,$ the latter choice is what we have used
thus far.
A second possibility is to leave $W(\D_t)$ unspecified and, if needed,
introduce extra parameters with weight into the DDE.

Let us explore these ideas for the AL lattice (\ref{ablowitzladiklattice}),
which we therefore replace by
\begin{eqnarray}
\label{ablowitzladiklatticebeta}
{\dot u} &=& \alpha \beta ( u_1 - 2 u + u_{-1} )
+ \beta u v ( u_1 + u_{-1} ),
\nonumber \\
{\dot v} &=& - \alpha \beta ( v_1 - 2 v + v_{-1} )
- \beta u v ( v_1 + v_{-1} ),
\end{eqnarray}
where $\beta$ is a second auxiliary parameter with weight.
Requiring uniformity in rank leads to
\begin{eqnarray}
\label{ablowitzladikweightequationsbeta}
W(u) + W(\D_t) &\!=\!& W(\alpha) + W(\beta) + W(u) = W(\beta) + 2 W(u) + W(v),
\nonumber \\
W(v) + W(\D_t) &\!=\!& W(\alpha) + W(\beta) + W(v) = W(\beta) + 2 W(v) + W(u),
\end{eqnarray}
which are satisfied when $W(\alpha) = W(u) + W(v)$ and
$W(\D_t) = W(u) + W(v) + W(\beta).$
Therefore, we can set $W(\D_t) = a, W(u) = b,$ and $W(\beta) = c$ with
$a,b,c$ rational numbers so that $W(v) = a - b - c$ and $W(\alpha) = a - c$
are strictly positive.
Thus (\ref{ablowitzladiklatticebeta}) is dilation invariant under a
three-parameter family of scaling symmetries,
\begin{equation}
\label{ablowitzladikgeneralscale}
(t, u, v, \alpha, \beta) \rightarrow
({\lambda}^{-a} t, \lambda^{b} u, {\lambda}^{a-b-c} v,
{\lambda}^{a-c} \alpha, {\lambda}^{c} \beta).
\end{equation}
Scaling symmetry (\ref{ablowitzladikscale}) corresponds to the case where
$a = 1, b = \frac{1}{2},$ and $c = 0,$ more precisely, $\beta = 1.$
The fact that (\ref{ablowitzladiklatticebeta}) is invariant under multiple
scales is advantageous.
Indeed, one can use the invariance under one scale to construct a
candidate density and, subsequently, use additional scale(s) to split
$\rho$ into smaller densities.
This ``divide and conquer" strategy drastically reduces the complexity of
the computations.
The use of multiple scales has proven to be successful in the computation
of conservation laws for PDEs with more than one space variable
\cite{WHetalbirkhauser2005}.
%
\vskip 0.0001pt
\noindent
\begin{table}[bht]
\begin{center}
\begin{tabular}{|l|l|l|l|l|}
\hline
\!i\! & Rank
    & Candidate ${\rho}_i$
    & Final ${\rho}_i$
    & Final ${\bf J}_i $
\\ \hline\hline
\!1\!
&\! $a + 2b - c$\!
&\!$ c_1\alpha u^2 + c_3 \alpha u u_1 + c_6 u^3 v + c_7 u^2 u_1 v
   + c_{16} u u_1^2 v_1$\!
& $ 0 $
& $ 0 $
\\ \hline
\!2\!
& \!$4 b$\!
& $c_2 u^4 $
& $ 0 $
& $ 0 $
\\ \hline
\!3\!
& \!$2 (a - c)$\!
& \!$
c_4 \alpha u v_{-1} + c_5 \alpha u v + c_8 u^2 v_{-1} v + c_{10} u^2 v^2 $\!
& $ c_4 \alpha u v_{-1} $
& \!$J^{(1)} \; {\rm in} \; (\ref{J1ablowitzladiklattice})$\!
\\
& $ $
& \!$ + c_{11} u u_1 v^2  + c_{15} \alpha u v_1 + c_{18} u^2 v v_1
    + c_{20} u u_1 v_1^2 $\!
& \!$ + c_{15} \alpha u v_1 $\!
& $ $
\\ \hline
\!4\!
& \!$3 a - 2 b - 3 c $\!
& \!$ c_9 \alpha v^2 + c_{12} u v_{-1} v^2 + c_{13} u v^3
+ c_{17} \alpha v v_1 + c_{19} u v^2 v_1 $\!
& $ 0 $
& $ 0 $
\\ \hline
\!5\!
& \!$4 (a - b - c)$\!
& $ c_{14} v^4 $
& $ 0 $
& $ 0 $
\\ \hline
\end{tabular}
\caption{
\label{ablowitzladikcandidatedensities}
Candidate densities for the Ablowitz-Ladik lattice
(\ref{ablowitzladiklatticebeta}). }
\end{center}
\end{table}

Candidate density (\ref{formrho3ablowitzladiklattice}) is uniform of rank $2$
under (\ref{ablowitzladikscale}) but can be split into smaller pieces,
$\rho_i,$ using (\ref{ablowitzladikgeneralscale}), even without specifying
values for $a,b,$ and $c.$
Indeed, as shown in Table~\ref{ablowitzladikcandidatedensities},
$\rho$ in (\ref{formrho3ablowitzladiklattice}) can be split into
$\rho_1$ through $\rho_5$ of distinct ranks under
(\ref{ablowitzladikgeneralscale}).
Steps 2 and 3 of the algorithm are then carried out for each of these
$\rho_i, \, (i=1, \cdots, 5)$ separately.
As shown in the table, all but one density lead to a trivial result.
The longest density, $\rho_3,$ with 8 terms, leads to $\rho^{(1)}$ and
$J^{(1)}$ in (\ref{rho1ablowitzladiklattice}) and
(\ref{J1ablowitzladiklattice}), respectively.
\vspace{-2mm}
\section{A New Method to Compute Conservation Laws of Nonlinear DDEs}
\label{newmethod}
In the continuous case, the total derivative $\D_x$ has weight one.
Consequently, any density is bounded with respect to the order in $x.$
For example, as shown in Section~\ref{mucPDEs} for the KdV case,
the candidate density $\rho$ of rank $6$ in (\ref{kdvcandidaterho3})
is of first order (after removing the second and fourth order terms.)
Obviously, a density of rank $6$ could never have leading order terms of
fifth or higher order in $x$ because the rank of such terms would exceed $6.$

In the discrete case, however, the shift operator $\D$ has no weight.
Thus, any shift $\D^k u = u_k$ of a dependent variable $u$ has the same
weight as that dependent variable.
Simply put, $W(u_k) = W(u)$ for any integer $k.$
Consequently, using the total derivative $\D_t$ as a tool to construct a
(candidate) density has a major shortcoming: the density may lack terms
involving sufficiently high shifts of the variables.
As shown in Section~\ref{applkvmlattice} for the KvM lattice, the candidate
density $\rho$ of rank $3$ in (\ref{kvmrho3candidate}) has leading order
term $ u u_1 u_2,$ i.e.\ the term with the highest shift $(2$ in this example).
It is {\em a priori} not excluded that $\rho$ might have terms involving
higher shifts.
For example, $u u_1 u_3$ has rank $3$ and so do infinitely many other cubic
terms.
Note that for this example we constructed $\rho$ starting from powers of $u,$
viz.\ $u^3, u^2, u;$ and, by repeated differentiation, worked our way
``down" towards the leading order term $u u_1 u_2.$
In this section we outline the key features of a new method which goes in
the opposite direction: (i) first compute the leading order term and
subsequently (ii) compute the terms involving lower shifts.
In step (ii) only the necessary terms are computed, nothing more,
nothing less.
This method is fast and powerful for it circumvents the use of the
dilation invariance and the method of undetermined coefficients.
More importantly, the new method is not restricted to densities and fluxes
of polynomial form.
%
\subsection{Leading Order Analysis}
\label{leadingorderanalysis}
Consider a density, $\rho,$ that depends on $q$ shifts.
Since $\D^q \,\rho$ is also a density, we may, without loss of generality,
assume that $\rho$ has canonical form
$\rho(\mathbf{u}, \ \mathbf{u}_1, \ \ldots, \ \mathbf{u}_q)$ with
$\frac{\partial^2 \rho}{\partial \mathbf{u} \, \partial \mathbf{u}_q} \neq 0.$
In \cite{HickmanJNMP2008}, Hickman derived necessary conditions on this
leading term (which, in the system case, is a matrix).
First all terms in the candidate density $\rho$ that contribute directly to
the flux are removed.
Rather than applying the Euler operator on the remaining terms in $\rho$,
the necessary condition \cite{MHandWHprsa2003},
\begin{equation}
\label{necEuler}
\frac{\partial^2 g}{\partial \mathbf{u}\, \partial \mathbf{u}_q} = 0,
\end{equation}
for $g$ to be a total difference is applied to obtain a system of equations
for the terms that depend on the maximal shift, $\mathbf{u}_q$, in $\rho.$
This system is rewritten as a matrix equation.
Solutions to this system will give us the form of the leading term in $\rho.$

We apply a splitting of the identity operator,
\begin{equation}
\label{Dsplit}
\Id = ( \D - \Id + \Id ) \, \D^{-1}
= \Delta \, \D^{-1} + \D^{-1},
\end{equation}
to the part, $\rho^*,$ of the candidate $\rho$ that is independent of the 
variables with the lowest order shift.
The first term ${\Delta \, {\rm D}^{-1}} \, \rho^*$ contributes to the flux
while the second term ${\rm D}^{-1} \, \rho^*$ has a strictly lower shift 
than $\rho^*.$
Applying this split repeatedly we get
\begin{equation}
\label{Dsplitrepeatedly}
{\Id} = (\D^k - {\Id} + {\Id} ) \, \D^{-k} = \Delta \,
\left(\D^{k-1}+\D^{k-2} + \cdots + \D + {\Id} \right) \D^{-k} +\D^{-k},
\end{equation}
where, again, the first term contributes to the flux and the remainder has
strictly lower shift.

This decomposition is repeatedly applied to terms that do not involve the
lowest order shifted variables.
Any terms that remain will involve the lowest order shifted variable.
These terms yield the constraints on the undetermined coefficients or unknown
functions in the density $\rho.$
As shown in \cite{HickmanJNMP2008}, the result of this split is that $\rho$
is a density if and only if
\begin{equation}
\label{sigma}
\sigma = \D^l{\left( \mathbf{F} \,\frac{\partial}{\partial \textbf{u}} \right)}
\sum_{j=0}^{l} \D^j{\rho} + \sum_{j=l+1}^{q} \D^j{ \left( \mathbf{F} \,
\frac{\partial}{\partial \textbf{u}}\right)} \rho
\end{equation}
is a total difference, where the operator
\begin{equation}
\label{Fpartialu}
\mathbf{F} \,\frac{\partial}{\partial \textbf{u}}
= \sum_\alpha F^\alpha \frac{\partial}{\partial u^\alpha}
\end{equation}
involves a summation over the components in the system of DDEs.
As before, in the examples we will use $u, v, w, $ etc.\ to denote the
dependent variables $u^{\alpha}.$

Now, $\sigma$ depends on the shifted variables
$\mathbf{u}, \ \ldots, \ \mathbf{u}_{q+L}$ where $L = \max \, (l,\ m)$.
For $q > L$, (\ref{necEuler}) gives
\begin{equation}
\label{conditionsigma}
\frac{\partial^2 \sigma}{\partial \mathbf{u} \,\partial \mathbf{u}_{q+L}}
= \D^L{\left( \frac{\partial^2\rho}{\partial \mathbf{u} \,
  \partial \mathbf{u}_q}\,
  \frac{\partial \mathbf{F} }{\partial \mathbf{u}_{-L}} \right)}
  + \D^q{\left(\frac{\partial \mathbf{F}}{\partial \mathbf{u}_L}\right)^{\rm T}}
 \frac{\partial^2 \rho}{\partial \mathbf{u} \, \partial \mathbf{u}_q} = 0,
\end{equation}
where ${\rm T}$ stands for transpose.
The system case is treated in detail in \cite{HickmanJNMP2008}.
For brevity, we continue with the scalar case.
Let
\begin{equation}
\label{generallambdamu}
\lambda = \frac{\partial F}{\partial u_{-L}}, \quad
\mu = \frac{\partial F}{\partial u_L},
\end{equation}
then the leading term will satisfy
\begin{equation}
\label{scalarcond}
 \mathcal{S} \, \frac{\partial^2 \rho}{\partial u \, \partial u_q} = 0,
\end{equation}
with
$\mathcal{S} = \D^L \, \lambda \, \D^L + \D^q \, \mu.$
We immediately see that if $l \neq m$ then either $\lambda$ or $\mu$ is zero
and (\ref{scalarcond}) has no non-trivial solutions.
Let $q = p L + r$ with $p$ and $r$ integers, $0 \leq r < L ,$ and
\begin{equation}
\label{c}
c = \left(\prod_{k=1}^{p-1} \D^{k L} \lambda \right) \D^{p L} \zeta.
\end{equation}
Then
\begin{equation}
\label{Sc}
 \mathcal{S} \, c = \prod_{k=1}^{p-1} \D^{k L} \lambda \, \D^{p L}
 \left(\lambda \, \D^L\zeta + \zeta \, \D^r \mu \right).
\end{equation}
Thus $c \neq 0$ lies in the kernel of $\mathcal{S}$ if and only if
\begin{equation}
\label{zetaeqn}
\lambda \, \D^L\zeta + \zeta \, \D^r \mu = 0
\end{equation}
has a non-zero solution $\zeta.$
Suppose (\ref{zetaeqn}) has two non-zero solutions,
say, $\zeta_1$ and $\zeta_2.$
Then,
\begin{equation}
\label{eqfractzetas}
\frac{\zeta_2}{\zeta_1} = \D^L \left( \frac{\zeta_2}{\zeta_1} \right).
\end{equation}
So, since $L \neq 0$, $\zeta_2 = a \zeta_1$ for some constant $a$.
Therefore, the kernel of $\mathcal{S}$ is, {\em at most, one dimensional}.
This implies that a scalar DDE can have, at most, one conserved density that
depends on $q$ shifts for $q > L.$
The leading term, $\tilde{\rho},$ will satisfy
\begin{equation}
\label{conditionforrho}
\frac{\partial^2 \rho}{\partial u \, \partial u_q} = c,
\end{equation}
which, upon integration, yields
\begin{equation}
\tilde{\rho} = \iint c \, du \, du_q.
\end{equation}
The density (if it exists) may now be computed by a ``split and shift"
strategy on this leading term.
Starting with
$ \rho = \tilde{\rho}. $
the objective is to successively compute the terms (of lower shift!)
that must be added to $\rho$ until $\D_t \,\rho \equiv 0.$
First, $\D_t \, \rho$ is computed and evaluated on the DDE.
Next, all terms are shifted so that the resulting expression depends on $u$
(and not on lower shifts of $u).$
Then the leading terms, $\xi,$ in $\D_t \,\rho$ are isolated.
Last, we solve
\begin{equation}
\label{leadingshift}
\D_t \, \rho^{(1)} = \xi  + \text{terms of lower shift}.
\end{equation}
If (\ref{leadingshift}) has no solution then a density with $q$ shifts does
not exist.
On the other hand, if (\ref{leadingshift}) has a solution,
the ``correction" term $\rho^{(1)}$ is then subtracted from $\rho$ and
we recompute $\D_t \, \rho.$
By construction, the highest shift that occurs in the result will now be
lower than before and we repeat the entire procedure to obtain a
new correction term $\rho^{(2)}.$
After a finite number of steps, we will either find an that the correction
term does not exist (and so the density does not exist) or we will obtain
$\D_t \,\rho \equiv 0$ and $\rho$ will be a density.

This algorithm has been coded \cite{HickmanDiscreteCode} in {\tt Maple}.
Further details and worked examples will be presented in
\cite{MHandWHnew2008}.
For now, we illustrate the algorithm with a simple example.
\subsection{A``Modified" Volterra Lattice}
\label{modifiedvolterralatticeappl}
Consider the DDE,
\begin{equation}
\label{modifiedvolterralattice}
\dot{u} = u^2 \, (u_{2} - u_{-2}),
\end{equation}
which is related to the well-known modified Volterra Lattice
\cite{Adleretal99}.
Here $L=2$ and, using (\ref{generallambdamu}),
\begin{equation}
\label{lambdamuspecial}
\lambda = \frac{\partial F}{\partial u_{-2}} = {} - u^2, \quad
\mu = \frac{\partial F}{\partial u_2} = u^2.
\end{equation}
Thus, the condition (\ref{zetaeqn}) for a non-trivial density becomes
$ \zeta \, u_r^2 = u^2 \, \D^2 \zeta $ for $r = 0, \ 1.$
For $r = 0,$ we have $\zeta = \D^2 \zeta$ and so we may choose $\zeta = 1.$
For $r = 1,$ we have $ \zeta \, u_1^2 = u^2 \, \D^2 \zeta,$ which has
no non-zero solutions.
Thus, densities only exist for $r=0.$
Since $q = p L = 2 p,$ with $p$ integer, we conclude that $q$ must be even.
In these cases the leading term will satisfy
\begin{equation}
\label{leadingcondition}
\frac{\partial^2 \rho}{\partial u \, \partial u_q}
= c
= \left(\prod_{k=1}^{p-1} u_{2k} \right)
= u \, u_2 \, \cdots \, u_{q-2}.
\end{equation}
Therefore, after scaling to remove constants, the leading term is
${\tilde{\rho}} = u^2 \, u_2 \, \cdots \, u_{q-2} \, u_{q}.$

For example, let us compute the density that depends on $q = 4$ shifts.
We set $ \rho = {\tilde{\rho}} = u^2 \, u_2 \, u_4.$
Using (\ref{modifiedvolterralattice}),
\begin{eqnarray}
\label{Dtrhomodifiedvolterralattice}
\D_t \, \rho &=& u^2 u_2^2  u_4 \left( u_2 - u_{-2} \right)
+ 2 u \, u_2^3 \, u_4 \left( u_4 - u \right)
+ u \, u_2^2 \, u_4^2 \left( u_6 - u_2 \right)
\nonumber \\
&\equiv& u \, u_2^3 \, u_4^2 - u^2 \, u_2^3 \, u_4.
\end{eqnarray}
The highest shift is $u_4$ and so the leading terms are
\begin{equation}
\label{ximodifiedvolterralattice}
 \xi = u\, u_{{2}}^{3}\, u_{{4}}^2 -u^2 \, {u_{{2}}}^{3}\, u_{{4}}.
\end{equation}
Note that the terms in $u_6$ {\em must} cancel by the construction of
$\tilde{\rho}$.
Next, we note that $u_4$ as a leading term must arise as either
${\dot u}_2 = u_2^2 \, (u_4 - u)$ or
\begin{equation}
\label{choice2}
u_2 \, {\dot u}
    = u_2 \, u^2 \, ( u_2 - u_{-2} ) \equiv u^2 \, u_2^2 - u \, u_2^2 \, u_4.
\end{equation}
The quadratic term must arise from (\ref{choice2})
(as we have already determined all terms that involve $u_4,$
so, we cannot have a quadratic term given by $u_4 {\dot u}_2).$
Now, we solve (\ref{leadingshift}) to get
\begin{equation}
\label{rho1modifiedvolterralattice}
\rho^{(1)} = - \tfrac{1}{2} \, u^2 \, u_2^2 + \cdots.
\end{equation}
Indeed,
\begin{eqnarray}
\label{finalDtrhomodifiedvolterralattice}
\D_t \left( - \tfrac{1}{2} \, u^2 \, u_2^2 \right)
&=& - u \, u_2^2 \, \dot{u} - u^2 \, u_2 \, \dot{u}_2
\nonumber \\
&\equiv& u \, u_2^3 \, u_4^2 - u^2\, u_2^3 \, u_4
  + \text{lower shift terms}
\nonumber \\
&=& \xi + \text{lower shift terms,}
\end{eqnarray}
with $\xi$ in (\ref{ximodifiedvolterralattice}).
Thus, we update $\rho = u^2 u_2 u_4$ by subtracting $\rho^{(1)}.$
This yields,
\begin{equation}
\label{rhowithq4}
\rho = u^2 \, u_2 \, u_4 + \tfrac{1}{2} \, u^2 \, u_2^2.
\end{equation}
We readily verify that $ \D_t \, \rho \equiv 0 $
and, thus, $\rho$ in (\ref{rhowithq4}) is the unique density of
(\ref{modifiedvolterralattice}) that depends on $4$ shifts.
\vspace{-2mm}
\section{The Gardner Lattice}
\label{gardnerlatticesection}
In this section we consider the DDE described in \cite{TTmcs1993},
\begin{eqnarray}
\label{gardnerlatticeoriginal}
\dot{u} &\!=\!& \left( 1 + \alpha h^2 u + \beta h^2 u^2 \right)
\left\{
\frac{1}{h^3} \left(
\tfrac{1}{2} u_{-2} - u_{-1} + u_1 - \tfrac{1}{2} u_2 \right)
+ \frac{\alpha}{2h} \left[ u_{-1} u_{-2} + u_{-1}^2
+ u ( u_{-1} - u_1 )
\right. \right.
\nonumber\\
& & \left. \left. - u_1^2  - u_1 u_2 \right]
+ \frac{\beta}{2h} \left[  u_{-1}^2 ( u_{-2} + u ) - u_1^2 ( u + u_2 )
\right] \right\},
\end{eqnarray}
which is a completely integrable discretization of the Gardner equation,
\begin{equation}
\label{gardner}
u_t + 6 \alpha u u_x + 6 \beta u^2 u_x + u_{3x} = 0.
\end{equation}
Therefore, we call (\ref{gardnerlatticeoriginal}) the Gardner lattice.
Note that (\ref{gardner}) is a combination of the KdV equation
($\beta = 0)$ and the mKdV equation $(\alpha = 0).$
Consequently, (\ref{gardnerlatticeoriginal}) includes the completely
integrable discretizations of the KdV and mKdV equations as special cases.
%
\subsection{Derivation of the Gardner Lattice}
\label{derivationgardnerlattice}
Based on work by Taha \cite{TTmcs1993}, 
we outline the derivation of the Gardner lattice from a discretized version
\cite{MAandJLjmp1975} of the eigenvalue problem of Zakharov and Shabat.
Consider the discrete system \cite{MASIAMrev1977,MAandHSbook1981},
\begin{eqnarray}
\label{discreteAKNS}
&& V_{1,n+1} = z V_{1,n} + Q_n(t) V_{2,n}, \quad\quad
V_{2,n+1} = \frac{1}{z} V_{2,n} + R_n(t) V_{1,n}, \\
&& {\dot V}_{1,n} = A_n(z) V_{1,n} + B_n(z) V_{2,n}, \quad
{\dot V}_{2,n} = C_n(z) V_{1,n} + D_n(z) V_{2,n},
\end{eqnarray}
where, in general, the coefficients $A_n$ through $D_n$ depend on the
potentials, $R_n$ and $Q_n.$
The eigenvalue $z$ is assumed to be time-independent.
\vskip 3pt
\noindent
{\bf Step 1}:
Expressing the compatibility conditions,
${\dot V}_{i,n+1} = \D {\dot V}_{i,n},$ for $i = 1, 2,$
and equating the coefficients of $V_{1,n}$ and $V_{2,n},$ we get
\cite{MAandHSbook1981}
\begin{eqnarray}
\label{compatibilitydifferenceeqs}
\!\!\!\!\!\!\!\!\!\!&&\!\!\!
C_n Q_n - B_{n+1} R_n = z (A_{n+1} - A_n), \quad\quad\quad
B_n R_n - C_{n+1} Q_n = \frac{1}{z} (D_{n+1} - D_n), \\
\label{compatibilityDDEs}
\!\!\!\!\!\!\!\!\!\!&&\!\!\!
{\dot Q}_n + D_n Q_n - A_{n+1} Q_n = \frac{1}{z} B_{n+1} - z B_n,
\;\,
{\dot R}_n + A_n R_n - D_{n+1} R_n      =      \frac{1}{z} C_{n+1} - z C_n.
\end{eqnarray}
\vskip 1pt
\noindent
{\bf Step 2}:
Substituting the expansions \cite{TTmcs1993},
\begin{eqnarray}
\label{expansionscoefficientsAD}
A_n \!&\!=\!&\! \sum_{j=-2}^{2} z^{2j} A_{n}^{(2j)},
\quad\quad\quad
D_n = \sum_{j=-2}^{2} z^{2j} D_{n}^{(2j)}, \\
\label{expansionscoefficientsBC}
B_n \!&\!=\!&\! \sum_{j=-1}^{2} z^{2j-1} B_{n}^{(2j-1)}, \quad
C_n = \sum_{j=-1}^{2} z^{2j-1} C_{n}^{(2j-1)},
\end{eqnarray}
into (\ref{compatibilitydifferenceeqs}) and (\ref{compatibilityDDEs}) and
setting the coefficients of like powers of $z$ equal to zero,
we obtain a system of twenty equations for the eighteen unknowns functions
$A_{n}^{(2j)}, D_{n}^{(2j)}$ with $j = -2,-1,0,1,2;$ and
$B_{n}^{(2j-1)},$ $C_{n}^{(2j-1)}$ with $j = -1,0,1,2.$
The simplest equations arise from the coefficients of the terms in
$z^{\pm 5}$ and $z^{\pm 4}:$
\begin{eqnarray}
\label{a4}
&& A_{n+1}^{(4)} - A_n^{(4)} = 0,
\quad\quad\quad\quad\quad\quad\quad\;\,
D_{n+1}^{(-4)} - D_n^{(-4)} = 0, \\
\label{b3cneg3}
&& Q_n ( D_n^{(4)}  - A_{n+1}^{(4)} )  + B_{n}^{(3)} = 0, \quad\quad\;
   R_n ( A_n^{(-4)} - D_{n+1}^{(-4)} ) + C_{n}^{(-3)} = 0, \\
\label{bneg3c3}
&& Q_n ( D_n^{(-4)} - A_{n+1}^{(-4)} ) - B_{n+1}^{(-3)} = 0, \quad
   R_n ( A_n^{(4)}  - D_{n+1}^{(4)} )  - C_{n+1}^{(3)} = 0,
\end{eqnarray}
\vskip 3pt
\noindent
{\bf Step 3}:
We outline the solution process which follows the strategy in
\cite[Section 2.2a]{MAandHSbook1981}.
From (\ref{a4}), we conclude that $A_n^{(4)}$ and
$D_n^{(-4)}$ are independent of $n.$
Hence, $A_n^{(4)} = {\tilde A}^{(4)}$ and $D_n^{(-4)} = {\tilde D}^{(-4)}$
are constants.
The tilde will remind us that we are dealing with {\em constants}.
Solving (\ref{b3cneg3}) and (\ref{bneg3c3}), we get
\begin{eqnarray}
\label{solb3c3}
B_{n}^{(3)}  \!&\!=\!&\! ({\tilde A}^{(4)} - D_n^{(4)}) Q_n, \quad\quad\quad
C_{n}^{(-3)} = ({\tilde D}^{(-4)} - A_n^{(-4)}) R_n, \\
\label{solbneg3cneg3}
B_n^{(-3)} \!&\!=\!&\!   ({\tilde D}^{(-4)} - A_n^{(-4)}) Q_{n-1}, \quad
C_n^{(3)}  = ({\tilde A}^{(4)} - D_n^{(4)}) R_{n-1}.
\end{eqnarray}
Substituting these results into two of the equations coming from $z^{\pm 3},$
we find that $A_n^{(-4)} = {\tilde A}^{(-4)}$ and
$D_n^{(4)} = {\tilde D}^{(4)}$ are constants.
From equations corresponding to $z^{\pm 3}, z^{\pm 1},$ we obtain
\begin{eqnarray}
\label{couplingA2}
\Delta A_n^{(2)} \!&\!=\!&\! A_{n+1}^{(2)} - A_n^{(2)}
       = {\tilde \alpha} ( Q_n R_{n-1} - Q_{n+1} R_n), \\
\label{couplingDneg2}
\Delta D_n^{(-2)} \!&\!=\!&\! D_{n+1}^{(-2)} - D_n^{(-2)}
       = {\tilde \beta}  (R_n Q_{n-1} - R_{n+1} Q_n), \\
\label{couplingAneg2}
\Delta A_n^{(-2)} \!&\!=\!&\! A_{n+1}^{(-2)} - A_n^{(-2)}
       = {\tilde \beta}  (Q_n R_{n+1} - Q_{n-1} R_n), \\
\label{couplingD2}
\Delta D_n^{(2)} \!&\!=\!&\! D_{n+1}^{(2)} - D_n^{(2)}
       = {\tilde \alpha} (R_n Q_{n+1} - R_{n-1} Q_n),
\end{eqnarray}
where ${\tilde \alpha} = {\tilde A}^{(4)} - {\tilde D}^{(4)}$ and
${\tilde \beta} = {\tilde D}^{(-4)} - {\tilde A}^{(-4)}.$
Equations (\ref{couplingA2})-(\ref{couplingD2}) are satisfied for
\begin{eqnarray}
\label{a2dneg2solutions}
A_n^{(2)} \!&\!=\!&\!
           {\tilde A}^{(2)}  - {\tilde \alpha} Q_n R_{n-1}, \quad\;\;
D_n^{(-2)}  =  {
           \tilde D}^{(-2)} - {\tilde \beta}  R_n Q_{n-1}, \\
\label{aneg2d2solutions}
A_n^{(-2)} \!&\!=\!&\!
           {\tilde A}^{(-2)} + {\tilde \beta}  Q_{n-1} R_n, \quad
D_n^{(2)}   =   {\tilde D}^{(2)}  + {\tilde \alpha} R_{n-1} Q_n,
\end{eqnarray}
where ${\tilde A}^{(2)}, {\tilde D}^{(-2)}$ and
${\tilde A}^{(-2)}, {\tilde D}^{(2)}$ are constants.
Next, we solve equations (from the terms in $z^{\pm 2})$
for $B_n^{(\pm 1)}$ and $C_n^{(\pm 1)},$ yielding
\begin{eqnarray}
\label{solb1}
B_n^{(1)} &=& {\tilde \delta} Q_n + {\tilde \alpha} Q_{n+1}
              - {\tilde \alpha} Q_n ( Q_{n+1} R_n + Q_n R_{n-1} ), \\
\label{solcneg1}
C_n^{(-1)} &=& {\tilde \gamma} R_n + {\tilde \beta} R_{n+1}
               - {\tilde \beta} R_n ( R_{n+1} Q_n + R_n Q_{n-1} ), \\
\label{solbneg1}
B_n^{(-1)} &=& {\tilde \gamma} Q_{n-1} + {\tilde \beta} Q_{n-2}
               - {\tilde \beta} Q_{n-1} ( R_{n-1} Q_{n-2} + R_n Q_{n-1}), \\
\label{solc1}
C_n^{(1)} &=& {\tilde \delta} R_{n-1} + {\tilde \alpha} R_{n-2}
              - {\tilde \alpha} R_{n-1} ( Q_{n-1} R_{n-2} + Q_n R_{n-1}),
\end{eqnarray}
where ${\tilde \delta} = {\tilde A}^{(2)} - {\tilde D}^{(2)}$ and
${\tilde \gamma} = {\tilde D}^{(-2)} - {\tilde A}^{(-2)}.$
Next, we solve equations coming from $z^{\pm 1},$ which involve
$\Delta A_n^{(0)}$ and $\Delta D_n^{(0)}.$
These equations, which are similar in structure to
(\ref{couplingA2})-(\ref{couplingD2}), yield
\begin{eqnarray}
\label{sola0}
\!\!\!\!\!\!\!\!A_n^{(0)}\!\!&\!=\!&\!\!{\tilde \alpha}
\left( - Q_{n+1} R_{n-1} - Q_n R_{n-2} + Q_n Q_{n+1} R_{n-1} R_n
+ Q_{n-1} Q_n R_{n-2} R_{n-1} + Q_n^2 R_{n-1}^2 \right)
\nonumber \\
&& + {\tilde A}^{(0)} - {\tilde \delta} R_{n-1} Q_n, \\
\label{sold0}
\!\!\!\!\!\!\!\!D_n^{(0)} \!\!&\!=\!&\!\! {\tilde \beta}
\left( - R_{n+1} Q_{n-1} - R_n Q_{n-2} + R_n R_{n+1} Q_{n-1} Q_n
+ R_{n-1} R_n Q_{n-2} Q_{n-1} + R_n^2 Q_{n-1}^2 \right)
\nonumber \\
&& + {\tilde D}^{(0)} - {\tilde \gamma} Q_{n-1} R_n,
\end{eqnarray}
where ${\tilde A}^{(0)}$ and ${\tilde D}^{(0)}$ are constants.
All the difference equations for the unknown coefficients are now satisfied.
We are left with the two DDEs (coming from the terms in $z^{0}),$
\begin{eqnarray}
\label{ddeforqn}
\!\!\!\!\!\!\!\!\!\!\!\!&&\!\!\!\!\!\!\!\!\!\!
{\dot Q}_n - {\tilde \kappa} Q_n
+ (1 - Q_n R_n) \left[ - {\tilde \alpha} Q_{n+2} - {\tilde \delta} Q_{n+1}
 + {\tilde \gamma} Q_{n-1} + {\tilde \beta} Q_{n-2}
+ {\tilde \alpha} Q_{n+1} (Q_{n+2} R_{n+1}  \right.
\nonumber \\
\!\!\!\!\!\!\!\!\!&&\!\!\!\!\!\!\left. + Q_{n+1} R_n )
- {\tilde \beta} Q_{n-1} ( Q_{n-2} R_{n-1} + Q_{n-1} R_n )
+ {\tilde \alpha} Q_n Q_{n+1} R_{n-1}
- {\tilde \beta}  Q_n Q_{n-1} R_{n+1} \right] = 0, \\
\label{ddeforrn}
\!\!\!\!\!\!\!\!\!\!\!\!&&\!\!\!\!\!\!\!\!\!\!
{\dot R}_n + {\tilde \kappa} R_n
+ (1 - R_n Q_n) \left[ - {\tilde \beta} R_{n+2} - {\tilde \gamma} R_{n+1}
 + {\tilde \delta} R_{n-1} + {\tilde \alpha} R_{n-2}
+ {\tilde \beta} R_{n+1} ( R_{n+2} Q_{n+1}  \right.
\nonumber \\
\!\!\!\!\!\!\!&&\!\!\!\!\!\!\left. + R_{n+1} Q_n )
- {\tilde \alpha} R_{n-1} (R_{n-2} Q_{n-1} + R_{n-1} Q_n)
+ {\tilde \beta}  R_n R_{n+1} Q_{n-1}
- {\tilde \alpha} R_n R_{n-1} Q_{n+1} \right] = 0,
\end{eqnarray}
where ${\tilde \kappa} = {\tilde A}^{(0)} - {\tilde D}^{(0)}.$
\vskip 3pt
\noindent
{\bf Step 4}:
The terms $-{\tilde \kappa} Q_n$ in (\ref{ddeforqn}) and ${\tilde \kappa} R_n$
in (\ref{ddeforrn}) could be removed with a suitable transformation.
Accomplishing the same, we set ${\tilde \kappa} = 0.$
Next, we substitute
\begin{equation}
Q_n = u_n, \quad R_n = - h^2 (\alpha + \beta u_n),
\end{equation}
into (\ref{ddeforqn}) and (\ref{ddeforrn}).
Next, we set ${\tilde \beta} = {\tilde \alpha}$ and
${\tilde \gamma} = {\tilde \delta}$ to remove all constant terms.
Doing so, (\ref{ddeforqn}) and (\ref{ddeforrn}) collapse into a single DDE,
\begin{eqnarray}
\label{singledotun}
\!\!\!\!\!\!\!\!&&\!\!\!\!\!\!\!\!\!{\dot u}_n
+ (1 + \alpha h^2 u_n + \beta h^2 u_n^2)
\!\left\{ {\tilde \alpha} ( u_{n-2} - u_{n+2} )
+ {\tilde \delta} ( u_{n-1} - u_{n+1} )
+ \alpha {\tilde \alpha} h^2 ( u_{n-2} u_{n-1} + u_{n-1}^2
\right. \nonumber \\
\!\!\!\!\!\!\!\!&&\!\!\!\!\!\!\!\!\!\left.
+u_n (u_{n-1}\!-\! u_{n+1})-\! u_{n+1}^2-\! u_{n+1} u_{n+2} )
+\!\beta {\tilde \alpha} h^2 ( u_{n-1}^2 ( u_{n-2} \!+\!u_n )
-\!u_{n+1}^2 (u_n \!+\!u_{n+2}) )\!\right\}\! = 0,
\end{eqnarray}
\vskip 1pt
\noindent
{\bf Step 5}:
To fix the scale on $t,$ as well as the constants
${\tilde \alpha},$ ${\tilde \beta},$ and ${\tilde \delta},$
we consider the limit of (\ref{singledotun}) for $h \rightarrow 0.$
Using, $\lim_{h \rightarrow 0} u_n(t) = u(x,t),$
$\lim_{h \rightarrow 0} {\dot u}_n(t) = u_t(x,t),$ and substituting
\begin{equation}
\label{limithtozero}
\lim_{h \rightarrow 0} u_{n + m}(t)
= u(x,t) + m h u_x(x,t) + \tfrac{1}{2} (m h)^2 u_{2x}(x,t) +
  \tfrac{1}{6} (m h)^3 u_{3x}(x,t) + \ldots,
\end{equation}
into (\ref{singledotun}), we obtain
\begin{eqnarray}
\label{continuouslimitdotun}
u_t - 2 h ( {\tilde \delta} + 2 {\tilde \alpha} ) u_x
- 2 h^3 ( {\tilde \delta} + 8 {\tilde \alpha} )
 \left( \alpha u u_x + \beta u^2 u_x + \tfrac{1}{6} u_{3x} \right)
+ {\cal O}(h^5) = 0.
\end{eqnarray}
The term in $h$ disappears when ${\tilde \delta} = - 2 {\tilde \alpha}.$
Substituting ${\tilde \alpha} = -\tfrac{1}{2}, {\tilde \delta} = 1,$ into
(\ref{singledotun}) and (\ref{continuouslimitdotun}), we get
\begin{eqnarray}
\label{gardnerlatticeoriginalscaled}
\dot{u} &\!=\!& \left( 1 + \alpha h^2 u + \beta h^2 u^2 \right)
\left\{ \tfrac{1}{2} u_{-2} - u_{-1} + u_1 - \tfrac{1}{2} u_2
+ \tfrac{1}{2} \alpha h^2 \left[ u_{-1} u_{-2} + u_{-1}^2
+ u ( u_{-1} - u_1 )
\right. \right.
\nonumber\\
& & \left. \left. - u_1^2  - u_1 u_2 \right]
+ \tfrac{1}{2} \beta h^2 \left[  u_{-1}^2 ( u_{-2} + u ) - u_1^2 ( u + u_2 )
\right] \right\},
\end{eqnarray}
and
\begin{equation}
\label{kdvmkdv}
u_t + h^3 (6 \alpha u u_x + 6 \beta u^2 u_x + u_{3x}) = 0,
\end{equation}
where the ${\cal O}(h^5)$ terms were ignored.
Using a scale, $t \rightarrow h^3 t,$ we get (\ref{gardnerlatticeoriginal})
and (\ref{gardner}).
\vskip 5pt
\noindent
{\bf Remarks.}
Step 2 is well suited for a CAS.
Solving the system, as outlined in Step 3, is a challenging task,
in particular, if attempted with pen and paper.
The system consists of two DDEs and eighteen difference equations.
None of the CAS has a build-in solver for such mixed systems.
A fully automated solution is therefore impossible.
We used a feedback mechanism which mimics what one would do by hand:
Solve the simplest difference equations; enter that partial solution;
let CAS simplify the entire system; repeat the process until all
difference equations are satisfied and the two DDEs are simplified as far
as possible.
Continuing with (\ref{ddeforqn}) and (\ref{ddeforrn}), steps 4 and 5 were
straightforward to implement.
The five steps have been implemented in {\tt Mathematica}
\cite{WHwebsite2004c}.
Starting from (\ref{discreteAKNS}), the code generates the Gardner 
lattice (\ref{gardnerlatticeoriginalscaled}).
The code could be modified to assist in the derivation of other completely
integrable DDEs, such as discrete versions of the nonlinear Schr\"odinger
and sine-Gordon equations.
%
\subsection{Dilation Invariance of the Gardner Lattice}
\label{gardnerlatticeinvariance}
Since (\ref{gardnerlatticeoriginal}) is not uniform in rank we must 
introduce auxiliary parameters with weight.
This can be done in several ways.
One of the possibilities is to replace (\ref{gardnerlatticeoriginal}) by
\begin{eqnarray}
\label{gardnerlattice}
\dot{u} &\!=\!& \left( \gamma + \alpha u + \beta u^2 \right)
\left\{
\gamma \left( \frac{1}{2} u_{-2} - u_{-1} + u_1 - \frac{1}{2} u_2 \right)
+ \frac{\alpha}{2} \left[ u_{-1} u_{-2} + u_{-1}^2 + u ( u_{-1} - u_1 )
\right. \right.
\nonumber\\
& & \left. \left. - u_1^2  - u_1 u_2 \right]
+ \frac{\beta}{2} \left[ u_{-1}^2 ( u_{-2} + u ) - u_1^2 ( u + u_2 ) \right]
\right\},
\end{eqnarray}
where we have set $h = 1$ (by scaling) and introduced a parameter $\gamma.$

Expressing uniformity of rank, setting $W(\D_t) = 1,$ and solving the
linear system for the weights, one finds that
$W(u) = W(\alpha) = \frac{1}{4},$ $ W(\beta) = 0,$ and
$W(\gamma) = \frac{1}{2}.$
So, we do not need a scale on $\beta$ and (\ref{gardnerlattice}) is invariant
under the scaling symmetry
\begin{equation}
\label{gardnerlatticescale}
(t, u, \alpha, \gamma)
\rightarrow ({\lambda}^{-1} t, \lambda^{\frac{1}{4}} u,
{\lambda}^{\frac{1}{4}} \alpha, {\lambda}^{\frac{1}{2}} \gamma).
\end{equation}
For $\beta = 0,$ (\ref{gardnerlattice}) reduces to
\begin{eqnarray}
\label{disckdv}
\dot{u}\!&\!=\!&\! \left( \gamma + \alpha u \right)
\left\{ \gamma \left(
\frac{1}{2} u_{-2} - u_{-1} + u_1 -\frac{1}{2} u_2 \right)
+ \frac{\alpha}{2} \left[ u_{-1} u_{-2} + u_{-1}^2
+ u ( u_{-1} - u_1 ) \right. \right.
\nonumber \\
&& \left. \left.\;\; - u_1^2 - u_1 u_2 \right] \right\},
\end{eqnarray}
which is a completely integrable discretization of the KdV equation,
$u_t + 6 \alpha u u_x + u_{3x} = 0.$
Computing the weights, one can set $W(\alpha) = 0,$ which leads to
$W(u) = W(\gamma) = \frac{1}{2}.$
So, (\ref{disckdv}) is invariant under the scaling symmetry
$ (t, u, \gamma) \rightarrow
({\lambda}^{-1} t, \lambda^{\frac{1}{2}} u, {\lambda}^{\frac{1}{2}} \gamma).$
For $\alpha = 0,$ (\ref{gardnerlattice}) reduces to
\begin{eqnarray}
\label{discmkdv}
\!\!\!\dot{u} \!&\!=\!&\! \left( \gamma + \beta u^2 \right)
\left\{
\gamma \left( \frac{1}{2} u_{-2} - u_{-1} + u_1 - \frac{1}{2} u_2 \!\right)
+ \frac{\beta}{2}\left[ u_{-1}^2 ( u_{-2} + u ) - u_1^2 ( u + u_2) \right]
\right\},
\end{eqnarray}
which is a completely integrable discretization of the mKdV equation,
$ u_t + 6 \beta u^2 u_x + u_{3x} = 0.$
In this case, one can set $W(\beta) = 0.$
Then $W(u) = \frac{1}{4}$ and $W(\gamma) = \frac{1}{2}.$
Thus, (\ref{discmkdv}) is invariant under the scaling symmetry
$ (t, u, \gamma) \rightarrow
({\lambda}^{-1} t, \lambda^{\frac{1}{4}} u, {\lambda}^{\frac{1}{2}} \gamma).$
%
\subsection{Conservation Laws of the Gardner Lattice}
\label{gardnerlatticeappl}
One can either apply the method of Section~\ref{newmethod} directly to
(\ref{gardnerlatticeoriginal}) or, alternatively, apply to 
(\ref{gardnerlattice}) the technique based on dilation invariance outlined
in Sections~\ref{applkvmlattice},~\ref{appltodalattice}, 
and~\ref{applALlattice}.
In particular, one can use the ``divide and conquer" strategy of 
Section~\ref{applALlatticedivideconquer} to split candidate densities into 
smaller pieces.
Computational details can be found in \cite{HEthesis2003} 
The results below were computed \cite{MHandWHnew2008} with the method in 
Section~\ref{newmethod}.
For $q=0$ shifts there are two (non-polynomial) density-flux pairs:
\begin{eqnarray}
\label{firstrhozerogardnerlattice}
\!\!\rho^{(0)}_1
\!\!&\!=\!&\!\! \ln \, (1 + \alpha h^2 \, u + \beta h^2 \, u^2), \\
\label{firstjzerogardnerlattice}
\!\!J^{(0)}_1 \!\!&\!=\!&\!\! - \tfrac{1}{2} \,
\Big\{
\frac{\alpha}{h} \left( u_{-2} - u_{-1} - u + u_1 \right)
+ \frac{\beta}{h} \left( 2 u_{-2} u - 4 u_{-1} u + 2 u_{-1} u_1  \right)
\nonumber \\
\!&\!&\!
+ \alpha^2 h \, \big( u_{-1} ( u_{-2} + u_{-1} + u )
+ u (u_{-1} + u + u_1) \big)
+ \alpha \beta \, h \, \big(
u_{-1}^{2} u_{-2} + 2 u_{-2} u_{-1} u + 3 u_{-1}^2 u
\nonumber \\
\!&\!&\!
+ 3 u_{-1} u^2 + 2 u_{-1} u u_1 + u^{2} u_1
\big)
+ 2 \beta^2 \, h \, u_{-1} u \big( u_{-2} u_{-1} + u_{-1} u + u u_1 \big)
\Big\},
\end{eqnarray}
and
\begin{eqnarray}
\label{secondrhozerogardnerlattice}
\rho^{(0)}_2 \!\!&\!=\!&\!\! \text{arctanh}
\left(\frac{h (\alpha + 2 \beta u)}{\sqrt{\alpha^2 \, h^2 -4 \beta}} \right),
\\
\label{secondjzerogardnerlattice}
J^{(0)}_2 \!\!&\!=\!&\!\!
\tfrac{1}{4} \sqrt{\alpha^2 h^2 - 4 \beta}
\Big\{ \tfrac{1}{h^2} \left( 2 u_{-1} - u_{-2} - 2 u_1 + u_2 \right)
+ \alpha \left( u_{-2} u_{-1} + u_{-1}^2 + 2 u_{-1} u + u^2 + u u_1 \right)
\nonumber \\
\!\!&\!&\!\!
+ \beta \left( u_{-1}^2 (u_{-2} + u) + u^2 (u_{-1} + u_1) \right) \Big\}.
\end{eqnarray}
The next two (of infinitely many polynomial) densities are
\begin{eqnarray}
\label{rho1gardnerlattice}
\rho^{(1)} \!&\!=\!&\! u u_1 + \frac{\alpha}{\beta} \, u,
\\
\label{rho2gardnerlattice}
\rho^{(2)} \!&\!=\!&\!
u u_2 \left( 1 + \alpha h^2 u_1 + \beta h^2 u_1^2 \right)
+ \alpha h^2 u u_1 (u + u_1) + \tfrac{1}{2} \beta h^2 u^2 u_1^2
\nonumber \\
\!&&\! + \frac{\alpha}{\beta} (1 - \frac{\alpha^2}{\beta} h^2 ) u
+ \frac{\alpha^2}{2 \beta} h^2 u^2,
\end{eqnarray}
where the associated fluxes have been omitted due to length.
\vskip 6pt
\noindent
{\bf Special Cases}.
\vskip 5pt
\noindent
We consider two important special cases.
The first few densities for (\ref{gardnerlatticeoriginal}) with
$\beta = 0$ are:
\begin{eqnarray}
\label{rhozero1disckdvlattice}
\rho^{(0)}_1 \!&\!=\!&\! \ln \, (1 + \alpha h^2 u), \quad\quad
\rho^{(0)}_2 = u, \\
\label{rho1disckdvlattice}
\rho^{(1)} \!&\!=\!&\! \tfrac{1}{2} \, u^2 + u u_1, \quad\quad\quad\,
\rho^{(2)} = u u_2 ( 1 + \alpha h^2 u_1)
             + \alpha h^2 u (u_1^2 + u u_1 + \tfrac{1}{3} \, u^2).
\end{eqnarray}
The first few densities for (\ref{gardnerlatticeoriginal}) with
$\alpha = 0$ are
\begin{eqnarray}
\label{rhozero1discmkdvlattice}
\rho^{(0)}_1 \!&\!=\!&\! \ln \, (1+ \beta h^2 u^2), \quad\;\,
\rho^{(0)}_2 = \arctan \, (\sqrt{\beta} h u), \\
\label{rho1discmkdvlattice}
\rho^{(1)} \!&\!=\!&\! u u_1, \quad\quad\quad\quad\quad\quad
\rho^{(2)} = u u_2 (1 + \beta h^2 u_1^2) + \tfrac{1}{2} \,\beta h^2 u^2 u_1^2.
\end{eqnarray}
\vspace{-6mm}
\section{Additional Examples of Nonlinear DDEs}
\label{additionalexamples}
\subsection{The Bogoyavlenskii Lattice}
\label{bogoyavlenskiilatticeappl}
The Bogoyavlenskii lattice \cite{Bogoyavlenskii1988} and
\cite[Eq.\ (17.1.2)]{YSbook2003},
\begin{equation}
\label{bogoyavlenskiilatticegeneral}
{\dot u} = u \left( \prod_{j=1}^{p} u_j - \prod_{j=1}^{p} u_{-j} \right),
\end{equation}
is a generalization of the KvM lattice (\ref{kvmlattice}).
For $p = 2,$ lattice (\ref{bogoyavlenskiilatticegeneral}) becomes
\begin{equation}
\label{bogoyavlenskiilatticep2}
{\dot u} = u ( u_1 u_2 - u_{-1} u_{-2} ),
\end{equation}
which is invariant under the following scaling symmetry
\begin{equation}
\label{bogoyavlenskiilatticep2scale}
(t, u)
\rightarrow ({\lambda}^{-1} t, \lambda^{\frac{1}{2}} u).
\end{equation}
Lattice (\ref{bogoyavlenskiilatticep2}) has the following density-flux pairs
(of infinitely many):
\begin{eqnarray}
\label{rhozerobogoyavlenskiilatticep2}
\rho^{(0)} &\!=\!& \ln u, \quad\quad
J^{(0)} = -(u_{-1} u_{-2} + u_{-1} u + u u_1), \\
\label{rho1J1bogoyavlenskiilatticep2}
\rho^{(1)} &\!=\!& u, \quad\quad\quad
J^{(1)} = - u u_{-1} ( u_{-2} + u_1), \\
\label{rho2J2bogoyavlenskiilatticep2}
\rho^{(2)} &\!=\!& u u_1, \quad\quad
J^{(2)} = - u_{-1} u u_1 ( u_{-2} + u + u_2), \\
\label{rho3bogoyavlenskiilatticep2}
\rho^{(3)} &\!=\!& u u_1 ( \tfrac{1}{2} u u_1 + u_1 u_2 + u_2 u_3), \\
\label{J3bogoyavlenskiilatticep2}
J^{(3)} &\!=\!& - u_{-1} u u_1
( u_{-2} u u_1 + u^2 u_1 + u_{-2} u_1 u_2 + 2 u u_1 u_2 + u_1 u_2^2
+ u_{-2} u_2 u_3 + u u_2 u_3 \nonumber \\
\!& &\! + u_2^2 u_3 + u_2 u_3 u_4).
\end{eqnarray}
For (\ref{bogoyavlenskiilatticep2}), we also computed the densities
$\rho^{(4)}$ through $\rho^{(9)}.$
Every time the rank increases by one, the number of terms in the density
increases by a factor three.
For example, $\rho^{(9)}$ has 2187 terms and the highest shift is $15.$
%
\subsection{The Belov-Chaltikian Lattice}
\label{belovchaltikianlatticeappl}
The Belov-Chaltikian lattice \cite[Eq.\ (12)]{BandC1993},
\begin{equation}
\label{belovchaltikianlattice}
{\dot u} = u ( u_1 - u_{-1} ) + v_{-1} - v, \quad
{\dot v} = v ( u_2 - u_{-1} ),
\end{equation}
in invariant under the scaling symmetry
\begin{equation}
\label{belovchaltikianlatticescale}
(t, u, v) \rightarrow ({\lambda}^{-1} t, \lambda u, \lambda^2 v).
\end{equation}
The first few density-flux pairs (of infinitely many) are
\begin{eqnarray}
\label{rho1J1belovchaltikianlattice}
\rho^{(1)} &=& u,
\quad\quad\quad\quad\quad\quad\quad
J^{(1)} = - u_{-1} u + v_{-1}, \\
\label{rho2J2belovchaltikianlattice}
\rho^{(2)} &=& \tfrac{1}{2} u^2 + u u_1 - v,
\quad\;\,
J^{(2)} = - u_{-1} u^2 - u_{-1} u u_1 + u v_{-1} + u_1 v_{-1} + u_{-1} v, \\
\label{rho3belovchaltikianlattice}
\rho^{(3)} &=& u ( \tfrac{1}{3} u^2 + u u_1 + u_1^2 + u_1 u_2 - v_{-2}
                    - v_{-1} - v - v_1),
\end{eqnarray}
where $J^{(3)}$ has been omitted due to length.
Our results match these in \cite{RSandSKjmp2001}.
%
\subsection{The Blaszak-Marciniak Lattices}
\label{BlaszakMarciniaklatticeappl}
In \cite{MBandKMjmp1994}, Blaszak and Marciniak used the $R$ matrix approach
to derive families of integrable lattices involving three and four fields.
Below we consider two cases involving three fields.
Examples based on four fields could be dealt with in a similar fashion
\cite{Zhuetalpla2002b}.

The Blaszak-Marciniak three field lattice I \cite[Eq.\ (2)]{RSandSKjmp2001},
\begin{equation}
\label{blaszakmarciniakthreefieldI}
{\dot u} = w_1 - w_{-1}, \quad
{\dot v} = u_{-1} w_{-1} - u w, \quad
{\dot w} = w ( v - v_1 ),
\end{equation}
is invariant under the scaling symmetry
\begin{equation}
\label{blaszakmarciniakthreefieldIscale}
(t, u, v, w) \rightarrow ({\lambda}^{-1} t, \lambda^{\frac{1}{2}} u,
\lambda v, \lambda^{\frac{3}{2}} w).
\end{equation}
We computed the following density-flux pairs of
(\ref{blaszakmarciniakthreefieldI}), which is a completely integrable
lattice:
\begin{eqnarray}
\label{rhozeroblaszakmarciniakthreefieldI}
\rho^{(0)} &=& \ln w, \quad\quad\quad\quad\! J^{(0)} = v, \\
\label{rho1J1blaszakmarciniakthreefieldI}
\rho^{(1)} &=& u,
\quad\quad\quad\quad\quad
J^{(1)} = - w_{-1} - w, \\
\label{rho2J2blaszakmarciniakthreefieldI}
\rho^{(2)} &=& v,
\quad\quad\quad\quad\quad
J^{(2)} = u_{-1} w_{-1}, \\
\label{rho3J3blaszakmarciniakthreefieldI}
\rho^{(3)} &=& \tfrac{1}{2} v^2 + u w,
\quad\;\,
J^{(3)} = u_{-1} v w_{-1} - w_{-1} w, \\
\label{rho4blaszakmarciniakthreefieldI}
\rho^{(4)} &=& \tfrac{1}{3} v^3 + u v w + u v_1 w - w w_1, \\
\label{J4blaszakmarciniakthreefieldI}
J^{(4)} &=& w_{-1} (u_{-1} v^2 + u_{-1} u w - v w - v_1 w).
\end{eqnarray}
Our results confirm those in \cite{RSandSKjmp2001} and \cite{Zhuetalpla2002a}.
%
%
The Blaszak-Marciniak three field lattice II 
\cite[Eq.\ (1.4)]{Zhuetalpla2002a},
\begin{equation}
\label{blaszakmarciniaklatticethreefieldII}
{\dot u} = v_1 - v + u ( w_{-1} - w ), \quad
{\dot v} = v ( w_{-2} - w ), \quad
{\dot w} = u_1 - u,
\end{equation}
is invariant under the scaling symmetry
\begin{equation}
\label{blaszakmarciniakthreefieldIIscale}
(t, u, v, w) \rightarrow
({\lambda}^{-1} t, \lambda^2 u, \lambda^3 v, \lambda w).
\end{equation}
The first few density-flux pairs for
(\ref{blaszakmarciniaklatticethreefieldII}), which is completely integrable,
are
\begin{eqnarray}
\label{rho1J1blaszakmarciniakthreefieldII}
\!\!\!\!\!\!\!\!\!\!\!\!\rho^{(1)} \!&\!=\!&\! w,
\quad\quad\quad\quad\quad\quad\quad\quad\quad\quad\quad\quad\,
J^{(1)} = - u, \\
\label{rho2J2blaszakmarciniakthreefieldII}
\!\!\!\!\!\!\!\!\!\!\!\!\rho^{(2)} \!&\!=\!&\! \tfrac{1}{2} w^2 - u,
\quad\quad\quad\quad\quad\quad\quad\quad\quad\;
J^{(2)} = v - u w_{-1}, \\
\label{rho3J3blaszakmarciniakthreefieldII}
\!\!\!\!\!\!\!\!\!\!\!\!\rho^{(3)} \!&\!=\!&\!
\tfrac{1}{3} w^3 + v - u w_{-1} - u w,
\quad\quad\quad\;
J^{(3)} = u_{-1} u + v w_{-2} + v w_{-1} - u w_{-1}^2, \\
\label{rho4blaszakmarciniakthreefieldII}
\!\!\!\!\!\!\!\!\!\!\!\!\rho^{(4)} \!&\!=\!&\!
\tfrac{1}{4} w^4 + \tfrac{1}{2} u^2 + u u_1 + v w_{-2}
               + v w_{-1} - u w_{-1}^2 + v w - u w_{-1} w - u w^2, \\
\label{J4blaszakmarciniakthreefieldII}
\!\!\!\!\!\!\!\!\!\!\!\!J^{(4)} \!&\!=\!&\!
- u_{-2} v - u v - u_{-1} v_1 + v w_{-2}^2 + 2 u_{-1} u w_{-1}
+ v w_{-2} w_{-1} + v w_{-1}^2 - u w_{-1}^3 + u_{-1} u w.
\end{eqnarray}
Our results agree with those in \cite{RSandSKjmp2001} and
\cite{Zhuetalpla2002a}.
%
%
\vspace{-2mm}
\section{Software to Compute Conservation Laws for PDEs and DDEs}
\label{software}
We first discuss our packages for conservation laws of PDEs and DDEs,
followed by a brief summary of symbolic codes developed by other researchers.
%
\subsection{Our Mathematica and Maple Software}
\label{oursoftware}
The package {\tt TransPDEDensityFlux.m} \cite{PAandWHsoftware},
automates the computation of conservation laws of nonlinear PDEs in
$(1+1)$ dimensions.
In addition to polynomial PDEs, the software can handle PDEs with
transcendental nonlinearities.
The results in Sections~\ref{continuousmuc} and~\ref{continuousexamples}
were computed with {\tt TransPDEDensityFlux.m} and cross-checked with 
the newest version of {\tt condens.m}, introduced in \cite{UGandWHjsc1997}.
We used {\tt TransPDEDensityFlux.m} to compute the density-flux pairs
for the examples in Sections~\ref{transcendentalexamples}
and~\ref{mixedderivativePDE}.
Details about the algorithm and a discussion of implementation issues
can be found in \cite{PAthesis2003}.

The code {\tt DDEDensityFlux.m} \cite{HEandWHsoftware} was used to compute
the conservation laws in Sections~\ref{discretemuc} and~\ref{systemsDDEs}.
The results were cross-checked with the latest version of
{\tt diffdens.m}, featured in \cite{UGandWHpd1998}.
Using multiple scales, the efficiency of {\tt DDEDensityFlux.m} was
drastically improved.
Nonetheless, the algorithms \cite{HEthesis2003} within {\tt DDEDensityFlux.m}
are impractical for finding densities and fluxes of high rank.
Therefore, we used the new {\tt Maple} library {\tt discrete} 
\cite{HickmanDiscreteCode} to compute the results in
Sections~\ref{gardnerlatticesection} and~\ref{additionalexamples}.

Some of the features of earlier versions of {\tt condens.m} and
{\tt diffdens.m} were combined into the {\tt InvariantsSymmetries.m}
\cite{UGandWHpd1998,UGandWHinvsoftware1997}, which allows one to
compute generalized symmetries as well as conserved densities (but no fluxes).
{\tt InvariantsSymmetries.m} is available from MathSource,
the {\tt Mathematica} program bank of Wolfram Research, Inc.

Our {\tt Mathematica} packages and notebooks are available at
\cite{WHwebsite2004} and Hickman's {\tt Maple} code is available at
\cite{HickmanDiscreteCode}.
Our {\tt Mathematica} codes for the continuous and discrete Euler and 
homotopy operators in one dimension are available at \cite{WHwebsite2004b}.
We are currently designing a comprehensive package to compute conservation
laws of PDEs in multiple space dimensions
\cite{WHIJQC2006,WHetalbirkhauser2005,DPphdthesis2008}.

Our codes have been used in a variety of research projects.
For example, {\tt condens.m} \cite{UGandWHjsc1997} was used by
Sakovich and Tsuchida \cite{SSandTTjpa2000,TTetaljpa1999}
to compute conservation laws of nonlinear Schr\"odinger equations.
In \cite{RSandSKjmp2001}, Sahadevan and Khousalya use the algorithms of
{\tt diffdens.m} \cite{UGetalpla1997} and {\tt InvariantsSymmetries.m}
\cite{UGandWHpd1998,UGandWHinvsoftware1997} to compute conserved densities
of the Belov-Chaltikian and Blaszak-Marciniak lattices.
Ergen\c{c} and Karas\"ozen \cite{TEandBK2006} used our software in the
design of Poisson integrators for Volterra lattices.
\subsection{Software Packages of Other Researchers}
\label{softwareothers}
Our {\tt Mathematica} code for conservation laws of PDEs has been
``translated" \cite{RYandZLijmp2004,RYandZLbirkhauser2005,RYandZLiamc2006}
into a {\tt Maple} package, called {\tt CONSLAW}, which only handles PDEs
in $(1+1)$ dimensions.
Based on the concept of dilation invariance and the method of
undetermined coefficients, similar software was developed by Deconinck
and Nivala \cite{BDandMVwebsite2004} and Yang {\em et al.\/}
\cite{XYandHRandSL2007}.
Our algorithms \cite{UGandWHjsc1997,UGetalpla1997} for DDEs have been
adapted to fully-discretized equations \cite{MGetalcpc2002,MGetal2004}.

There are several algorithms (see e.g.\ \cite{TW2002}) to symbolically
compute conservation laws of nonlinear PDEs but few have been fully
implemented in CAS.
Wolf's package {\tt ConLaw} \cite{TW2002,TWwebsite2005,TWetal2003}
computes first integrals of ODEs and conservation laws of PDEs.
{\tt ConLaw} uses the {\tt REDUCE} package {\tt CRACK}
\cite{TWcrackwebsite2005,crackwolfbrand1992,crackwolfbrand1995},
which contains tools to solve overdetermined systems of PDEs.
Wolf's application packages heavily rely on the capabilities of
{\tt CRACK}, which took years to develop and perfect.
Unfortunately, no such package is available in {\tt Mathematica}.

A common approach is to use the link between conservation laws and
symmetries as stated in Noether's theorem
\cite{IAbook2004,IKandAVbook1998,PObook1993}.
Among the newest software based on that approach is the {\tt Maple} code 
{\tt GeM} \cite{ACwebsite2008} by Cheviakov \cite{ACcpc2007,ACissac2008},
which allows one to compute conservation laws of systems of ODEs and PDEs
based on the knowledge of generalized symmetries.
However, the computation of such symmetries \cite{WHbook1996} is as
difficult a task as the direct computation of conservation laws for it
requires solving systems of overdetermined PDEs with, e.g., the {\tt Rif}
package \cite{WHbook1996,AWandGR2005}}.
Some methods circumvent the existence of a variational principle (required
by Noether's theorem) \cite{SAandGB2002a,GBandSA2002,TW2002,TWetal2003}
but they still rely on software to solve ODEs or PDEs.

The package \verb|DE|$\_$\verb|APPLS| \cite{IAvessiot2004,EMvessiot1999}
also offers commands for constructing conservation laws from (variational)
symmetries by Noether's theorem, but the computation is not fully automated.
Likewise, the package {\tt Noether} \cite{PGandDTwebsite2005} in
{\tt Maple} allows one to compute conservation laws from infinitesimal
symmetry generators corresponding to (simple) Lagrangians.

Based on the formal symmetry approach, Sokolov and Shabat
\cite{VSandASinbook1984},
Mikhailov {\em et al.\/} \cite{AMetalinbook1991,AMetal1987}, and 
Adler {\em et al.\/} \cite{VAetaltmp2000} classified completely 
integrable PDEs and DDEs in $(1+1)$ dimensions.
Unfortunately, the software used (see \cite{UGandWHjsc1997}) in the
classification is obsolete.

For completeness, we also mention the packages {\tt Jets} by Marvan
\cite{MMjets2003} and {\tt Vessiot} by Anderson
\cite{IAvessiot2004,EMvessiot1999}.
Both are general purpose suites of {\tt Maple} packages for computations on
jet spaces.
The commands within {\tt Jets} and {\tt Vessiot} use differential forms and
advanced concepts from differential geometry.
By avoiding differential forms, our codes were readily adaptable to nonlinear
DDEs (not covered in {\tt Jets} and {\tt Vessiot}).

Finally, Deconinck and Nivala \cite{BDandMNmcs2008} developed 
{\tt Maple} software for the continuous and discrete homotopy operators.
Their code is available at \cite{BDandMNwebsite2008}.
\vspace{-2mm}
\section{Summary}
\label{conclusions}
We presented methods to symbolically compute conservation laws of nonlinear
polynomial and transcendental systems of PDEs in $(1+1)$ dimensions and
polynomial DDEs in one discrete variable.

The first part of this chapter dealt with nonlinear PDEs for which we showed
the computation of densities and fluxes in detail.
Using the dilation invariance of the given PDE, candidate polynomial densities
are constructed as linear combinations with undetermined coefficients of
scaling invariant building blocks.
For polynomial PDEs, the undetermined coefficients are constants which must
satisfy a linear system of algebraic equations.
That system will be parameterized by constants appearing in the PDE, if any.
For transcendental PDEs, the undetermined coefficients are functions which
much satisfy a linear system which is a mixture of algebraic equations 
and ODEs.

The continuous homotopy operator is a powerful, algorithmic tool to compute 
fluxes explicitly.
Indeed, the homotopy operator handles integration by parts which allowed us
to invert the total derivative operator.
The methods for polynomial PDEs are illustrated with classical examples such
as the KdV and Boussinesq equations and the Drinfel'd-Sokolov-Wilson system.
The computation of conservation laws of system with transcendental
nonlinearities is applied to sine-Gordon, sinh-Gordon, and Liouville equations.

In the second part we dealt with the symbolic computation of conservation
laws of nonlinear DDEs.
Again, we used the scaling symmetries of the DDE and the method of
undetermined coefficients to find densities and fluxes.
In analogy with the continuous case, to compute the flux one could use the 
discrete homotopy operator, which handles summation by parts and inverts 
the forward difference operator.
However, in comparison with the ``splitting and shifting" technique,
the discrete Euler and homotopy operators are inefficient tools for the
symbolic computation of conservation laws of DDEs.
The undetermined coefficient method is illustrated with classical examples
such as the Kac-van Moerbeke, Toda and Ablowitz-Ladik lattices.

There is a fundamental difference between the continuous and discrete cases
in the way densities (of selected rank) are constructed.
The total derivative has a weight but the shift operator does not.
Consequently, a density of a PDE is bounded in order (with respect to $x).$
Unfortunately, there is no {\em a priori} bound on the number of shifts in the
density, unless a leading order analysis is carried out.
To overcome this difficulty and other shortcomings of the undetermined
coefficient method, we presented a new method to compute conserved densities
of DDEs.
That method no longer uses dilation invariance and is no longer restricted to
polynomial conservation laws.
Instead of building a candidate density with undetermined coefficients,
one first computes the leading order term in the density and, second,
generates the correction terms of lower order.
The method is fast and efficient since no unnecessary terms are computed.
The new method was illustrated using a modified Volterra lattice as
an example, and applied to lattices due to Bogoyavlenskii, Belov-Chaltikian,
Blaszak-Marciniak, and Gardner.
A derivation of the latter lattice was also given.
%
\section*{Acknowledgements}
\label{acknowledgements}
This material is based in part upon work supported by the National Research
Foundation (NRF) of South-Africa under Grant No.\ FA2007032500003.

Jan Sanders (Free University of Amsterdam) and Bernard Deconinck 
(University of Washington) are thanked for valuable discussions.
Loren ``Douglas" Poole is thanked for proof reading the manuscript.

Willy Hereman is grateful for the hospitality and support of the Department of
Mathematics and Statistics of the University of Canterbury
(Christchurch, New Zealand) and the Applied Mathematics Division of the
Department of Mathematical Sciences of the University of Stellenbosch
(Stellenbosch, South Africa) during his sabbatical visits in AY 2007-2008.
\vskip 5pt
\noindent
%
%

\end{document}